\documentclass[11pt]{article}
\usepackage[english]{babel}
\usepackage{graphicx}
\usepackage{hyperref}
\usepackage{amsmath}
\usepackage{amssymb}
\usepackage{amsthm}
\usepackage{natbib}
\usepackage{enumerate}
\usepackage{float}
\usepackage{bbm}
\usepackage{rotating}
\usepackage{dcolumn}
\usepackage{booktabs}
\usepackage{subcaption}
\usepackage{textcomp}
\usepackage{pdfpages} 

\allowdisplaybreaks[2]

\usepackage{color}
\definecolor{DarkBlue}{rgb}{0,0.18,0.55}
\usepackage{hyperref}
\hypersetup{pdfauthor={D. Kim},colorlinks=true,citecolor=DarkBlue,filecolor=DarkBlue,linkcolor=DarkBlue,urlcolor=DarkBlue,pdftex}

\theoremstyle{definition}
\usepackage{xargs}
\usepackage{array}
\usepackage{booktabs}
\usepackage[colorinlistoftodos,prependcaption,textsize=tiny]{todonotes}
\newcommandx{\change}[2][1=]{\todo[linecolor=blue,backgroundcolor=blue!25,bordercolor=blue,#1]{#2}}

\theoremstyle{plain}

\date{}
\begin{document}
	
	\title{Vaccination strategies and transmission of COVID-19: evidence across advanced countries\thanks{We thank our editor, Owen O'Donnell, two anonymous referees, and also Alexander Karaivanov and Hiro Kasahara for excellent comments and suggestions.}}
	\author{Dongwoo Kim\thanks{Corresponding author. Department of Economics, Simon Fraser University, 8888 University Drive, Burnaby, BC, V5A 1S6, Canada. Email: dongwook@sfu.ca}\\ Simon Fraser University \and Young Jun Lee\thanks{Department of Economics, University of Copenhagen, Øster Farimagsgade 5, Building 26, DK-1353 Copenhagen K, Denmark. Email: yjl@econ.ku.dk.}\\ University of Copenhagen}
	\date{\today}
	\maketitle
	
	\thispagestyle{empty}
	\begin{abstract}
		Given limited supply of approved vaccines and constrained medical resources, design of a vaccination strategy to control a pandemic is an economic problem. We use time-series and panel methods with real-world country-level data to estimate effects on COVID-19 cases and deaths of two key elements of mass vaccination – time between doses and vaccine type. We find that new infections and deaths are both significantly negatively associated with the fraction of the population vaccinated with at least one dose. Conditional on first-dose coverage, an increased fraction with two doses appears to offer no further reductions in new cases and deaths. For vaccines from China, however, we find significant effects on both health outcomes only after two doses. Our results support a policy of extending the interval between first and second doses of vaccines developed in Europe and the US. As vaccination progresses, population mobility increases, which partially offsets the direct effects of vaccination. This suggests that non-pharmaceutical interventions remain important to contain transmission as vaccination is rolled out.
	\end{abstract}
	
	
	 
	\newpage
	
	\section{Introduction}
	
	Vaccination has been perceived as key to reaching herd immunity since the COVID-19 pandemic was declared in 2020. This paper investigates the effects of two key elements of mass vaccination--allocation and timing of first and second doses and the type of vaccine implemented. Mass vaccination programs have been launched in advanced countries since the UK first approved the use of Comirnaty (previously the Pfizer–BioNTech vaccine) in December 2020. Due to the very limited supply of approved vaccines in the early stage, however, even the world's wealthiest countries had difficulties rolling out vaccine doses quickly. Most COVID-19 vaccines require two doses to be administered per person within a time window recommended by manufacturers. Three or four weeks are recommended for Comirnaty and Spikevax (previously the Moderna vaccine) and 8--12 weeks for Vaxzevria (previously the AstraZeneca-Oxford vaccine). Under these supply constraints, countries adopted different vaccine roll-out strategies. The UK and Canada extended the interval between first and second doses up to 12--16 weeks in order to maximize the share of (at least partially) vaccinated people, whereas many other countries, such as Israel and the US, followed the recommended dosing interval. Some countries, such as Bahrain, Chile, Turkey, the UAE, and Uruguay, relied on less widely approved vaccines developed by Chinese firms.
	
	Implementing a longer interval between doses sparked a heated debate among medical experts and policy makers. In the UK, the British Medical Association (BMA) stood strongly against the idea of delaying the second dose (\cite{bma_second_dose}) as there is no clinical evidence that this strategy works. A similar criticism arose in Canada (\cite{cbcnews2021}). Nonetheless, both governments implemented longer intervals to vaccinate as many people as possible. There have been a handful of papers in the medical literature that suggest extending the interval between two doses might be beneficial. \cite{moghadas2021evaluation} use agent-based modeling to compare vaccination strategies. They found that delaying the second dose can avert more infections, hospitalizations, and deaths than the standard dosing if the efficacy of the first dose does not wane before the delayed second dose. Similar results using simulation-based methods are found by \cite{Saad_Roy_2021}, \cite{romero2021public} and \cite{tuite2021alternative}. To the best of our knowledge, however, there is no research on real-world outcomes across different countries. This paper aims to provide observational cross-country evidence on the effectiveness of vaccination strategies using real-world data.
	
	The choice of vaccine is potentially even more critical to the effectiveness of a mass vaccination program than the timing of vaccine doses. After inoculating more than a half of the population, a few countries, such as Chile, the UAE, and Seychelles, still experienced high rates of infection. These countries heavily relied on Chinese vaccines (Sinopharm, Sinovac Biotech, and CanSino).\footnote{The recommended dosing interval for these vaccines is 2--4 weeks, according to the \cite{world2021background}).} Vaccines from China can be a useful tool to fight the pandemic for low- and middle-income countries, as the supply of vaccines developed in the US and Europe is limited and low- and middle-income countries cannot compete with the wealthiest countries to secure enough stocks of those vaccines. We include countries relying on Chinese vaccines in our analysis to see whether and how much those vaccines contributed to mitigating disease transmission.
	
	The choice of a vaccination strategy is essentially a general economic problem. The resources for mass vaccination are constrained. Not only is the supply of vaccines limited, medical facilities and healthcare workers are capacity constrained. Moreover, the effectiveness of vaccines tends to wane over time. Therefore, policy makers should devise an optimal allocation strategy under these constraints to maximize the overall effect of mass vaccination (\cite{kitagawa2021should}). In this regard, our analysis can provide lessons that go beyond the context of the COVID-19 pandemic. We deliver evidence on whether the interval between first and second doses matters to contain the transmission of COVID-19. We also take the types of vaccines used in mass vaccination into account.
	
	We employ standard time-series and panel data frameworks with cross-country data to estimate the effects of first- and second-dose vaccination on the COVID-19 transmission conditional on potential confounders. We first estimate time-series models of new infections and deaths for 8 selected countries with high vaccination rates (Canada, Israel, the US, the UK, Chile, Uruguay, the UAE, and Bahrain). Then, we closely follow the approaches employed by \cite{chernozhukov2021causal} and \cite{karaivanov2021face} for panel data analysis in which we investigate the impact of vaccination on new cases and deaths, as well as behavior i.e. people's mobility. Member countries in the OECD and the European Union that publicly release daily epidemiological data are considered. Bahrain and Uruguay are added to the country panel to evaluate the effects of vaccines from China, along with Chile and Turkey, which are already included in the OECD.\footnote{Other countries that heavily relied on Chinese vaccines, such as the UAE and Seychelles, are omitted in the panel data analysis as they do not provide daily vaccination statistics.}
	
	The main findings from our empirical analysis are as follows. First, growth of new infections and deaths are significantly negatively associated with progress in administering at least one dose. Larger reductions in both these health outcomes are found for countries delaying the second dose (Canada and the UK). Conditional on first-dose progress, two-dose coverage does not give further reductions in new cases and deaths. These findings are consistent in both time-series and panel data analyses. No reductions in health outcomes are found for first-dose progress in countries heavily relying on vaccines from China, whereas two-dose progress offers significant effects on new cases and deaths in these countries. Second, we find that progress in vaccination induces people to be more mobile. As higher mobility leads to more new infections and deaths, this indirect effect of vaccination partially offsets its direct effect. Last, our counterfactual simulations for selected countries suggest that substantially extending the interval between two doses reduces new cases and deaths.
	
	Our findings can be useful to draw policy implications for low-income countries that fall behind in mass vaccination. Only 11.84 vaccine doses per 100 people were administered overall in low-income countries as of December 31, 2021, according to \textit{Our World in Data}.\footnote{\url{https://ourworldindata.org/grapher/cumulative-covid-vaccinations-income-group?country=High+income~Low+income~Lower+middle+income~Upper+middle+income}.} As many advanced countries have started administering booster shots (third doses), the supply of mRNA-based vaccines and alternative vaccines will continue to be limited. Thus, it is important to vaccinate as many people as possible with one dose using secured vaccine stocks. Our analysis, alongside the evidence found in the medical literature, suggests that extending the dosing interval is an effective strategy to mitigate the transmission of COVID-19. Especially in low-income countries, not only is the supply of vaccines constrained, the medical capacity to administer the vaccine doses is also limited. Thus, this strategy can further help low-income countries maximize the level of protection among the population more quickly. As observed in many countries, the progress in single-dose vaccination starts to flatten out at a certain point. Thus, countries can begin to vaccinate with the second dose as resources become available. Public health measures are still very important amid vaccination progress as they can further contain the spread of the virus by reducing mobility and potential interactions among people. There is a caveat in our findings for the use of vaccines from China: single-dose vaccination is not enough to alleviate the spread of the virus. Therefore, the importance of second-dose vaccination should be emphasized if those vaccines are used in a significant portion of mass vaccination.
	
	The remainder of this paper is structured as follows. Section 2 describes the rationale behind the idea of delaying the second dose. Section 3 provides empirical analysis using time series and panel data models. Section 4 discusses the robustness of our findings. Section 5 presents counterfactual simulations for selected countries using hypothetical vaccine allocations. Section 6 concludes. Additional empirical results, robustness analysis using alternative data frequency/model specifications, and additional counterfactual simulations are provided in Supplementary Materials for online publication.
	
	\section{Rationale Behind Delaying the Second Dose}
	The UK government decided to extend the interval between the first and second doses of Comirnaty up to 12 weeks at the end of December 2020. Two letters sent to health professionals (\cite{nhsletter} and \cite{dhsc}) lay out the rationale for delaying the second dose. The main reason is that the great majority of protection comes after the first dose. The second dose is important for the duration of protection, but in the short run, the additional protection afforded by the second dose is likely to be marginal. Later, in Canada, the \cite{naci} recommended that the mRNA-based vaccines and Vaxzevria should be given to as many people as possible by extending the dosing interval up to 4 months.
	
	These decisions are based on three assumptions: 1. the first dose provides good enough protection (the first dose's efficacy is higher than the marginal efficacy gain from the second dose); 2. protection after the first dose does not wane too quickly; 3. delaying the second dose does not lower the efficacy of full vaccination. Under these assumptions, we conduct a very simple thought experiment, without a complicated epidemiological model. Suppose that a country secures an amount of mRNA-based vaccines that can inoculate $70$\% of the population with one dose. There will be a second batch of vaccines delivered to this country in 12--16 weeks. The country has to allocate its vaccine stock to the first and second doses before the delivery of the second batch. We consider two cases in which the level of protection from the first dose substantially varies. In Case 1, the first dose provides 90\% vaccine efficacy (VE), which does not wane in 12--16 weeks, whereas first-dose VE is only 50\% effective in Case 2. In both cases, the VE after the second dose reaches 95\%, so that the marginal efficacy gain from the second dose is 5\% and 45\% respectively. Case 1 resembles the vaccine efficacy results for the mRNA-based vaccines from the clinical trials.\footnote{\cite{polack2020safety} reported that the vaccine efficacy of Comirnaty between the first and second doses is $52.4\%.$ However, they used data collected during the first two weeks after the first dose in their calculation. \cite{NEJMc2036242} reanalyzed Comirnaty efficacy between first and second doses using data submitted to the Food and Drug Administration (FDA) and found that first-dose efficacy is $92.6\%$ from two weeks after the first dose and before the second dose. This is similar to the first-dose efficacy of Spikevax ($92.1\%$).} Case 2 supposes a situation where variants of the virus substantially lower first-dose VE.
	
	\begin{table}[ht!]
		\caption{Vaccine allocations and overall protection}
		\label{tab: vaccine efficacy calculation}
		\begin{center}
		\begin{tabular}{ccccccc}
			\\[-1.8ex]\cline{1-3} \cline{5-7}
			\multicolumn{3}{c}{Case 1} & &\multicolumn{3}{c}{Case 2}\\
			\multicolumn{3}{c}{\small(first dose VE: 90\%, second dose VE: 95\%)} & &\multicolumn{3}{c}{\small(first dose VE: 50\%, second dose VE: 95\%)}\\
			\cline{1-3} \cline{5-7}
			first dose & second dose & protection & & first dose & second dose & protection\\
			\cline{1-3} \cline{5-7}
			70\% & 0\% & 63.0\% & & 70\%& 0\% & 35.0\%\\
			60\% & 10\%& 54.5\% & & 60\%& 10\% & 34.5\%\\
			50\% & 20\%& 46.0\% & & 50\%& 20\% & 34.0\%\\
			40\% & 30\%& 37.5\% & & 40\%& 30\% & 33.5\%\\
			35\% & 35\%& 33.3\% & & 35\%& 35\% & 33.25\%\\
			\cline{1-3} \cline{5-7}
		\end{tabular}
		\end{center}
	\footnotesize \textbf{Note:} The table shows the average protection levels given different vaccine allocations. The protection levels are computed by (first dose VE $\times$ first dose fraction) + (second dose marginal VE $\times$ second dose fraction).
	\end{table}
	
	We compute the average protection levels among the population given different allocations of the first and second doses in Table \ref{tab: vaccine efficacy calculation}. In both cases, allocating all the vaccine stock to the first dose is the dominant strategy. Delaying the second dose is obviously much more effective when the first dose gives stronger protection, as shown in Case 1. The allocation strategy does not matter if first-dose efficacy is identical to the marginal efficacy gained from the second dose. This simple calculation illustrates why the longer interval between doses can be beneficial under supply constraints if the underlying assumptions hold. Recent findings in the medical literature support our assumptions. \cite{carazosingle} found that first-dose mRNA-based vaccines provide solid protection until 16 weeks after vaccination using data from Quebec, Canada. There is also medical evidence that the longer dosing interval increases the overall efficacy. \cite{voysey2021safety} showed that the longer interval ($\ge 12$ weeks) between doses of Vaxzevria provided a greater efficacy than the shorter interval ($< 6$ weeks). A recent study in the UK (\cite{payne2021immunogenicity}) found that longer dosing intervals (6--14 weeks) resulted in stronger immune responses than the standard regimen (3--4 weeks) for Comirnaty. \cite{tauzin2021strong} also support that a longer (16 weeks) interval induced more robust immune responses than a shorter interval (4 weeks) for Cominarty. \cite{pouwels2021effect}, on the other hand, observed no effect of the dosing interval on vaccine effectiveness. Another possibility is that one dose of vaccine could provide stronger protection for people previously infected. \cite{angyal2021t} reported that previously infected healthcare workers in the UK showed strong immune responses after only one dose, equivalent to those of people receiving two doses with no past history. Similar findings are shown in recent studies (\cite{stamatatos2021mrna}, \cite{samanovic2021poor}, and \cite{tauzin2021strong}). In the latter two studies, the authors find that the second dose did not offer significant antibody responses for individuals with prior exposure. This means that single-dose vaccination can give full protection to a significant part of the population for countries with a large number of cumulative cases.
	
	\begin{figure}[ht!]
		\begin{center}
			\caption{Vaccination strategies and overall protection over time}\label{fig: strategies}
			\vspace{-3mm}
			
			\begin{subfigure}[h!]{\textwidth}
			\begin{center}
			\caption[]%
			{{\small Three vaccination strategies}}
			\includegraphics[width=\textwidth]{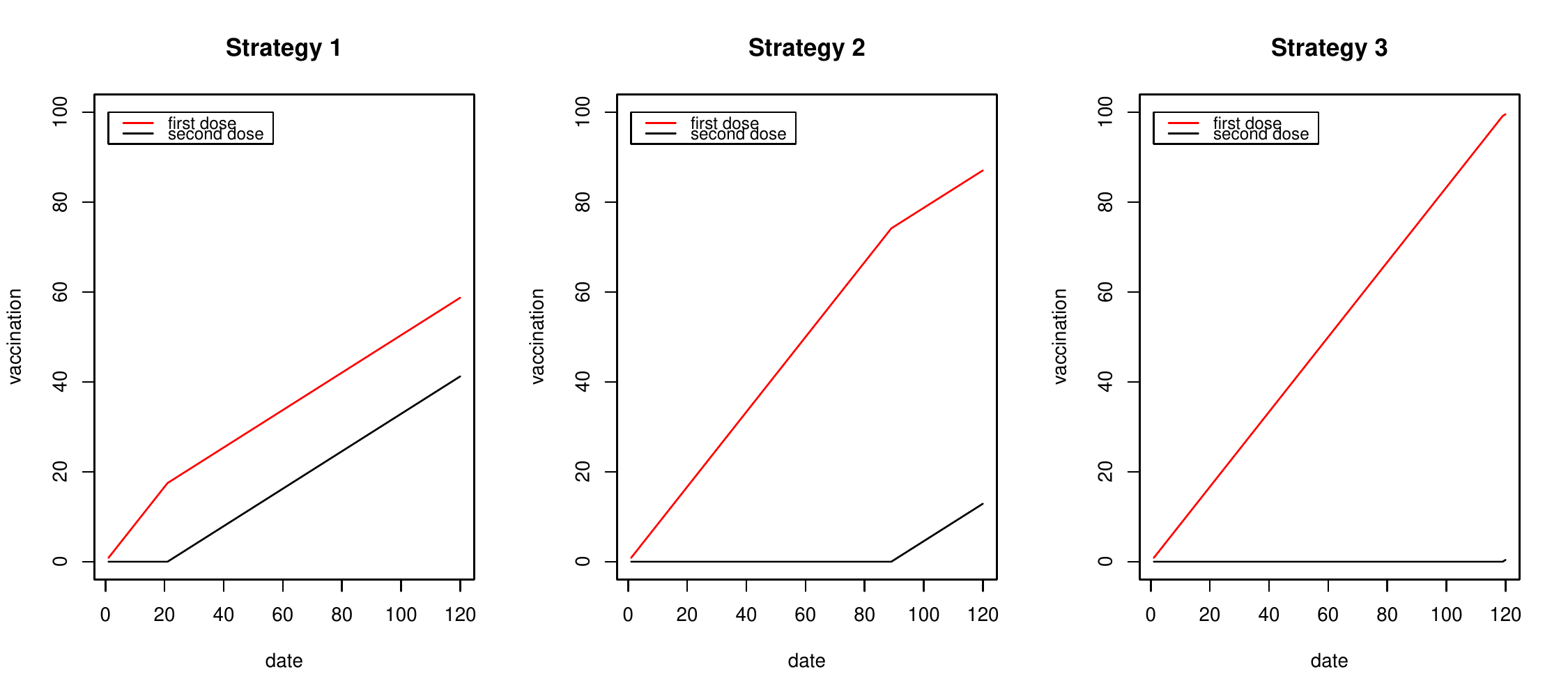}
			\end{center}
			\footnotesize \textbf{Note:} The figures plot the number of people vaccinated with at least one dose per hundred (red line) and the number of people vaccinated with two doses per hundred (black line) over time given different allocation strategies. 
			\end{subfigure}
			\vspace{5mm}
			
			\begin{subfigure}[h!]{\textwidth}
			\begin{center}
			\caption[]%
			{{\small Overall protection over time}}
			\includegraphics[width=\textwidth]{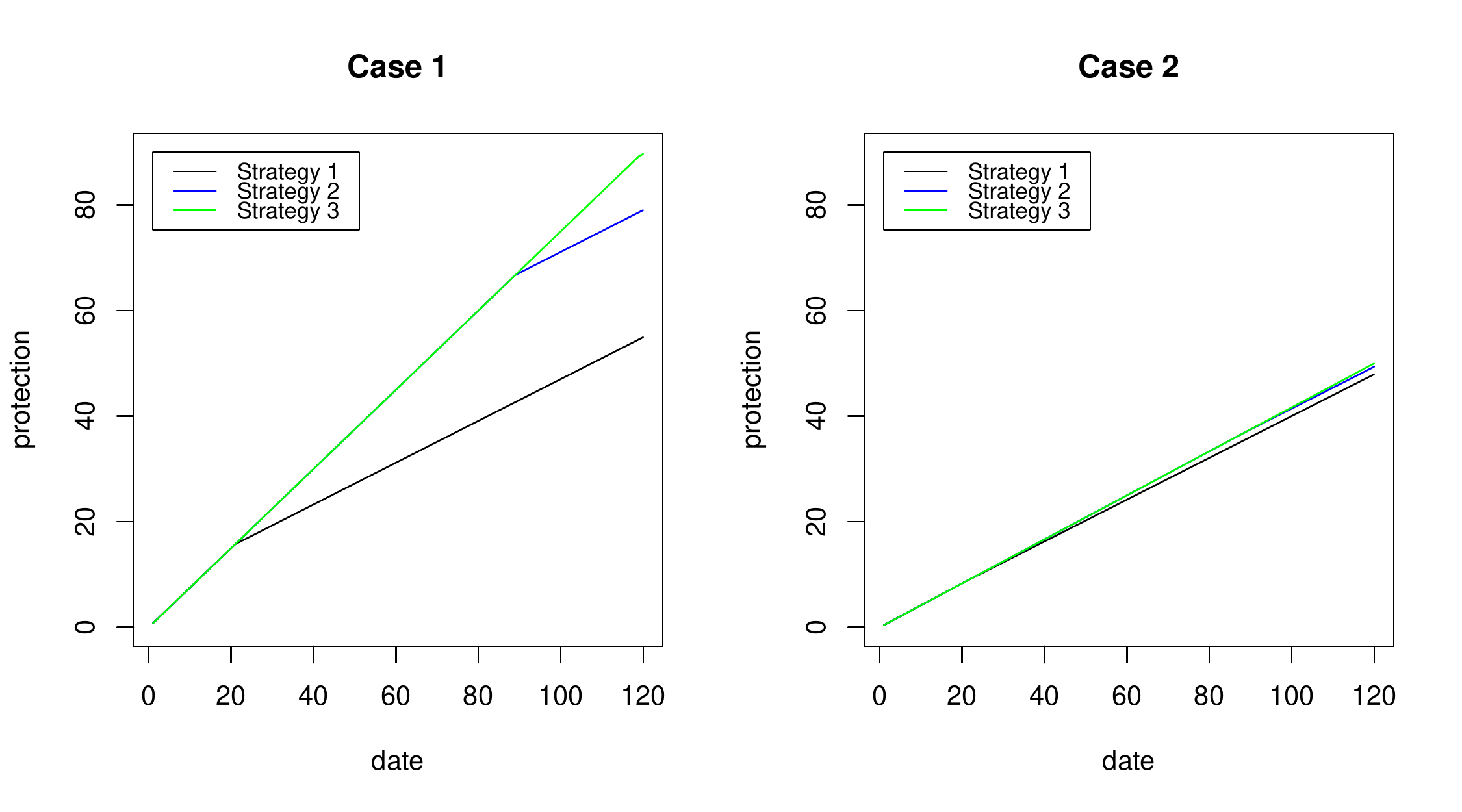}
			\end{center}
			\footnotesize \textbf{Note:} The figures plot the average protection levels over time given different allocation strategies for Cases 1 and 2 described in Table \ref{tab: vaccine efficacy calculation}. 
			\end{subfigure}
		\end{center}
	\end{figure}

	Consider another thought experiment in which a country secures 100 doses per hundred people and will be inoculating its population with the vaccine stock over 120 days. On top of the limited supply of vaccines, health care capacity limits the number of daily doses that can be administered. We assume that medical capacity allows to deliver $100/120$ doses per hundred people daily. Three vaccination strategies on the dosing interval are considered. In Strategy 1, the dosing interval is standard 3 weeks, so the country delivers first doses from day 1 and starts to inoculate second doses on day 21. From this time, the stocks are equally allocated to the first and second doses. In strategies 2 and 3, the dosing intervals are 3 months (90 days) and 4 months (120 days) respectively. We compute the overall protection levels under these strategies for both cases in Table \ref{tab: vaccine efficacy calculation}. As shown in Figure \ref{fig: strategies}, longer dosing intervals help achieve higher protection levels in the given time period. It is obvious that Strategy 3 gives the best results in both cases, followed by Strategy 2.
	
	In sum, there is evidence in the medical literature from which one can expect that the idea of delaying the second dose may work in practice. Allocating the available vaccine stocks to first doses could help maximize the level of protection in the population given the limited supply. Furthermore, there might be healthcare capacity constraints. Focusing on the first doses by extending the interval between doses could also help reach a higher level of protection more quickly. Extending the dosing interval is particularly more effective when the first-dose vaccine efficacy is much higher than the marginal efficacy gain from the second dose. This could be the case in low-income countries where a significant part of the population is previously infected. However, it is still uncertain whether this strategy actually works in the real-world setting, as the previous studies rely on small-scale experiments or simulations. Therefore, it is important to investigate the performances of different vaccination strategies using observational data across countries employing different vaccination strategies.
	
	\section{Empirical Analysis}
	\subsection{Data}
	Our empirical analysis mainly relies on the country-level epidemiological data from \textit{Our World in Data} (\url{https://ourworldindata.org/}) which is a collaborative project between the University of Oxford and \textit{Global Change Data Lab}. The worldwide country-level database on many aspects of the COVID-19 pandemic is in its GitHub repository (\url{https://github.com/owid/covid-19-data/tree/master/public/data}). We use the daily counts of new cases, total cumulative cases, new tests, total cumulative tests, vaccinated people with at least one dose per hundred, fully vaccinated people per hundred, and total vaccine doses administered per hundred.\footnote{The Janssen (Johnson and Johnson) vaccine only requires a single dose. If people were immunized using Janssen vaccines, they are counted as fully vaccinated. However, this vaccine was used only for a very small fraction of mass vaccination in most countries (3.8\% in the US, 2.3\% in the European Union, and 0\% for Canada, Israel, Chile, and Uruguay as of July 8, 2021), so we suppose the effect of vaccination mainly comes from the other vaccines, on which countries heavily relied.} For government policy responses, we employ the Containment and Health Index (\url{https://ourworldindata.org/policy-responses-covid#containment-and-health-index}) developed by the Oxford COVID-19 Government Response Tracker (OxCGRT).\footnote{This index is built upon OxCGRT's Stringency Index which is a composite measure based on nine policy response indicators including school closures, workplace closures, public event bans, restrictions on gatherings, public transit closure, public information campaigns, stay at home order, domestic and international travel bans rescaled to a value from 0 to 100 (100 = strictest). Containment and Health Index further includes testing policy, contact tracing, face coverings, and vaccine availability. If policies vary at the sub-national level, the index is shown as the response level of the strictest sub-region.} There are some missing values in this data set. We impute the missing values using linear interpolation for weekdays; if values in the weekends are missing, we take the previous value to impute the missing values. We also use the country-level data from Google COVID-19 Community Mobility Reports (\url{https://www.google.com/covid19/mobility/}) to  take people's behavioral responses into account. These data show mobility trends over time in each country across different categories of places compared to the baseline period (Jan 3--Feb 6, 2020) before the pandemic. Following \cite{karaivanov2021face}, we construct a mobility index by calculating the arithmetic average of three indexes (`retail,' `grocery and pharmacy,' and `workplace').
	
	We use the period from June 1, 2020, to July 8, 2021, for our analysis. The first wave of the pandemic was almost over across most countries around the start date chosen. Many countries did not have enough testing--tracing infrastructure early in the pandemic, so the data would have missed a significant amount of infection cases. After the first wave, most countries were well equipped with testing--tracing facilities, as one can see from the substantially increased number of daily new tests. This period includes the second wave, and the wide spread of the Delta variant first identified in India is captured in some countries, such as the UK and Israel.
	
	We estimate time-series and panel data models to investigate the effects of vaccination on health outcomes controlling for policy and behavioral factors. In our time-series analysis, we use a sample of countries with high vaccination rates. We select eight nations, four of which (Canada, Israel, the US, and the UK) have used vaccines developed in the US and Europe, while the others (Chile, Uruguay, the UAE, and Bahrain) have been heavily reliant on vaccines produced by China.\footnote{Canada, Israel, and the US relied on mRNA-based vaccines (Comirnaty and Spikevax) to inoculate most of the population (the whole population for Israel). The UK has used both mRNA-based vaccines and Vaxzevria. Chile, Uruguay, the UAE, and Bahrain have vaccinated a significant part of the population with vaccines from China. According to \textit{Our World in Data}, As of July 8, 2021, Chile administered $76.3\%$ of its total doses with Chinese vaccines (17.72 million doses of Sinovac and 0.47 million doses of CanSino out of total 23.93 million doses), and Uruguay's share of Sinovac is $73.6\%$ (3.12 million doses of Sinovac out of total 4.24 million doses). The UAE and Bahrain also inoculated a majority of the population using the Sinopharm vaccine, but the exact numbers are not publicly available.} These countries have achieved the highest vaccination rates in the world, and relevant data are well kept and gathered at daily frequency. Then, we employ a larger panel of high-income countries (OECD and EU members, Bahrain and Uruguay) to further investigate the impact of vaccination progress exploiting both time-series and cross-country variations.
	
	\begin{figure}[p]
		\begin{center}
			\caption{Trends in new cases, vaccinations, and policy and mobility indexes}
			\label{fig: data trends}
			\includegraphics[width=0.9\textwidth]{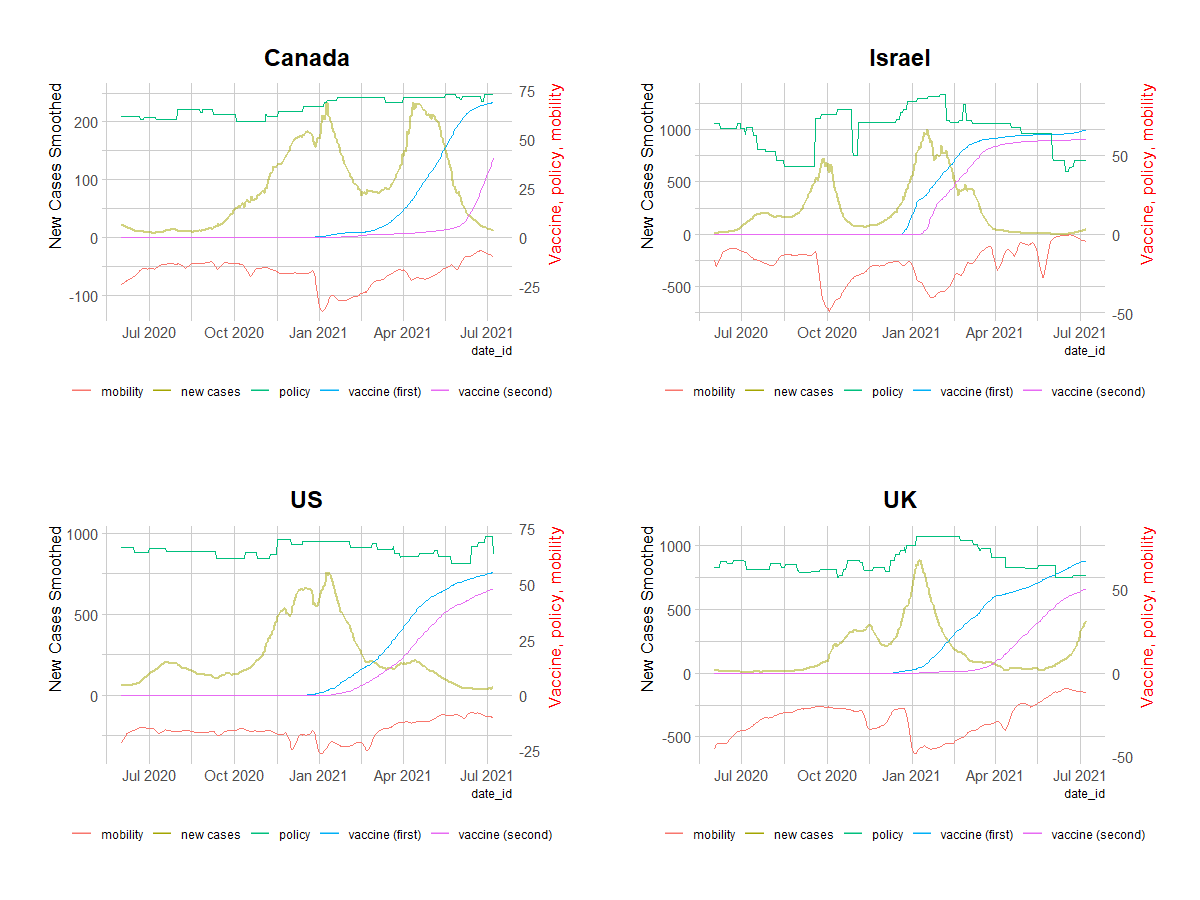}
			\includegraphics[width=0.9\textwidth]{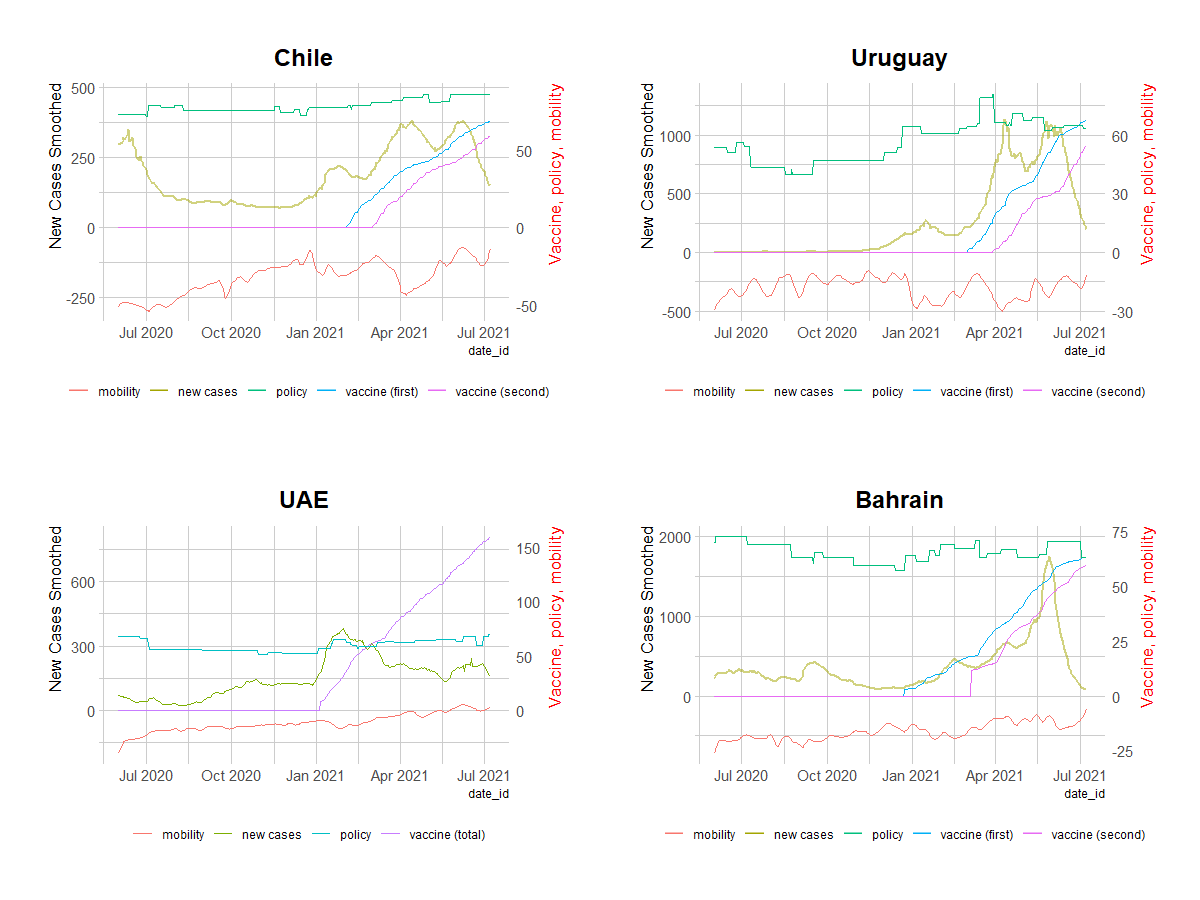}
		\end{center}
		\footnotesize\textbf{Note}: Time period -- June 1, 2020 to July 8, 2021. The UAE does not publicly report daily numbers of first and second doses administered until early July, 2021.
	\end{figure}
	
	We draw trends in variables of primary interest (new confirmed cases per million, vaccination progress, Containment and Health Index, and the mobility index) in Figure \ref{fig: data trends} for eight selected countries we use in our time-series analysis. Among those countries, only Canada and the UK extended the dosing interval for mRNA-based vaccines up to 12--16 weeks. The other countries recommended the standard regimen (3--4 weeks; the US allowed up to a 6-week interval in exceptional circumstances). As these countries adopted different vaccination strategies, the effectiveness of mass vaccination could vary across countries. In the upper four panels (Canada, Israel, the US, and the UK), the daily new cases per million (smoothed by a 7-day moving average) decreased as the vaccination program progressed. The government policy index responded to the spike in new cases. All four countries eased restrictions as vaccination progressed and the case number declined. The mobility index showed the opposite trend, as it dropped in line with the surge in new cases and recovered up as the case number decreased. It is not clear whether people's behavior responded to government restrictions or case numbers. Mass vaccination obviously led to lower new infections, but at the same time, government restrictions were eased and people's mobility increased. Therefore, the estimated effects of vaccination on new infections are likely to be attenuated if policy and behavioral factors are not taken into account.
	
	Different trends were observed in the lower four panels (Chile, Uruguay, the UAE, and Bahrain). These countries heavily relied on vaccines from China and experienced the surge of new cases while mass vaccination continued through the period at a steady pace. Declines in the growth of new cases were seen only after very high vaccination rates were achieved and stronger restrictions were implemented. These trends cast some doubt on the effectiveness of Chinese vaccines. Mobility and policy indexes responded to new cases similarly to the way they did in the countries in the upper four panels. This may suggest that countries relying on vaccines from China have to employ a different dosing strategy.
	
	\subsection{Time series analysis}
	We analyze the evolution of COVID-19 transmission using standard time-series models. We use the ARIMA model (autoregressive integrated moving average models with exogenous variables) to estimate the effect of first and second doses on new infections. The outcome variable is log of daily new cases per million. The number of people vaccinated per hundred (at least one dose given) and the number of people fully vaccinated per hundred (two doses) are included as exogenous variables. To control for test intensity and the weekend effect, the log of the number of new tests and a weekend dummy are also included.\footnote{For Israel, the UAE, and Bahrain, the weekend dummy takes 1 if the day is either Friday or Saturday. For the other countries, the dummy variable equals 1 when the day is either Saturday or Sunday.} Government policies and people's behavioral changes are also taken into account by including the OxCGRT Containment and Health Index and the mobility index from Google Community Mobility Reports.
	
	As vaccinations, government policies, and behavioral responses do not immediately affect infections, it is important to choose appropriate lags for exogenous variables in the ARIMA models. For first and second vaccine doses, we choose 21-day and 7-day lags, respectively. These lags are chosen based on scientific evidence from clinical trials and real-world outcomes. The first dose of approved vaccines begins to provide substantial protection against infection after 2--3 weeks, and at least 14 days are required to see protection start (see \cite{naci} for a detailed survey and \cite{hunter2021estimating} for evidence on Comirnaty from Israel, where the estimated vaccine effectiveness reached its peak at day 21 after the first dose). Then, 7 days after the second dose, the vaccine efficacy further enhances as shown in \cite{polack2020safety} and clinical trials. Medical studies such as \cite{hall2021covid} also used 21-day and 7-day thresholds for the first and second dose, respectively, to evaluate the vaccine effectiveness. We choose 14-day lags for the government policy index and the mobility index, as the same lags were used in \cite{chernozhukov2021causal} and \cite{karaivanov2021face} after careful investigation.
	
\begin{figure}[t]
		\begin{center}
			\caption{Daily new cases and fitted values}
			\label{fig: ARIMAX model fit}
			\includegraphics[width=\textwidth]{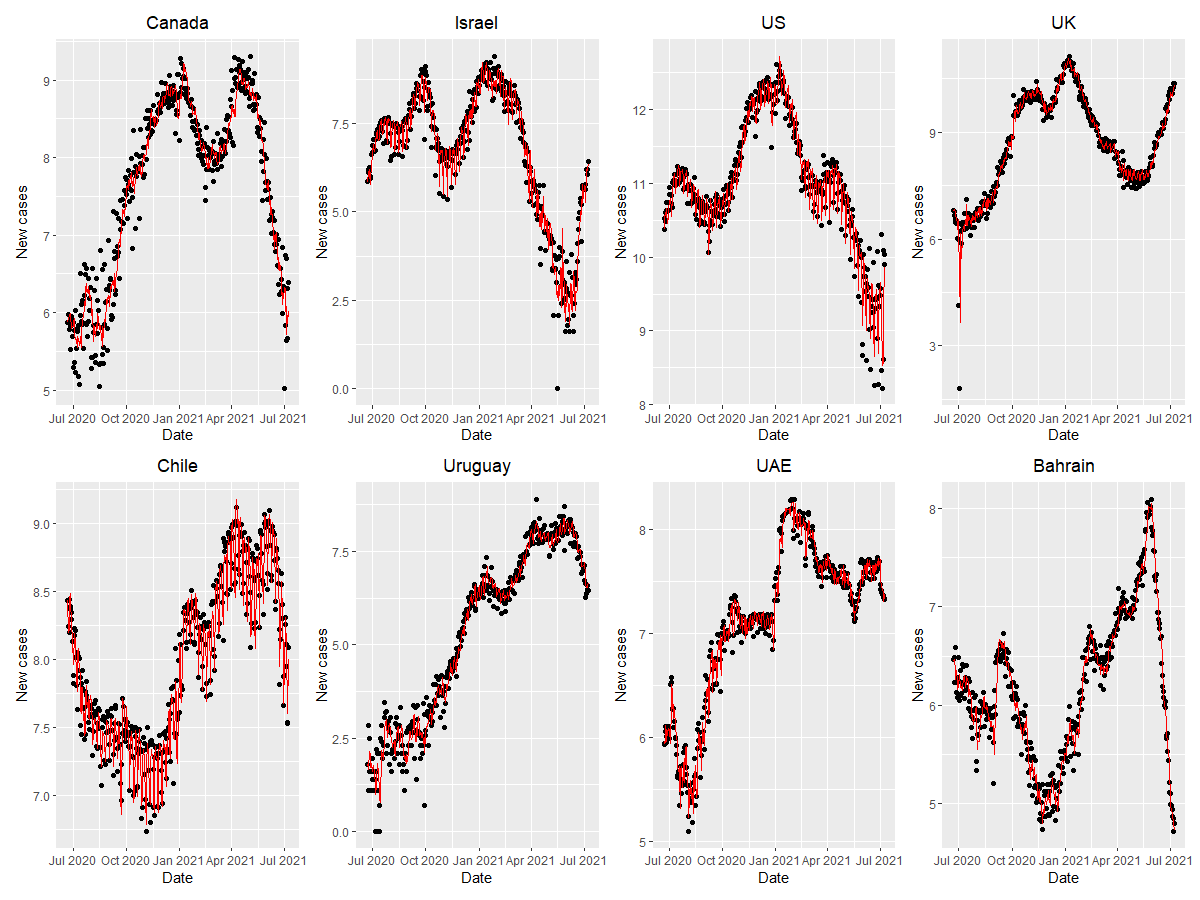}
		\end{center}
		\footnotesize\textbf{Note}: Time period -- June 1, 2020 to July 8, 2021. The black dots are logs of daily new cases. The red solid lines are fitted values computed by optimally chosen ARIMA models.
	\end{figure}

	The order $(p, d, q)$ of the ARIMA model is selected by \textsf{R}'s \texttt{forecast} package (\cite{Rforcast}) in which the command \texttt{auto.arima} chooses the optimal order based on various criteria while ensuring that the chosen model behaves well numerically. The outcome variable for each country becomes stationary after the first difference, so the order $d$ is set at $1$. Note that all the exogenous variables are also first differenced in the estimation procedure. The econometric model we estimate is
	\begin{align}
		\Delta\log Y_t = & c + \beta_1 \Delta V1_{t-21} + \beta_2 \Delta V2_{t-7} + \beta_3 \Delta \log T_t + wkd_t + \Delta P_{t-14} + \Delta M_{t-14} + n_t, \\
		n_t = & \sum_{i=1}^{p} \phi_i n_{t-i} + \varepsilon_t - \sum_{i=1}^{q} \theta_i \varepsilon_{t-i},
	\end{align}
	where $Y_t$ is daily new cases per million, $V1_t$ is population vaccinated with at least one dose per hundred, $V2_t$ is population fully vaccinated per hundred, $T_t$ is log daily new tests, $wkd_t$ is a weekend dummy, $P_t$ is OxCGRT Containment and Health Index, and $M_t$ is the mobility index. There certainly is day-to-day noise in the data, but the selected ARIMA models fit this daily fluctuation very well as shown in Figure \ref{fig: ARIMAX model fit}. Note that the UAE does not provide daily first- and second-dose progress, so we use the number of total dose administered per hundred. The resulting regression residuals are all mean-zero stationary processes.
	
	\begin{table}[t!]
		\caption{ARIMA model results for new cases -- countries relying on vaccines developed in the US and Europe}
		\label{tab: ARIMAX estimation results - Canada, Israel, UK, US}
		\begin{center}
		\begin{tabular}{@{\extracolsep{5pt}}lcccc}
			\\[-1.8ex]\hline
			\hline \\[-1.8ex]
			&  Canada & Israel & US & UK \\
			\cline{2-5}
			\\[-1.8ex] $(p,d,q)$& $(2,1,3)$ & $(5,1,3)$ & $(3,1,3)$ & $(2,1,1)$\\
			\hline \\[-1.8ex]
			$\Delta V1_{t-21}$ & $-$0.062$^{***}$ & $-$0.041 & $-$0.018 & $-$0.086$^{***}$ \\
			& (0.018) & (0.043) & (0.041) & (0.032) \\
			& & & & \\
			$\Delta V2_{t-7}$ & $-$0.011 & $-$0.006 & $-$0.020 & 0.040 \\
			& (0.020) & (0.045) & (0.046) & (0.033) \\
			& & & & \\
			$\Delta \log T_t$ & 0.060 & 0.557$^{***}$ & 0.826$^{***}$ & 0.0004 \\
			& (0.048) & (0.057) & (0.057) & (0.061) \\
			& & & & \\
			$wkd_t$ & $-$0.104$^{***}$ & $-$0.135$^{***}$ & 0.007 & $-$0.101$^{***}$ \\
			& (0.031) & (0.044) & (0.009) & (0.037) \\
			& & & & \\
			$\Delta P_{t-14}$ & $-$0.031$^{**}$ & $-$0.006 & 0.014$^{*}$ & 0.004 \\
			& (0.014) & (0.005) & (0.007) & (0.013) \\
			& & & & \\
			$\Delta M_{t-14}$ & 0.004 & 0.005 & 0.008 & 0.018 \\
			& (0.007) & (0.008) & (0.007) & (0.011) \\
			& & & & \\
			\hline \\[-1.8ex]
		\end{tabular}
	\end{center}
	\footnotesize \textbf{Note:} Time period -- June 1, 2020 to July 8, 2021. The results are produced using daily data. $(p,d,q)$ is the optimal ARIMA order chosen by \texttt{auto.arima} in \textsf{R}. Standard errors are in parentheses. $^{***}$, $^{**}$, and $^{*}$ denote the 99\%, 95\% and 90\% confidence level respectively.  
	\end{table}
	
	The estimation results are displayed in Tables \ref{tab: ARIMAX estimation results - Canada, Israel, UK, US}--\ref{tab: ARIMAX estimation results - Chinese vaccine}. For the countries that mainly vaccinated their populations with widely approved vaccines developed in the US and Europe, the growth of new cases is negatively associated with progress in vaccination. It is noticeable that the spread of COVID-19 is particularly strongly associated with the share of people vaccinated with at least one dose for Canada and the UK, where the delaying strategy was employed. A $1$\% increase of the share of people vaccinated with at least one dose is associated with around $6.2$\% and $8.6$\% reductions in the daily growth of new cases, respectively. In both countries, the effect of full vaccination progress is not significant. For Israel and the US, where the standard dosing schedule was followed, an insignificant negative association between vaccination progress and new infections is found. The weekend dummy is in general negatively associated with new infections, whereas the number of new tests is positively associated. The government policy index and the mobility index are not significant, though the signs of their coefficients are generally consistent with expected directions.
	
	\begin{table}[t!] 
		\caption{ARIMA model results for new cases -- countries relying on vaccines from China}
		\label{tab: ARIMAX estimation results - Chinese vaccine}
		\begin{center}
		\begin{tabular}{@{\extracolsep{5pt}}lcccc}
			\\[-1.8ex]\hline
			\hline \\[-1.8ex]
			& Chile & Uruguay & UAE & Bahrain\\
			\cline{2-5}
			\\[-1.8ex] $(p,d,q)$& $(2,1,4)$ & $(2,1,2)$ & $(0,1,3)$ & $(1,1,2)$\\
			\hline \\[-1.8ex]
			$\Delta V1_{t-21}$ & 0.011 & $-$0.020 &  & $-$0.020 \\
			& (0.010) & (0.035) &  & (0.023) \\
			& & & & \\
			$\Delta V2_{t-7}$ & $-$0.008 & $-$0.041 &  & $-$0.012 \\
			& (0.010) & (0.040) &  & (0.010) \\
			& & & & \\
			$\Delta V_{t-21}$ &  &  & $-$0.013$^{**}$ &  \\
			&  &  & (0.006) &  \\
			& & & & \\
			$\Delta \log T_t$ & 0.686$^{***}$ & 0.004 & 0.189$^{***}$ & 0.001 \\
			& (0.017) & (0.005) & (0.039) & (0.003) \\
			& & & & \\
			$wkt_t$ & $-$0.074$^{***}$ & $-$0.242$^{***}$ & $-$0.014 & $-$0.032$^{**}$ \\
			& (0.010) & (0.040) & (0.017) & (0.014) \\
			& & & & \\
			$\Delta P_{t-14}$ & $-$0.007 & 0.012 & 0.011 & 0.009 \\
			& (0.004) & (0.010) & (0.007) & (0.007) \\
			& & & & \\
			$\Delta M_{t-14}$ & 0.004$^{*}$ & $-$0.003 & $-$0.006 & $-$0.011 \\
			& (0.002) & (0.011) & (0.013) & (0.012) \\
			& & & & \\
			\hline \\[-1.8ex]
		\end{tabular}
		\end{center}
	\footnotesize \textbf{Note:} Time period -- June 1, 2020 to July 8, 2021. The results are produced using daily data. $(p,d,q)$ is the optimal ARIMA order chosen by \texttt{auto.arima} in \textsf{R}. Standard errors are in parentheses. $^{***}$, $^{**}$, and $^{*}$ denote the 99\%, 95\% and 90\% confidence level respectively. The UAE does not publicly announce first and second dose progress separately so we include total doses administered ($V$) instead.
	\end{table}
	
	The results are mixed for the countries relying on vaccines from China. These countries had already vaccinated a significant part of the population (as of July 8, 2021, the shares of fully vaccinated people are the UAE 65\%, Chile 58.8\%, Uruguay 54.7\%  Bahrain 59.6\%). However, significant negative association between new cases and vaccine coverage is only found in the UAE, where the total dose administered per hundred ($V$) is included instead of $V1$ and $V2.$ One additional dose per hundred is associated with a $1.3$\% reduction in new cases in the UAE.
	
	We also estimate the ARIMA models for new death counts. The results in Supplementary Materials \ref{appen: time-series death} (Tables \ref{tab: ARIMAX estimation death - Canada, Israel, UK, US} and \ref{tab: ARIMAX estimation death - Chinese vaccine}) are in general very similar to the results for new cases, except Israel where we find a significant effect of full vaccination progress.\footnote{In Israel, death counts started sharply dropping shortly after the mass vaccination program had launched. The number of daily new deaths was close to 0 between Apr 1 and Jul 8, 2021. Therefore, the outcome, log differenced daily new deaths, becomes very noisy.}
	
	Our results from time-series analysis suggest that extending the interval between first and second doses might be an effective strategy for the vaccines developed in the US and Europe. Larger negative effects of vaccination progress are found in the countries adopting this strategy. It is not clear whether this strategy works for vaccines from China. The purpose of estimating time-series models is to obtain some limited descriptive evidence of how the choice of the dosing interval and types of vaccines matter to contain the spread of COVID-19. We only use 8 countries in the analysis and have relied on daily fluctuations, which are noisy due to the stationarity concern. To investigate relationships between vaccination and new infections, we will further exploit cross-country variations in vaccination, policies, and mobility using multi-country panel data models.
	
	\subsection{Panel data analysis}
	We use OECD and EU member countries to construct our country panel, adding Bahrain and Uruguay in order to examine the effect of vaccines from China. Countries that do not publish daily COVID-19-related statistics are excluded. In total, there are 37 countries in our data set.\footnote{Included countries are Australia, Austria, Belgium, Bulgaria, Bahrain, Canada, Switzerland, Chile, Colombia, Germany, Denmark, Spain, Estonia, Finland, France, the UK, Greece, Croatia, Hungary, Israel, Italy, Japan, South Korea, Lithuania, Luxembourg, Latvia, Mexico, Malta, Norway, New Zealand, Poland, Portugal, Romania, Slovenia, Turkey, Uruguay, and the US.} We closely follow the estimation strategies used in \cite{chernozhukov2021causal} and \cite{karaivanov2021face}. 
	
	Our regression equations are motivated by a variant of the SIRD model introduced in \cite{chernozhukov2021causal} where new infections are only partially detected via testing. Let $S,I,R$ and $D$ denote the numbers of susceptible, infected, recovered, and deceased individuals in a given state. Each of these variables is a function of time. We modify the model by adding vaccination factors, $V1$ and $V2$. We assume that vaccinated individuals obtain marginal immunity gain against infection at the rates $\delta_{1}$ and $\delta_{2}$ from the first and second doses, with which they exit the susceptible class and enter the recovered class. The laws of motion for these variables are specified as
	\begin{align*}
		\dot{S}\left(t\right) & =-\left(\delta_{1}\dot{V1}\left(t-21\right)+\delta_{2}\dot{V2}\left(t-7\right)\right)-\frac{S\left(t\right)}{N}\beta\left(t\right)I\left(t\right)\\
		\dot{I}\left(t\right) & =\frac{S\left(t\right)}{N}\beta\left(t\right)I\left(t\right)-\gamma I\left(t\right),\\
		\dot{R}\left(t\right) & =\left(1-\kappa\right)\gamma I\left(t\right)+\left(\delta_{1}\dot{V1}\left(t-21\right)+\delta_{2}\dot{V2}\left(t-7\right)\right),\\
		\dot{D}\left(t\right) & =\kappa\gamma I\left(t\right),
	\end{align*}
	where $N$ is the population, $\beta\left(t\right)$ is the rate of infection spread between $S_{t-\ell}$ and $I_{t-\ell}$, $\gamma$ is the rate of recovery or death, and $\kappa$ is the probability of death conditional on infection. The total number of confirmed cases, $C_{t}$, evolves as
	\begin{align*}
		\dot{C}\left(t\right) & =\tau\left(t\right)I\left(t\right)\left(=\frac{\tau\left(t\right)}{\kappa\gamma}\dot{D}\left(t\right)\right),
	\end{align*}
	where $\tau_{j}\left(t\right)$ is the rate that infections are detected. Note that we only observed $C\left(t\right)$ and $D\left(t\right)$,
	but not $I\left(t\right)$. The unobserved $I\left(t\right)$ can	be eliminated in two ways:
	\begin{align*}
		\frac{\ddot{C}\left(t\right)}{\dot{C}\left(t\right)}  =\frac{S\left(t\right)}{N}\beta\left(t\right)-\gamma+\frac{\dot{\tau}\left(t\right)}{\tau\left(t\right)}, \quad
		\frac{\ddot{D}\left(t\right)}{\dot{D}\left(t\right)}  =\frac{S\left(t\right)}{N}\beta\left(t\right)-\gamma.
	\end{align*}
	We note that the rate of infection, $\beta\left(t\right)$, can be affected by individual behaviors and observed policies through social distancing and lockdown. We specify $\frac{S\left(t\right)}{N}\beta\left(t\right)$ as a linear function of vaccination, policies, behavioral responses, information, and confounders other than testing.
	
	Let $C_{it}$ denote the cumulative number of confirmed cases per million in country $i$ at time $t$. We define $\Delta C_{it}$ as the 7-day new COVID-19 cases reported at time $t$:
	\begin{equation}
		\Delta C_{it}:=C_{it}-C_{i,t-7}.
	\end{equation}
	Our dependent variable
	\begin{equation}
		Y_{it}^{C}=\Delta\log\left(\Delta C_{it}\right)=\log\left(\Delta C_{it}\right)-\log\left(\Delta C_{i,t-7}\right)
	\end{equation}
	approximates the weekly growth rate in new cases in country $i$ from $t-7$ to $t$. Similarly, we denote $\Delta\log\left(\Delta T_{it}\right)$
	as the 7-day growth rate in new tests with the cumulative number of tests, $T_{it}$.
	
	To analyze the impact of first- and second-dose vaccination, government policy responses, and people's behavioral changes on $Y_{it}^{C}$, we estimate
	\begin{eqnarray}\label{eq: case regression equation}
		Y_{it}^{C} & = & \alpha_{0i}+\alpha_{V1}V1_{i,t-21}+\alpha_{V2}V2_{i,t-7}+\alpha_{P}P_{i,t-14}+\alpha_{M}M_{i,t-14}\nonumber\\
		&  & +\alpha_{C1}\Delta\log\left(\Delta C_{i,t-14}\right)+\alpha_{C2}\log\left(\Delta C_{i,t-14}\right)+\alpha_{T}\Delta\log\left(\Delta T_{it}\right)+\varepsilon_{it}^{C},
	\end{eqnarray}
	where $V1_{it}$ is the number of people vaccinated with at least one dose per hundred, $V2_{it}$ is the number of fully vaccinated people per hundred, and $P_{it}$ and $M_{it}$ are policy and behavioral variables. We use the same lag specifications for the exogenous variables as in our time series analysis. For the growth rate in new deaths, $Y_{it}^{D}=\Delta\log\left(\Delta D_{it}\right)$, we use lags that are 14 days behind those in the case equation following \cite{karaivanov2021face}:
	\begin{eqnarray}\label{eq: death regression equation}
		Y_{it}^{D} & = & \beta_{0i}+\beta_{V1}V1_{i,t-35}+\beta_{V2}V2_{i,t-21}+\beta_{P}P_{i,t-28}+\beta_{M}M_{i,t-28}\nonumber\\
		&  & +\beta_{D1}\Delta\log\left(\Delta D_{i,t-28}\right)+\beta_{D2}\log\left(\Delta D_{i,t-28}\right)+\varepsilon_{it}^{D}.
	\end{eqnarray}
	To examine the potentially distinct effects of vaccines from China, we further include Chinese vaccine variables in the above regression equations \eqref{eq: case regression equation} and \eqref{eq: death regression equation} with appropriate lags. We first define a Chinese vaccine dummy, $D_{i},$ which takes value 1 if the country $i$ is Chile, Uruguay, Turkey, or Bahrain.\footnote{Turkey started its mass vaccination program using the Sinovac vaccine from China. According to Bridge Beijing (https://bridgebeijing.com/), 31.4 million doses of the Sinovac vaccine has been delivered to Turkey as of Sep 6, 2021. Turkey shifted to Comirnaty as shipments of the Sinovac doses kept delayed but until Jul 8, 2021, the Sinovac vaccine accounted for more than a half of total inoculated doses.}  Then the Chinese vaccine variables $V1^{CHN}_{it}$ and $V2^{CHN}_{it}$ are generated by $D_{i} \times V1_{it}$ and $D_{i} \times V2_{it}$ respectively.

	\begin{table}[htbp]
		\global\long\def\sym#1{\ifmmode^{#1}\else$^{#1}$\fi}%
		\caption{The direct effects of vaccination, policy and behavior on new cases}\label{tab: panel results - case}
		\begin{center}
		\begin{tabular}{lr@{\extracolsep{0pt}.}lr@{\extracolsep{0pt}.}lr@{\extracolsep{0pt}.}lr@{\extracolsep{0pt}.}l}
			\hline\hline
			\multicolumn{1}{l}{} & \multicolumn{8}{c}{Dependent variable: $\Delta\log\Delta C_{t}$}\tabularnewline
			\cline{2-9}
			& \multicolumn{2}{c}{(1)} & \multicolumn{2}{c}{(2)} & \multicolumn{2}{c}{(3)} & \multicolumn{2}{c}{(4)}\tabularnewline
			\midrule
			$V1_{t-21}$ & -0&0074\sym{{*}{*}{*}} & -0&0077\sym{{*}{*}{*}} & -0&0173\sym{{*}{*}{*}} & -0&0185\sym{{*}{*}{*}}\tabularnewline
			& (0&0021) & (0&0022) & (0&0043) & (0&0041)\tabularnewline
			\addlinespace
			$V2_{t-7}$ & 0&0029 & 0&0024 & 0&0038 & 0&0066\tabularnewline
			& (0&0025) & (0&0029) & (0&0057) & (0&0064)\tabularnewline
			$V1_{t-21}^{CHN}$ & \multicolumn{2}{c}{} & 0&0149\sym{{*}{*}{*}} & \multicolumn{2}{c}{} & 0&0108\tabularnewline
			& \multicolumn{2}{c}{} & (0&0033) & \multicolumn{2}{c}{} & (0&0087)\tabularnewline
			\addlinespace
			$V2_{t-7}^{CHN}$ & \multicolumn{2}{c}{} & -0&0138\sym{{*}{*}{*}} & \multicolumn{2}{c}{} & -0&0180\sym{{*}{*}}\tabularnewline
			& \multicolumn{2}{c}{} & (0&0042) & \multicolumn{2}{c}{} & (0&0074)\tabularnewline
			\addlinespace
			$P_{t-14}$ & -0&0020 & -0&0019 & -0&0027\sym{*} & -0&0027\sym{*}\tabularnewline
			& (0&0012) & (0&0013) & (0&0015) & (0&0015)\tabularnewline
			\addlinespace
			$M_{t-14}$ & 0&0052\sym{{*}{*}{*}} & 0&0053\sym{{*}{*}{*}} & 0&0101\sym{{*}{*}{*}} & 0&0098\sym{{*}{*}{*}}\tabularnewline
			& (0&0012) & (0&0012) & (0&0014) & (0&0013)\tabularnewline			
			$\Delta\log\Delta C_{t-14}$ & 0&1102\sym{{*}{*}{*}} & 0&1056\sym{{*}{*}} & 0&0901\sym{{*}{*}} & 0&0908\sym{{*}{*}}\tabularnewline
			& (0&0395) & (0&0394) & (0&0350) & (0&0343)\tabularnewline
			\addlinespace
			$\log\Delta C_{t-14}$ & -0&0374\sym{{*}{*}{*}} & -0&0403\sym{{*}{*}{*}} & -0&1496\sym{{*}{*}{*}} & -0&1509\sym{{*}{*}{*}}\tabularnewline
			& (0&0077) & (0&0077) & (0&0214) & (0&0214)\tabularnewline
			\addlinespace
			$\Delta\log\Delta T_{t}$ & 0&5094\sym{{*}{*}{*}} & 0&5064\sym{{*}{*}{*}} & 0&4090\sym{{*}{*}{*}} & 0&4079\sym{{*}{*}{*}}\tabularnewline
			& (0&1775) & (0&1767) & (0&1499) & (0&1496)\tabularnewline
			\midrule
			Country fixed effects & \multicolumn{2}{c}{yes} & \multicolumn{2}{c}{yes} & \multicolumn{2}{c}{yes} & \multicolumn{2}{c}{yes}\tabularnewline
			Country specific trend in days & \multicolumn{2}{c}{no} & \multicolumn{2}{c}{no} & \multicolumn{2}{c}{quadratic} & \multicolumn{2}{c}{quadratic}\tabularnewline
			\midrule
			${\rm R}^{2}$ & 0&2838 & 0&2868 & 0&3987 & 0&3999\tabularnewline
			Adjusted ${\rm R}^{2}$ & 0&2834 & 0&2864 & 0&3953 & 0&3964\tabularnewline
			Number of countries & \multicolumn{2}{c}{37} & \multicolumn{2}{c}{37} & \multicolumn{2}{c}{37} & \multicolumn{2}{c}{37}\tabularnewline
			Obs. per country & \multicolumn{2}{c}{382} & \multicolumn{2}{c}{382} & \multicolumn{2}{c}{382} & \multicolumn{2}{c}{382}\tabularnewline
			\midrule
		\end{tabular}
		\end{center}
	\footnotesize \textbf{Note:} Time period -- June 1, 2020 to July 8, 2021. The results are produced using daily data. Standard errors in parentheses are clustered at the country level. $^{***}$, $^{**}$, and $^{*}$ denote the 99\%, 95\% and 90\% confidence level respectively. 
	\end{table}
	
	\begin{table}[htbp]
		\global\long\def\sym#1{\ifmmode^{#1}\else$^{#1}$\fi}%
		\caption{The direct effects of vaccination, policy, and behavior on new deaths}\label{tab: panel results - death}
		\begin{center}
		\begin{tabular}{lr@{\extracolsep{0pt}.}lr@{\extracolsep{0pt}.}lr@{\extracolsep{0pt}.}lr@{\extracolsep{0pt}.}l}
			\hline\hline
			\multicolumn{1}{l}{} & \multicolumn{8}{c}{Dependent variable: $\Delta\log\Delta D_{t}$}\tabularnewline
			\cline{2-9}
			& \multicolumn{2}{c}{(1)} & \multicolumn{2}{c}{(2)} & \multicolumn{2}{c}{(3)} & \multicolumn{2}{c}{(4)}\tabularnewline
			\midrule
			$V1_{t-35}$ & -0&0096\sym{{*}{*}{*}} & -0&0093\sym{{*}{*}{*}} & -0&0125\sym{{*}{*}{*}} & -0&0132\sym{{*}{*}{*}}\tabularnewline
			& (0&0021) & (0&0023) & (0&0040) & (0&0041)\tabularnewline
			\addlinespace
			$V2_{t-21}$ & 0&0036 & 0&0014 & 0&0040 & 0&0061\tabularnewline
			& (0&0029) & (0&0033) & (0&0043) & (0&0046)\tabularnewline
			$V1_{t-35}^{CHN}$ & \multicolumn{2}{c}{} & 0&0169\sym{{*}{*}{*}} & \multicolumn{2}{c}{} & 0&0045\tabularnewline
			& \multicolumn{2}{c}{} & (0&0034) & \multicolumn{2}{c}{} & (0&0076)\tabularnewline
			$V2_{t-21}^{CHN}$ & \multicolumn{2}{c}{} & -0&0116\sym{{*}{*}} & \multicolumn{2}{c}{} & -0&0097\tabularnewline
			& \multicolumn{2}{c}{} & (0&0050) & \multicolumn{2}{c}{} & (0&0059)\tabularnewline
			\addlinespace
			$P_{t-28}$ & -0&0008 & -0&0005 & -0&0013 & -0&0013\tabularnewline
			& (0&0014) & (0&0015) & (0&0013) & (0&0013)\tabularnewline
			\addlinespace
			$M_{t-28}$ & 0&0057\sym{{*}{*}{*}} & 0&0057\sym{{*}{*}{*}} & 0&0077\sym{{*}{*}{*}} & 0&0075\sym{{*}{*}{*}}\tabularnewline
			& (0&0016) & (0&0015) & (0&0016) & (0&0015)\tabularnewline
			$\Delta\log\Delta D_{t-28}$ & 0&0584\sym{{*}{*}} & 0&0560\sym{*} & 0&0574\sym{{*}{*}} & 0&0567\sym{{*}{*}}\tabularnewline
			& (0&0281) & (0&0280) & (0&0258) & (0&0259)\tabularnewline
			\addlinespace
			$\log\Delta D_{t-28}$ & -0&0429\sym{{*}{*}{*}} & -0&0479\sym{{*}{*}{*}} & -0&1308\sym{{*}{*}{*}} & -0&1308\sym{{*}{*}{*}}\tabularnewline
			& (0&0096) & (0&0085) & (0&0112) & (0&0112)\tabularnewline
			\addlinespace
			\midrule
			Country fixed effects & \multicolumn{2}{c}{yes} & \multicolumn{2}{c}{yes} & \multicolumn{2}{c}{yes} & \multicolumn{2}{c}{yes}\tabularnewline
			Country specific trend in days & \multicolumn{2}{c}{no} & \multicolumn{2}{c}{no} & \multicolumn{2}{c}{quadratic} & \multicolumn{2}{c}{quadratic}\tabularnewline
			\midrule
			${\rm R}^{2}$ & 0&0966 & 0&1016 & 0&1747 & 0&1749\tabularnewline
			Adjusted ${\rm R}^{2}$ & 0&0962 & 0&1010 & 0&1698 & 0&1699\tabularnewline
			Number of countries & \multicolumn{2}{c}{37} & \multicolumn{2}{c}{37} & \multicolumn{2}{c}{37} & \multicolumn{2}{c}{37}\tabularnewline
			Obs. per country & \multicolumn{2}{c}{368} & \multicolumn{2}{c}{368} & \multicolumn{2}{c}{368} & \multicolumn{2}{c}{368}\tabularnewline
			\midrule
		\end{tabular}
		\end{center}
	\footnotesize \textbf{Note:} Time period -- June 1, 2020 to July 8, 2021. The results are produced using daily data. Standard errors in parentheses are clustered at the country level. $^{***}$, $^{**}$, and $^{*}$ denote the 99\%, 95\% and 90\% confidence level respectively. 
	\end{table}
	
	We estimate our regression equations with country fixed effects to account for country-specific heterogeneity. We do not include week or month fixed effects, as each country experienced heterogeneous evolution of the pandemic. Including country-specific time-fixed effects results in too many parameters to estimate, which lead to imprecise estimates. Instead, country-specific time trends are included to control for heterogeneous trends across countries. We consider quadratic time trends as a baseline specification. Alternative specifications (linear and cubic time trends) are also considered, and our main findings hold under those specifications, as shown in Supplementary Materials \ref{appen: alternative time trends}. 
	
	The results are displayed in Tables \ref{tab: panel results - case}--\ref{tab: panel results - death} for new infections and deaths respectively. It is notable that weekly growth rates of both new infections and deaths are significantly negatively associated at the $99\%$ confidence level with first-dose vaccination progress across all the specifications considered. With no time trends, the estimates suggest that a $1\%$ increase in the share of people vaccinated with at least one dose leads to around $0.77\%$ reduction in the weekly case growth rate and around $0.93\%$ reduction in the weekly death growth rate. When quadratic time trends are included, the magnitudes of first-dose vaccination effect estimates increase to $1.85\%$ and $1.32\%$ for new cases and deaths respectively. The estimated effect of second-dose vaccination progress is not significantly different from $0$ for both health outcomes under any of the specifications considered.
	
	The estimated effects of first and second doses of Chinese vaccines are the sums of coefficients on $V1$ and $V1^{CHN}$, and $V2$ and $V2^{CHN}$, respectively. Table \ref{tab: panel results - CHN} reports these estimates. Contrary to the results for mRNA-based and adenoviral vaccines, first-dose progress in countries relying on Chinese vaccines is not effective in reducing either case or death counts. Only full vaccination progress leads to significant reductions in new cases and deaths, although the effect on deaths becomes insignificant when quadratic time trends are controlled for. The estimated magnitude of the vaccine effectiveness is smaller than that of vaccines from the US and Europe. These results are in line with the evidence from clinical trials that these vaccines have a lower vaccine efficacy than mRNA-based counterparts and Vaxzevria.\footnote{The \cite{world2021background} approved the Sinovac vaccine for emergency use and stated that the vaccine efficacy of this vaccine is 51\% for symptomatic infection. \cite{dashdorj2021direct} also report that the Sinopharm vaccine induced weaker antibody responses than Comirnaty and Vaxzevria.} These findings suggest that keeping the recommended dosing interval is a better strategy for countries relying on Chinese vaccines.
	
	\begin{table}[h!]
		\global\long\def\sym#1{\ifmmode^{#1}\else$^{#1}$\fi}%
		\caption{Test for the effects of the Chinese vaccines on confirmed cases and deaths}
		\label{tab: panel results - CHN}
		\begin{center}
		{\small{}}%
		\begin{tabular}{lr@{\extracolsep{0pt}.}lr@{\extracolsep{0pt}.}llr@{\extracolsep{0pt}.}lr@{\extracolsep{0pt}.}l}
			\hline\hline
			& \multicolumn{4}{c}{$\Delta\log\Delta C_{t}$} & & \multicolumn{4}{c}{$\Delta\log\Delta D_{t}$}\tabularnewline
			\cline{2-5} \cline{7-10}
			& \multicolumn{2}{c}{(2)} & \multicolumn{2}{c}{(4)} &  & \multicolumn{2}{c}{(2)} & \multicolumn{2}{c}{(4)}\tabularnewline
			\midrule
			$V1_{t-21}+V1_{t-21}^{CHN}$ & 0&0072\sym{{*}{*}{*}} & -0&0077 & $V1_{t-35}+V1_{t-35}^{CHN}$ & 0&0075\sym{{*}{*}} & -0&0087\tabularnewline
			& (0&0025) & (0&0080) &  & (0&0028) & (0&0064)\tabularnewline
			$V2_{t-7}+V2_{t-7}^{CHN}$ & -0&0114\sym{{*}{*}{*}} & -0&0114\sym{{*}{*}{*}} & $V2_{t-21}+V2_{t-21}^{CHN}$ & -0&0102\sym{{*}{*}} & -0&0036\tabularnewline
			& (0&0027) & (0&0028) &  & (0&0038) & (0&0035)\tabularnewline
			\midrule
		\end{tabular}
		\end{center}
	\footnotesize \textbf{Note:} Time period -- June 1, 2020 to July 8, 2021. The results are produced using the estimates in Tables \ref{tab: panel results - case} and \ref{tab: panel results - death} under specifications (2) and (4). Standard errors in parentheses are clustered at the country level. $^{***}$, $^{**}$, and $^{*}$ denote the 99\%, 95\% and 90\% confidence level respectively.
	\end{table}
	
	The signs and magnitudes of the other coefficient estimates are, in general, very consistent across the specifications. The weekly growth rate of new cases is positively related to the growth of new tests. The weekly case growth rate is negatively associated with government policies, whereas both cases and deaths are positively associated with the mobility index. Note that government policy and mobility indexes are insignificant in our time series analysis for each country. These factors are now significant when time trends are taken into account, because we further exploit the cross-country variations in government policies and people's behavioral changes. Our results mean that less stringent policy measures and more mobile people can be translated into a higher $\beta(t)$ in the SIRD model, which results in more cases and deaths.
	
	\begin{table}[htbp]
		\global\long\def\sym#1{\ifmmode^{#1}\else$^{#1}$\fi}%
		\caption{The direct effects of vaccination, policy, and information on mobility}\label{tab: panel results - mobility}
		\begin{center}
		\small
		\begin{tabular}{lr@{\extracolsep{0pt}.}lr@{\extracolsep{0pt}.}lr@{\extracolsep{0pt}.}lr@{\extracolsep{0pt}.}l}
			\hline\hline
			\multicolumn{1}{l}{} & \multicolumn{8}{c}{Dependent variable: $M_{t}$}\tabularnewline
			\cline{2-9}
			& \multicolumn{2}{c}{(1)} & \multicolumn{2}{c}{(2)} & \multicolumn{2}{c}{(3)} & \multicolumn{2}{c}{(4)}\tabularnewline
			\midrule
			$\Delta V1_{t}$ & 0&5839\sym{{*}{*}{*}} & 0&6642\sym{{*}{*}{*}} & 0&5260\sym{{*}{*}{*}} & 0&5937\sym{{*}{*}{*}}\tabularnewline
			& (0&1260) & (0&1623) & (0&1266) & (0&1636)\tabularnewline
			$\Delta V2_{t}$ & 0&2641\sym{*} & 0&3941\sym{{*}{*}} & 0&3033\sym{{*}{*}} & 0&4396\sym{{*}{*}}\tabularnewline
			& (0&1316) & (0&1636) & (0&1439) & (0&1865)\tabularnewline
			\addlinespace
			$V1_{t-7}$ & 0&0494 & 0&0314 & 0&0377 & 0&0204\tabularnewline
			& (0&0324) & (0&0372) & (0&0318) & (0&0373)\tabularnewline
			\addlinespace
			$V2_{t-7}$ & -0&0084 & 0&0032 & 0&0120 & 0&0220\tabularnewline
			& (0&0349) & (0&0379) & (0&0315) & (0&0346)\tabularnewline
			$\Delta V1_{t}^{CHN}$ & \multicolumn{2}{c}{} & -0&3956\sym{*} & \multicolumn{2}{c}{} & -0&3474\tabularnewline
			& \multicolumn{2}{c}{} & (0&2144) & \multicolumn{2}{c}{} & (0&2233)\tabularnewline
			$\Delta V2_{t}^{CHN}$ & \multicolumn{2}{c}{} & -0&4834\sym{*} & \multicolumn{2}{c}{} & -0&5110\sym{*}\tabularnewline
			& \multicolumn{2}{c}{} & (0&2568) & \multicolumn{2}{c}{} & (0&2762)\tabularnewline
			$V1_{t-7}^{CHN}$ & \multicolumn{2}{c}{} & 0&0693 & \multicolumn{2}{c}{} & 0&0736\tabularnewline
			& \multicolumn{2}{c}{} & (0&0676) & \multicolumn{2}{c}{} & (0&0641)\tabularnewline
			$V2_{t-7}^{CHN}$ & \multicolumn{2}{c}{} & -0&0380 & \multicolumn{2}{c}{} & -0&0414\tabularnewline
			& \multicolumn{2}{c}{} & (0&0742) & \multicolumn{2}{c}{} & (0&0652)\tabularnewline
			$\Delta P_{t}$ & -0&3450\sym{{*}{*}{*}} & -0&3464\sym{{*}{*}{*}} & -0&3583\sym{{*}{*}{*}} & -0&3590\sym{{*}{*}{*}}\tabularnewline
			& (0&0449) & (0&0448) & (0&0428) & (0&0428)\tabularnewline
			$P_{t-7}$ & -0&1120\sym{{*}{*}{*}} & -0&1141\sym{{*}{*}{*}} & -0&1028\sym{{*}{*}{*}} & -0&1049\sym{{*}{*}{*}}\tabularnewline
			& (0&0173) & (0&0175) & (0&0174) & (0&0175)\tabularnewline
			$\Delta\log\Delta C_{t}$ & 0&4354\sym{*} & 0&4702\sym{*} & \multicolumn{2}{c}{} & \multicolumn{2}{c}{}\tabularnewline
			& (0&2520) & (0&2560) & \multicolumn{2}{c}{} & \multicolumn{2}{c}{}\tabularnewline
			\addlinespace
			$\log\Delta C_{t}$ & -0&7026\sym{{*}{*}{*}} & -0&6950\sym{{*}{*}{*}} & \multicolumn{2}{c}{} & \multicolumn{2}{c}{}\tabularnewline
			& (0&0849) & (0&0924) & \multicolumn{2}{c}{} & \multicolumn{2}{c}{}\tabularnewline
			$\Delta\log\Delta D_{t}$ & \multicolumn{2}{c}{} & \multicolumn{2}{c}{} & -0&0742 & -0&0624\tabularnewline
			& \multicolumn{2}{c}{} & \multicolumn{2}{c}{} & (0&1715) & (0&1717)\tabularnewline
			$\log\Delta D_{t}$ & \multicolumn{2}{c}{} & \multicolumn{2}{c}{} & -0&8039\sym{{*}{*}{*}} & -0&7976\sym{{*}{*}{*}}\tabularnewline
			& \multicolumn{2}{c}{} & \multicolumn{2}{c}{} & (0&1150) & (0&1211)\tabularnewline
			$M_{t-7}$ & 0&7635\sym{{*}{*}{*}} & 0&7637\sym{{*}{*}{*}} & 0&7469\sym{{*}{*}{*}} & 0&7472\sym{{*}{*}{*}}\tabularnewline
			& (0&0220) & (0&0219) & (0&0239) & (0&0239)\tabularnewline
			\midrule
			Country fixed effects & \multicolumn{2}{c}{yes} & \multicolumn{2}{c}{yes} & \multicolumn{2}{c}{yes} & \multicolumn{2}{c}{yes}\tabularnewline
			Time effects & \multicolumn{2}{c}{no} & \multicolumn{2}{c}{no} & \multicolumn{2}{c}{no} & \multicolumn{2}{c}{no}\tabularnewline
			\midrule
			${\rm R}^{2}$ & 0&8363 & 0&8367 & 0&8366 & 0&8369\tabularnewline
			Adjusted ${\rm R}^{2}$ & 0&8362 & 0&8366 & 0&8365 & 0&8368\tabularnewline
			Number of countries & \multicolumn{2}{c}{37} & \multicolumn{2}{c}{37} & \multicolumn{2}{c}{37} & \multicolumn{2}{c}{37}\tabularnewline
			Obs. per country & \multicolumn{2}{c}{396} & \multicolumn{2}{c}{396} & \multicolumn{2}{c}{396} & \multicolumn{2}{c}{396}\tabularnewline
			\midrule
		\end{tabular}
		\end{center}
	\footnotesize \textbf{Note:} Time period -- June 1, 2020 to July 8, 2021. The results are produced using daily data. Standard errors in parentheses are clustered at the country level. $^{***}$, $^{**}$, and $^{*}$ denote the 99\%, 95\% and 90\% confidence level respectively.
	\end{table}
	
	We also examine how vaccination, policies, and the evolution of pandemic affect people's mobility which is related to social distancing behaviors by estimating the following equation:
	\begin{eqnarray}\label{eq: mobility regression equation}
		M_{it} & = & \gamma_{0i}+\gamma_{d1}\Delta V1_{it}+\gamma_{d2}\Delta V2_{it} +\gamma_{1}V1_{i,t-7}+\gamma_{2}V2_{i,t-7}+\gamma_{dP}\Delta P_{it}+\gamma_{P}P{}_{i, t-7}\nonumber\\
		& & + \gamma_{dC}\Delta\log\Delta C_{it}+\gamma_{C}\log\Delta C_{it} + \gamma_{M} M_{i,t-7} + \varepsilon_{it}^{Mob}.
	\end{eqnarray}
	This regression equation includes the weekly case growth rate and the number of weekly new infections as information variables. We estimate the regression coefficients using many different specifications, all of which include country fixed effects. Time trends are omitted as they are in general insignificant (Results with country-specific linear time trends are provided in Supplementary Materials \ref{appen: alternative time trends}). In an alternative specification, we replace the weekly case number and growth rate with the weekly death count and growth rate. Table \ref{tab: panel results - mobility} displays the estimation results. Most explanatory variables are highly significant across all the specifications. The values of R-squared are close to $0.84$ in all the specifications. Our results imply that vaccination and government policies have large effects on people's behaviors. Weekly progresses in first- and second-dose vaccination lead to a large observed increase in people's mobility. When the weekly case counts, weekly case growth rates, and the Chinese vaccine dummy are included, a $1\%$ weekly growth in first- and second-dose vaccination increases the mobility index by $0.66$ and $0.39$, respectively. These results are consistent with findings in \cite{anderssonjheforth}.\footnote{Swedish survey data collected before vaccination was launched (Dec 10--13, 2020) show that providing vaccine information, such as the safety, effectiveness, and availability, reduces people's voluntary social distancing. On the other hand, more stringent government responses lead to a substantial reduction in people's mobility.} Information also matters as people's mobility is negatively associated with both weekly growth and counts of new infections and deaths. The magnitudes of estimates for the weekly growth and counts of deaths are larger than those for new cases. This indicates that people responded more sensitively to new death counts than new case numbers, which conforms to \cite{chernozhukov2021causal}. Interestingly, no significant increases in mobility due to vaccination are observed in countries relying on vaccines from China. This may be related to the perceived credibility of the effectiveness of those vaccines.
	
	The magnitudes of vaccination and policy effects are not very sensitive to whether or not we include cases or deaths as information variables and across different specifications. As we show in Tables \ref{tab: panel results - case}--\ref{tab: panel results - death}, people's mobility in the past is significantly positively associated with new infections and deaths. Therefore, there is an indirect effect of vaccination, which partially offsets its direct effects on the transmission of COVID-19 by inducing people to be more mobile. Many countries with high vaccination rates loosened public health measures amid high vaccination rates. Our findings imply that this could be a risky move, as loosened restrictions and vaccination progress collectively push mobility up. Public health measures are still key to containing the spread of the virus, by restraining people's mobility and social interactions. Note that our focus is solely given to epidemiological outcomes here. When overall social welfare is taken into account, it may be still optimal to relax non-pharmaceutical interventions to support economic activity in the presence of effective vaccination.
	
	\section{Robustness Checks}
	
	\subsection{Alternative lags}
	We conduct sensitivity analysis with different lag specifications for vaccination centered around the baseline lags of 21 and 7 days for first- and second-dose progress respectively. In alternative specifications, we use the baseline lags $\pm \ d$ ($d = 1, 2, 3$), for both vaccination variables. The results are provided in Supplementary Materials \ref{appen: alternative lags} (Figure \ref{fig: alternative lags}). All the estimates across the alternative lag specifications are consistent with our baseline findings. The magnitudes of estimates slightly vary with lags, and the coefficient estimates of first-dose progress always remain highly significant. The effect of the share of vaccinated people with at least one dose becomes slightly larger when the lag employed is longer. Full vaccination progress is insignificant with all the alternative lags. The effects of vaccines from China are not sensitive to the lag specifications.
    	
	\subsection{Alternative periods}
	We also estimate our regression coefficients for alternative initial and end dates. First, we change the initial date from June 1 to July 1, 2020, so that the first month in the full period is omitted. During the omitted period, the cases numbers were stable in many countries. Secondly, we exclude the data after May 31, 2021, from the full period. In the omitted period, some countries experienced a surge in new cases due to the spread of the Delta variant. The results are reported in Supplementary Materials \ref{appen: alternative periods} (Table \ref{tab: Early and late sub-periods}). In the first sub-period (July 1, 2020--July 8, 2021), the coefficients estimates remain very similar to Table \ref{tab: panel results - case}. The magnitude of the effect of first-dose progress shrinks in the second sub-period (June 1, 2020--May 31, 2021), but it is still negative and significant. This would be naturally the case, because most countries in our sample did not achieve high enough vaccination rates to effectively contain transmission until May 31, 2021.
		
	\subsection{Data frequency}
	A few countries (Costa Rica, Cyprus, Iceland, and the Netherlands) release key epidemiological variables weekly. We re-estimate our regression coefficients by aggregating the daily-level data to the weekly level to include these countries. In total, 41 countries are included in our weekly country panel. The results are almost identical to our findings in the daily frequency estimation in terms of the magnitudes and significance of coefficient estimates. We provide the results in Supplementary Materials \ref{appen: panel weekly frequency estimation} (Tables \ref{tab: panel weekly cases}--\ref{tab: panel weekly mobility}).
	
	\subsection{Interaction between vaccine distribution and mobility}
	Our baseline specifications do not allow for interactions between included variables. We extend these specifications to include the interaction terms between vaccine variables and the mobility index to test whether the effects of mobility on case/death growth are reduced when more people are vaccinated. The results are provided in Supplementary Materials \ref{appen: interaction} (Tables \ref{fig: interaction_cases}--\ref{fig: interaction_deaths}).\footnote{We use the same lags (14-day and 28-day lags respectively for cases and deaths) for vaccine variables and the mobility index to produce the interaction terms. This specification allows us to interpret the coefficients on interaction terms as differential effects of mobility conditional on vaccination rates.} All the coefficients on interaction terms are insignificant at the 95\% confidence level when the quadratic time trends are controlled for.
	
	\subsection{Countries adopting Chinese vaccines}
	The baseline specifications include only four countries that use Chinese vaccines. Hungary, Mexico, and Colombia also adopted Chinese vaccines but those countries are not included because they did not use Chinese vaccines for the majority of total doses.\footnote{According to Our World in Data, Wilson Center (https://www.wilsoncenter.org/), and Observer Research Foundation (https://www.orfonline.org/), the ratios of Chinese vaccines to total vaccine doses for three countries are as follows: 20.4\% (Hungary, as of Jul 2, 2021), 30.1\% (Mexico, as of Jul 1, 2021), and 53.6\% (Colombia, as of Jul 5, 2021).} We re-estimate including these three countries among those using Chinese vaccines. The results in Supplementary Materials \ref{appen: CHN} (Table \ref{fig: CHN}) are consistent with our baseline findings.
	
	\section{Counterfactual Vaccine Allocations}
	We use our estimates in Tables \ref{tab: panel results - case}--\ref{tab: panel results - death} to evaluate counterfactuals where the actual vaccine allocations between first and second doses are replaced by alternative hypothetical allocations. \cite{Saad_Roy_2021} consider the rate of first-dose progress as an increasing function of the dosing interval. In a similar spirit, we also assume that the total amount of available vaccines does not change:
	\begin{align*}
		V1_{it}+V2_{it}=V1_{it}^{*}+V2_{it}^{*},
	\end{align*}
	where $V1_{it}^{*}$ and $V2_{it}^{*}$ are sequences of counterfactual first- and second-dose vaccination rates for country $i$.
	
\begin{figure}[t]
		\begin{center}
			\caption{Actual and hypothetical vaccination rates (8-week dosing interval)}
			\label{fig: counterfactual vaccine scenario}
			\begin{subfigure}[b]{0.475\textwidth}
				\centering
				\caption[]%
				{{\small Canada}}
				\label{fig: counterfactual vaccine scenario - Canada}
				\includegraphics[width=\textwidth]{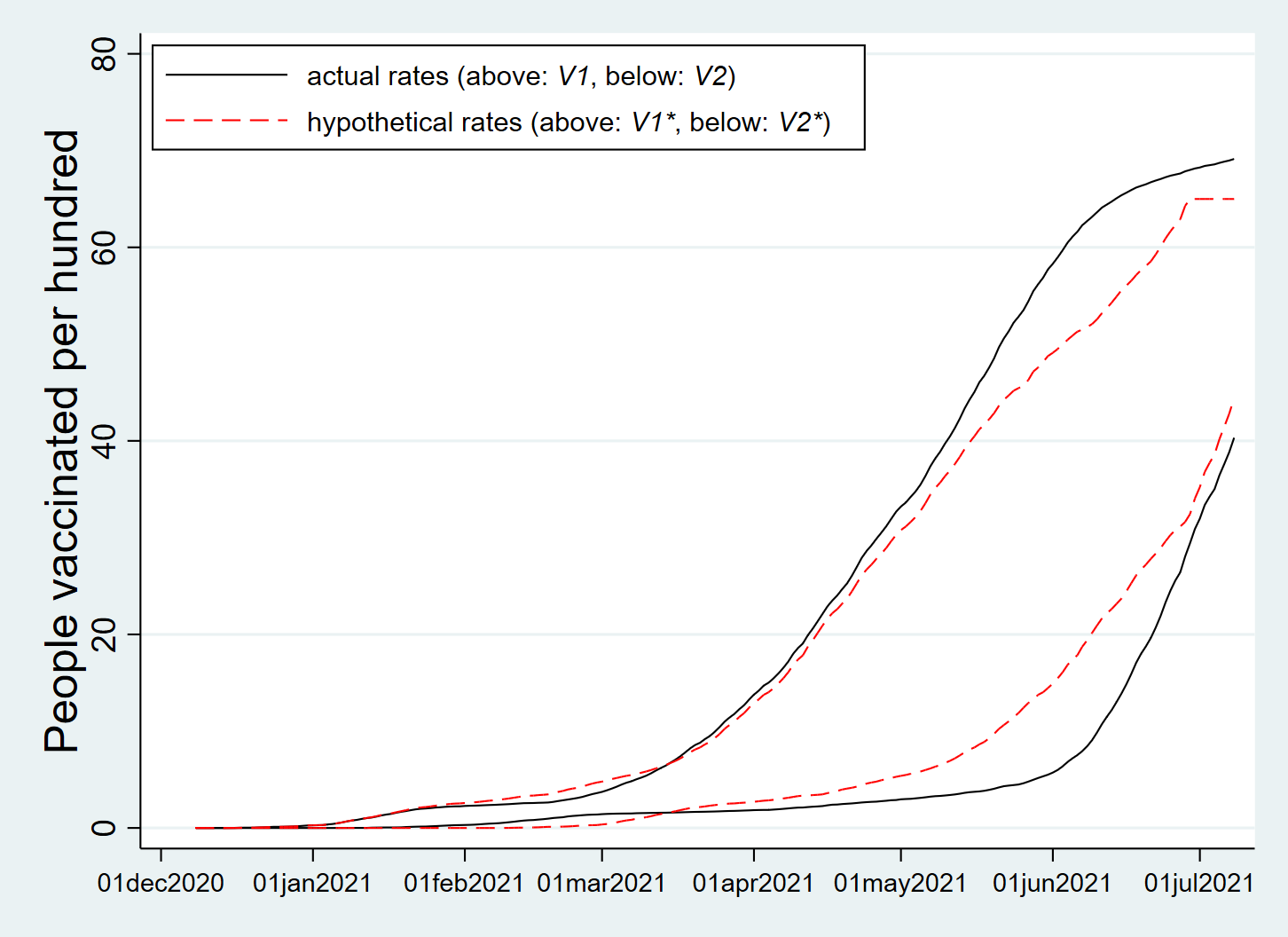}
			\end{subfigure}
			\hfill
			\begin{subfigure}[b]{0.475\textwidth}
				\centering
				\caption[]%
				{{\small US}}
				\label{fig: counterfactual vaccine scenario - USA}
				\includegraphics[width=\textwidth]{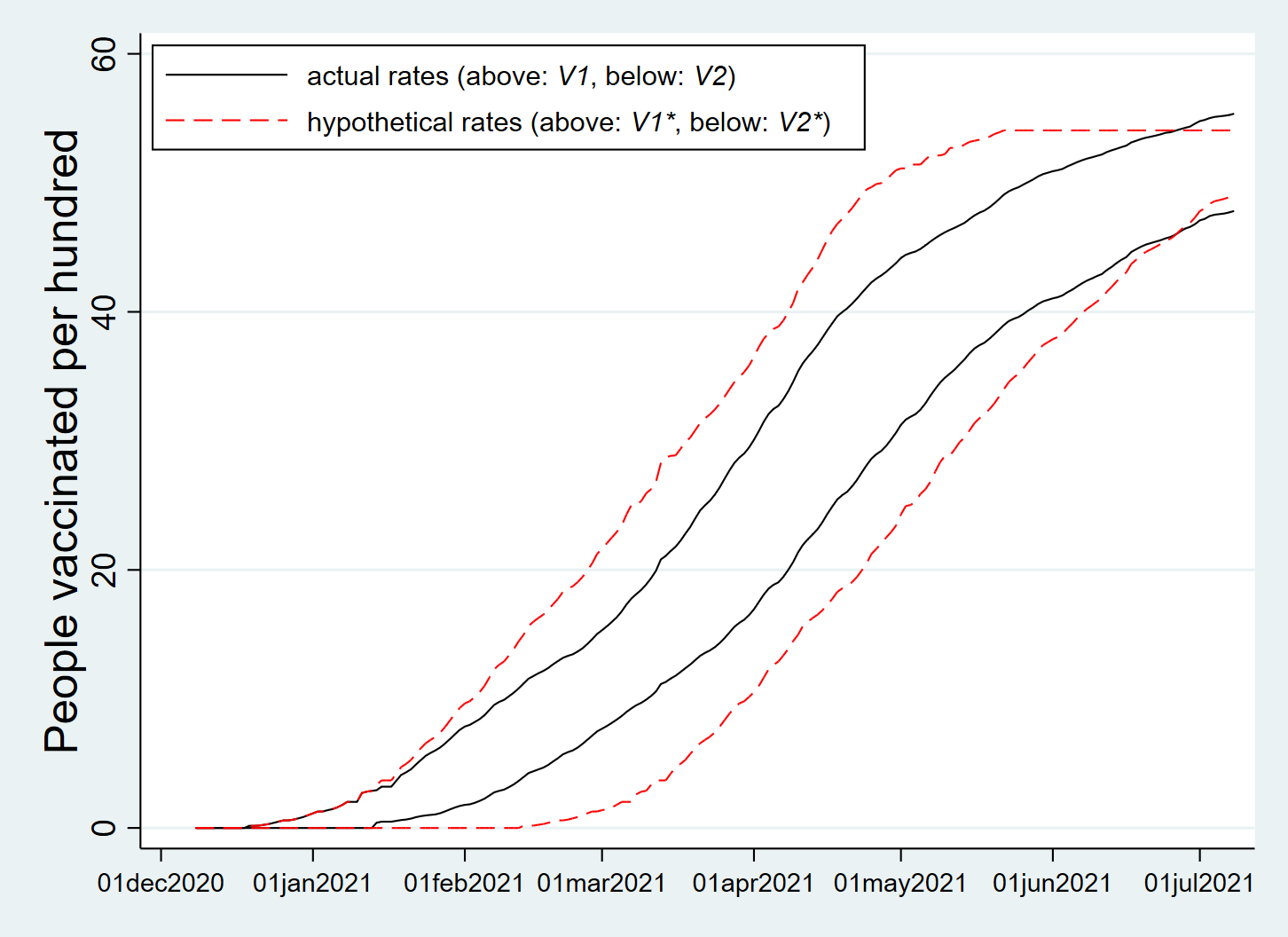}
			\end{subfigure}
		\end{center}
		\footnotesize\textbf{Note}: Black solid lines are actual vaccination rates, the share of people vaccinated and the share of people fully vaccinated per hundred. Red dotted lines are hypothetical vaccination rates we consider in our counterfactual experiments.
	\end{figure}

	We focus on Canada and the US for our counterfactual experiments as both countries mainly relied on mRNA-based vaccines but used very different dosing intervals. When it comes to the US, we extend the dosing interval to 8 weeks, resulting in faster growth of first-dose vaccination. For Canada, we reduce the interval between two doses to 8 weeks. This means that first-dose progress rises less steeply and full vaccination progress grows more quickly than the actual data. Following the actual vaccination rates for the two countries, we set the maximum rate of first-dose vaccination to 65\% for Canada and 55\% for the US. The hypothetical sequences of vaccination rates are shown in Figure \ref{fig: counterfactual vaccine scenario}.
	
	We compute the counterfactual outcomes using Equations \eqref{eq: case regression equation}--\eqref{eq: mobility regression equation} for new cases, deaths and mobility. We assume that government policies and all other variables remain fixed at their observed values in the data. As recent vaccination rates, case counts and growth affect people's behavior, we compute the counterfactual case growth and mobility iteratively conditional on the lagged values of counterfactual outcomes and the hypothetical sequences of vaccination allocations. Therefore, the information effects of counterfactual case and death counts and the indirect effects of vaccination via mobility are taken into account in such calculations. The 90\% confidence bands are computed using simulations.
	
	To compute counterfactual outcomes, Equations \eqref{eq: mobility regression equation} and \eqref{eq: case regression equation} are rewritten in terms of $\log\Delta C_{i,t}$:
	\begin{align*}
		M_{i,t-14}  = & \gamma_{0i}+\gamma_{M} M_{i,t-21} + \left(\gamma_{dC}+\gamma_{C}\right)\log\Delta C_{i,t-14}-\gamma_{dC}\log\Delta C_{i,t-21}\nonumber\\
		&   + \gamma_{d1} V1_{i,t-14}+\gamma_{d2} V2_{i,t-14} +\left(\gamma_{1}-\gamma_{d1}\right)V1_{i,t-21}+\left(\gamma_{2}-\gamma_{d2}\right)V2_{i,t-21} \\
		&   + \gamma_{dP}P_{i,t-14}+\left(\gamma_{P}-\gamma_{dP}\right)P_{i, t-21}+\varepsilon_{i,t-14}^{Mob}, \\
		\log\Delta C_{i,t}  = & \log\Delta C_{i,t-7} + \alpha_{0i} + \left(\alpha_{C1}+\alpha_{C2}\right)\log\left(\Delta C_{i,t-14}\right)-\alpha_{C1}\log\left(\Delta C_{i,t-21}\right)\nonumber\\
		&   + \alpha_{V1}V1_{i,t-21}+\alpha_{V2}V2_{i,t-7}+\alpha_{P}P_{i,t-14}+\alpha_{M}M_{i,t-14} +\alpha_{T}\Delta\log\left(\Delta T_{it}\right)+\varepsilon_{it}^{C},
	\end{align*}
	We use these two equations to iteratively compute counterfactual outcomes, $\{M_{i,t}^{*}\}$ and $\{\log\Delta C_{i,t}^{*}\}$, given $\{V1_{t}^{*},V2_{t}^{*}\}$ and $\{P_{i,t},\varepsilon_{i,t}^{Mob}\}$. We draw the parameters, $\tilde{\alpha}_{j}$, in Equations \eqref{eq: case regression equation} and \eqref{eq: death regression equation} from their asymptotic distribution and compute the associated residuals, $\tilde{\varepsilon}_{j}^{C}$, while the parameters for country specific time trends, country fixed effects in \eqref{eq: case regression equation} and all parameters in \eqref{eq: mobility regression equation} remain fixed. We then compute the counterfactual outcomes with $\left(\tilde{\alpha}_{j},\tilde{\varepsilon}_{j}^{C}\right)$ for $j=1,\ldots,200$. The point estimate of each counterfactual outcome is the mean across 200 replications. We plot the 90\% point-wise confidence bands by taking 5th and 95th percentiles across the replications at each $t$.
	
	From the estimates in Tables \ref{tab: panel results - case}--\ref{tab: panel results - death}, we expect that converting second doses administered to partially vaccinated people to first doses for unvaccinated individuals will lead to substantially lower numbers of new cases and deaths in the US. Analogously, reducing the interval between doses in Canada is expected to result in higher counts for both health outcomes. Effects of changing the vaccination strategy on cases and deaths are expected to be partially moderated via effects on mobility. 
	
	\begin{figure}[h!]
		\begin{center}
			\caption{Counterfactual case counts}
			\label{fig: counterfactual case}
			\vspace{-3mm}
			\begin{subfigure}[b]{\textwidth}
				\centering
				\caption[]%
				{{\small Canada}}
				\label{fig: counterfactual case - Canada}
				\includegraphics[width=0.45\textwidth,height=0.18\textheight]{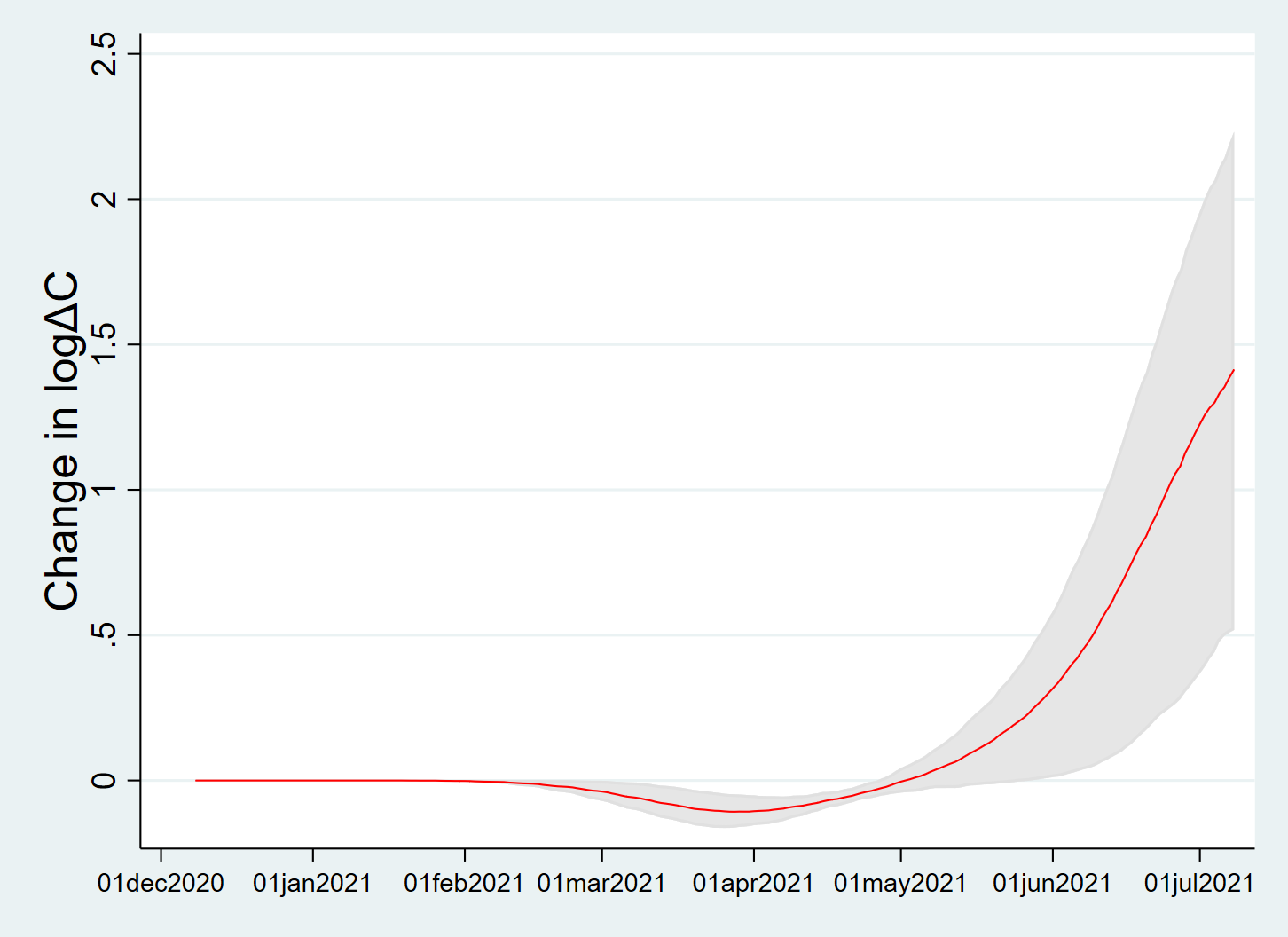}
				\includegraphics[width=0.45\textwidth,height=0.18\textheight]{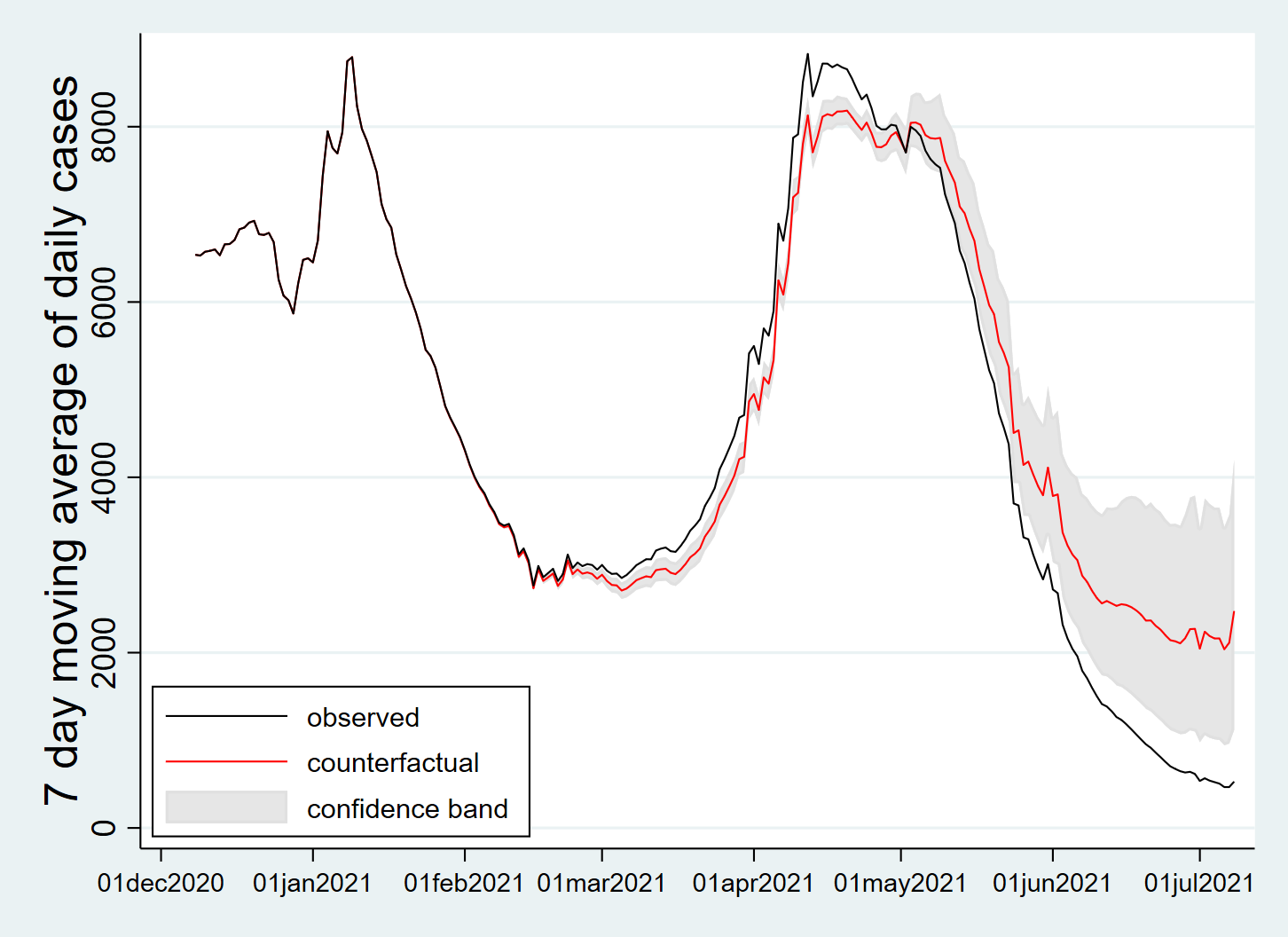}
			\end{subfigure}
			\vspace{0mm}
			%
			%
			%
			\vspace{1mm}
			\begin{subfigure}[b]{\textwidth}
				\centering
				\caption[]%
				{{\small US}}
				\label{fig: counterfactual case - USA}
				\includegraphics[width=0.45\textwidth,height=0.18\textheight]{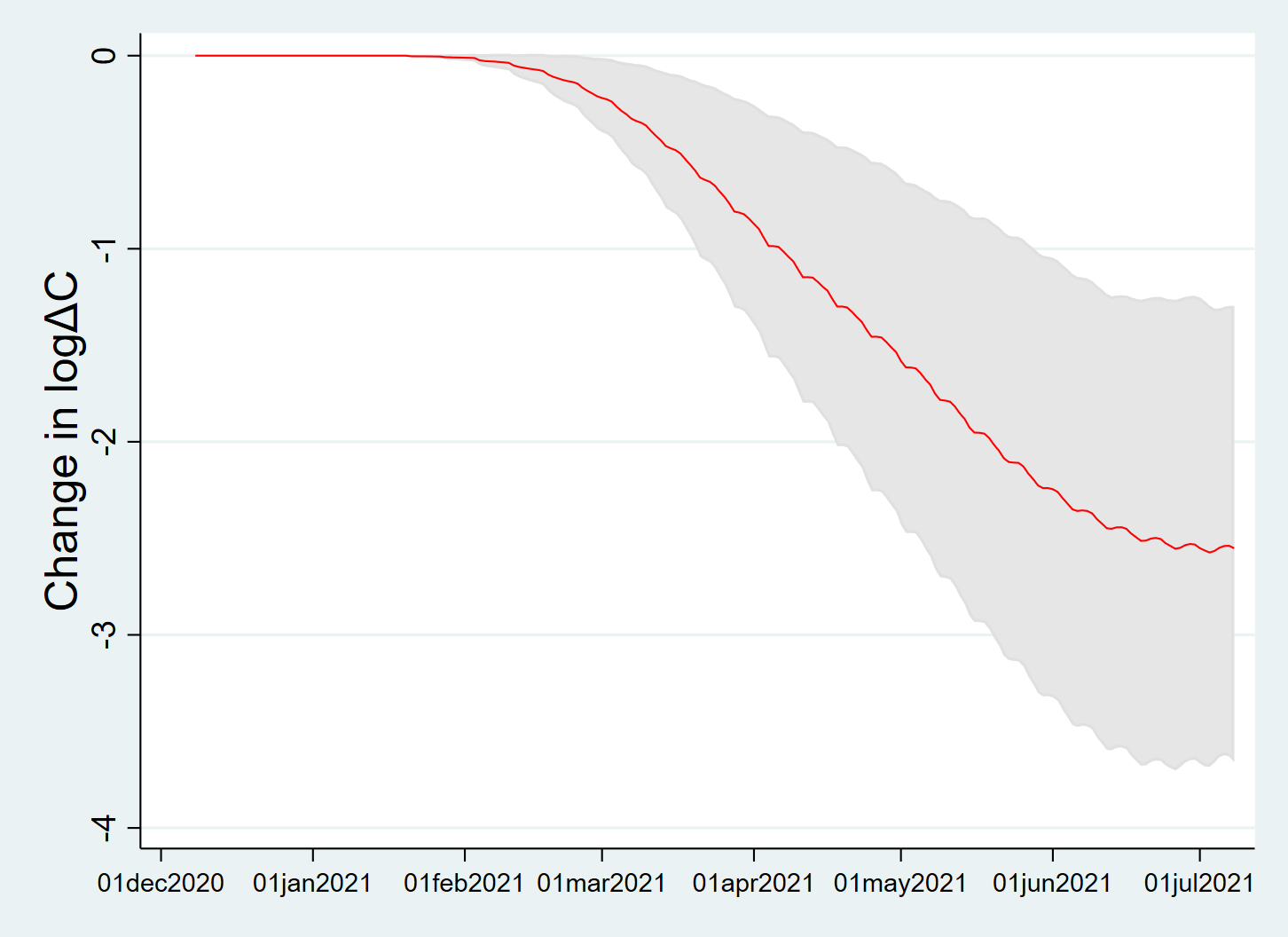}
				\includegraphics[width=0.45\textwidth,height=0.18\textheight]{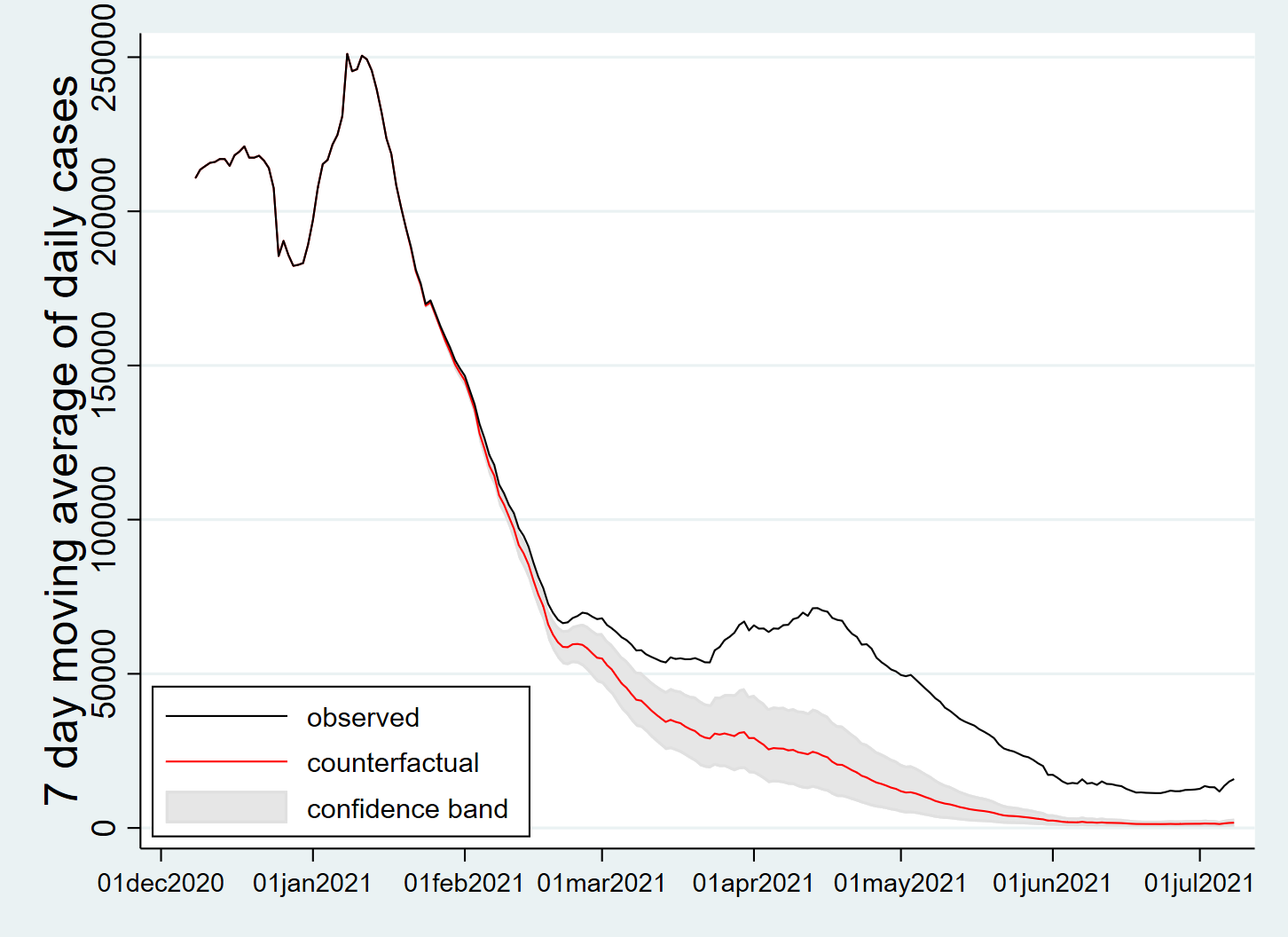}
			\end{subfigure}
		\end{center}
		\footnotesize\textbf{Note}: In the left panels, red solid lines are changes in log of weekly case counts due to altered vaccine schedules. In the right panels, black solid lines are actual case numbers and red solid lines are counterfactual counts. Shaded areas are 90\% confidence bands.
	\end{figure}
	
	\begin{figure}[b!]
		\begin{center}
			\caption{Counterfactual death counts}
			\label{fig: counterfactual death}
			\vspace{-3mm}
			\begin{subfigure}[b]{\textwidth}
				\centering
				\caption[]%
				{{\small Canada}}
				\label{fig: counterfactual death - Canada}
				\includegraphics[width=0.45\textwidth,height=0.18\textheight]{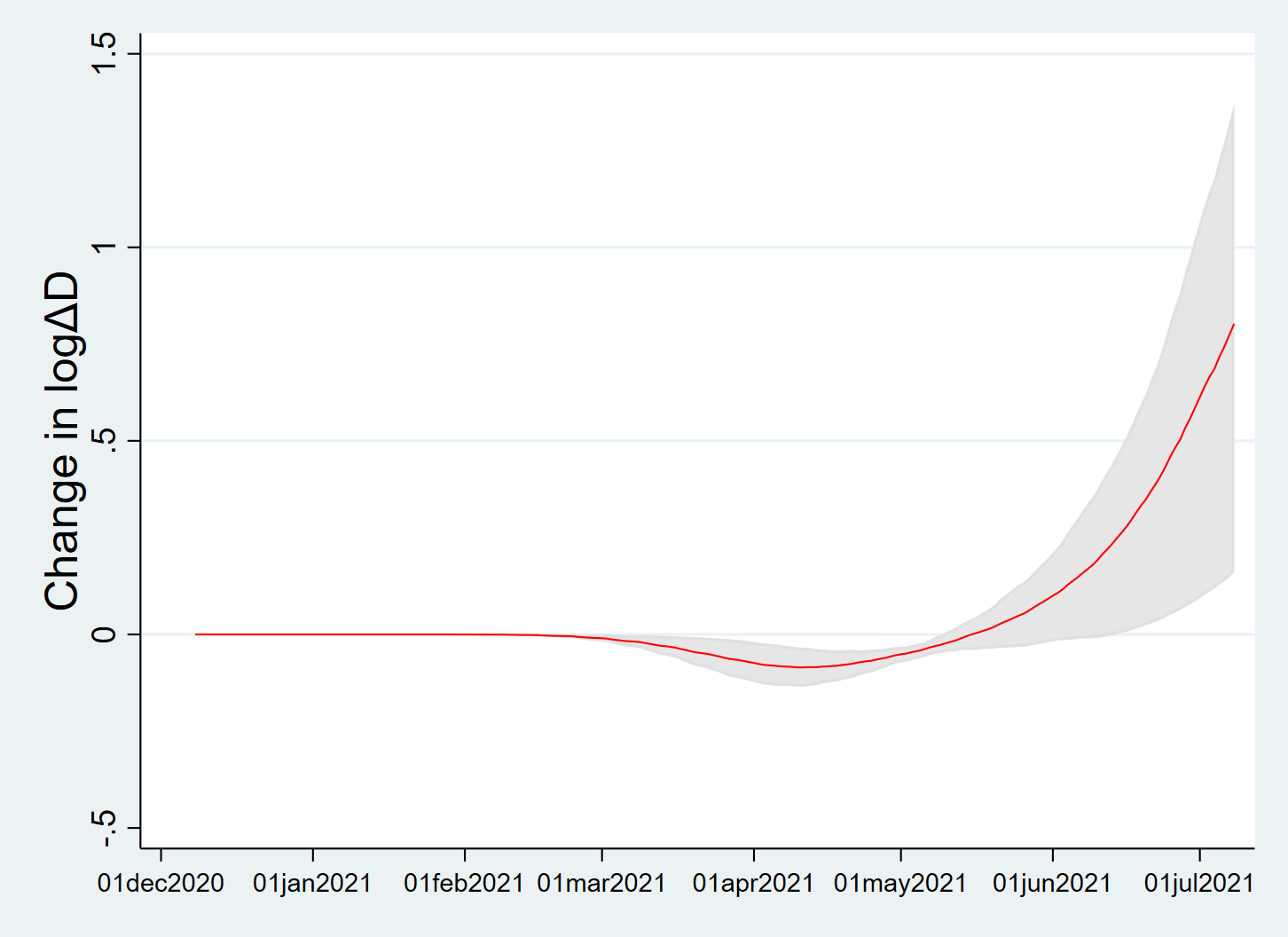}
				\includegraphics[width=0.45\textwidth,height=0.18\textheight]{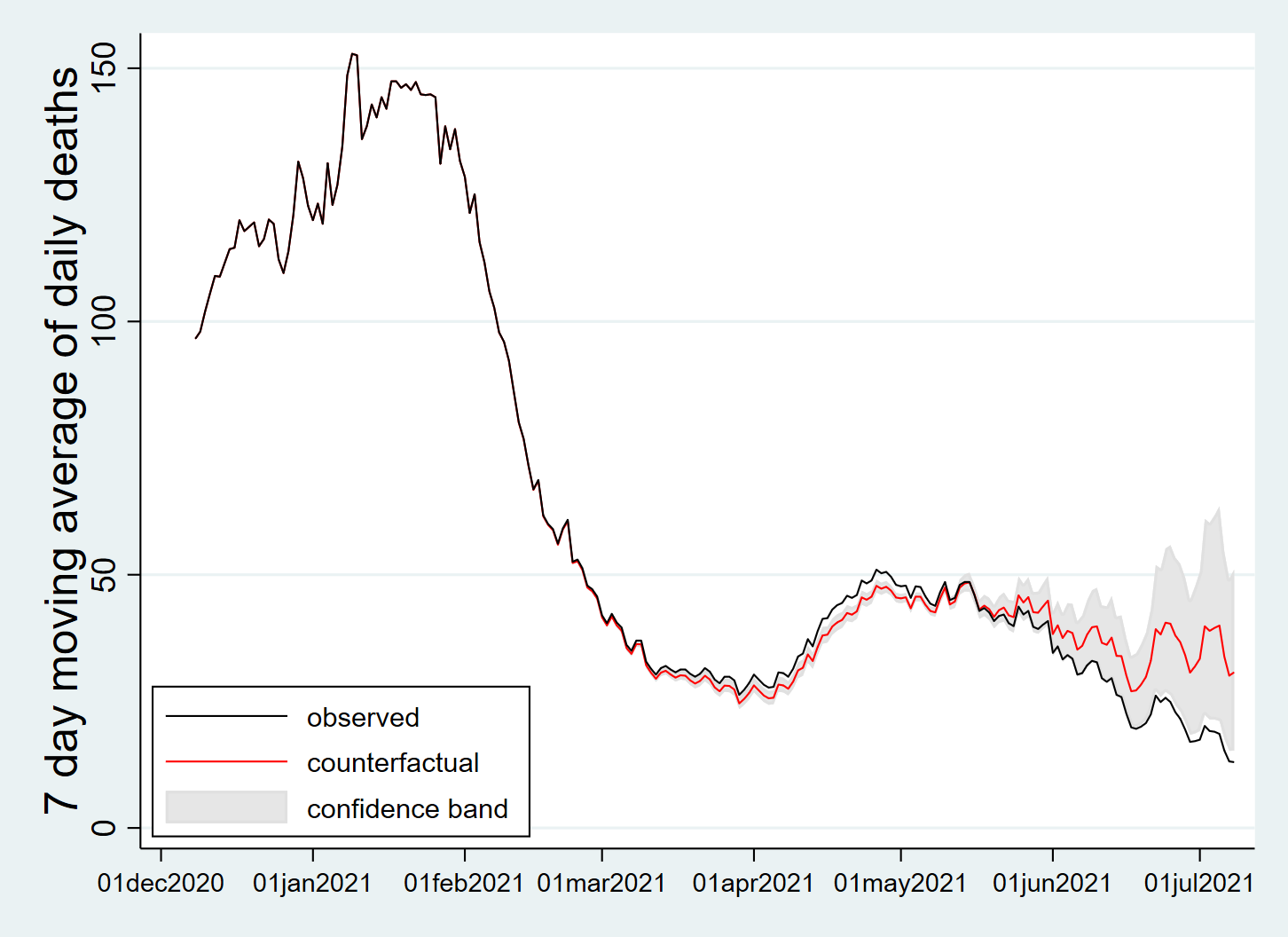}
			\end{subfigure}
			\vspace{0mm}
			%
			%
			%
			\vspace{1mm}
			\begin{subfigure}[b]{\textwidth}
				\centering
				\caption[]%
				{{\small US}}
				\label{fig: counterfactual death - USA}
				\includegraphics[width=0.45\textwidth,height=0.18\textheight]{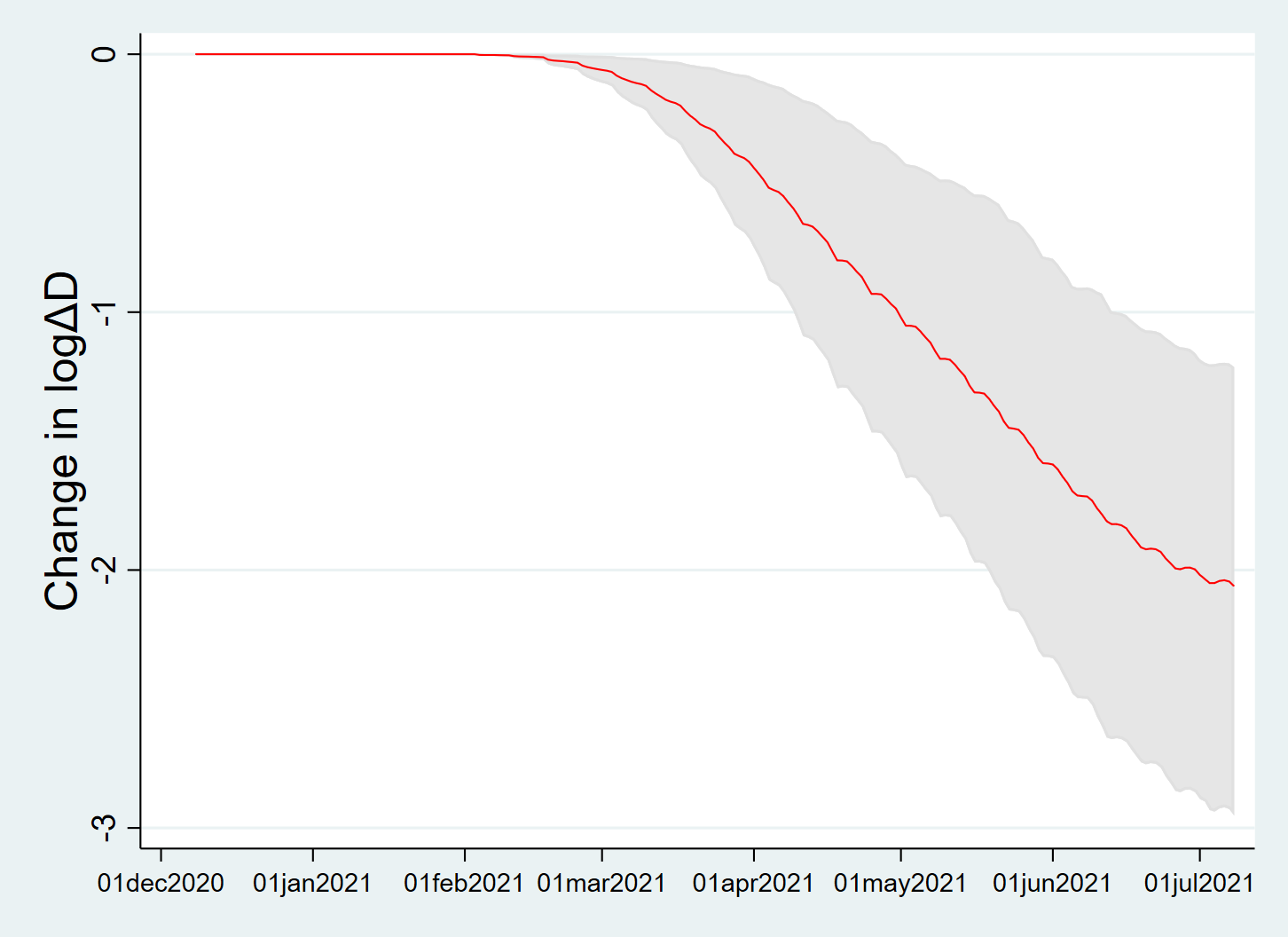}
				\includegraphics[width=0.45\textwidth,height=0.18\textheight]{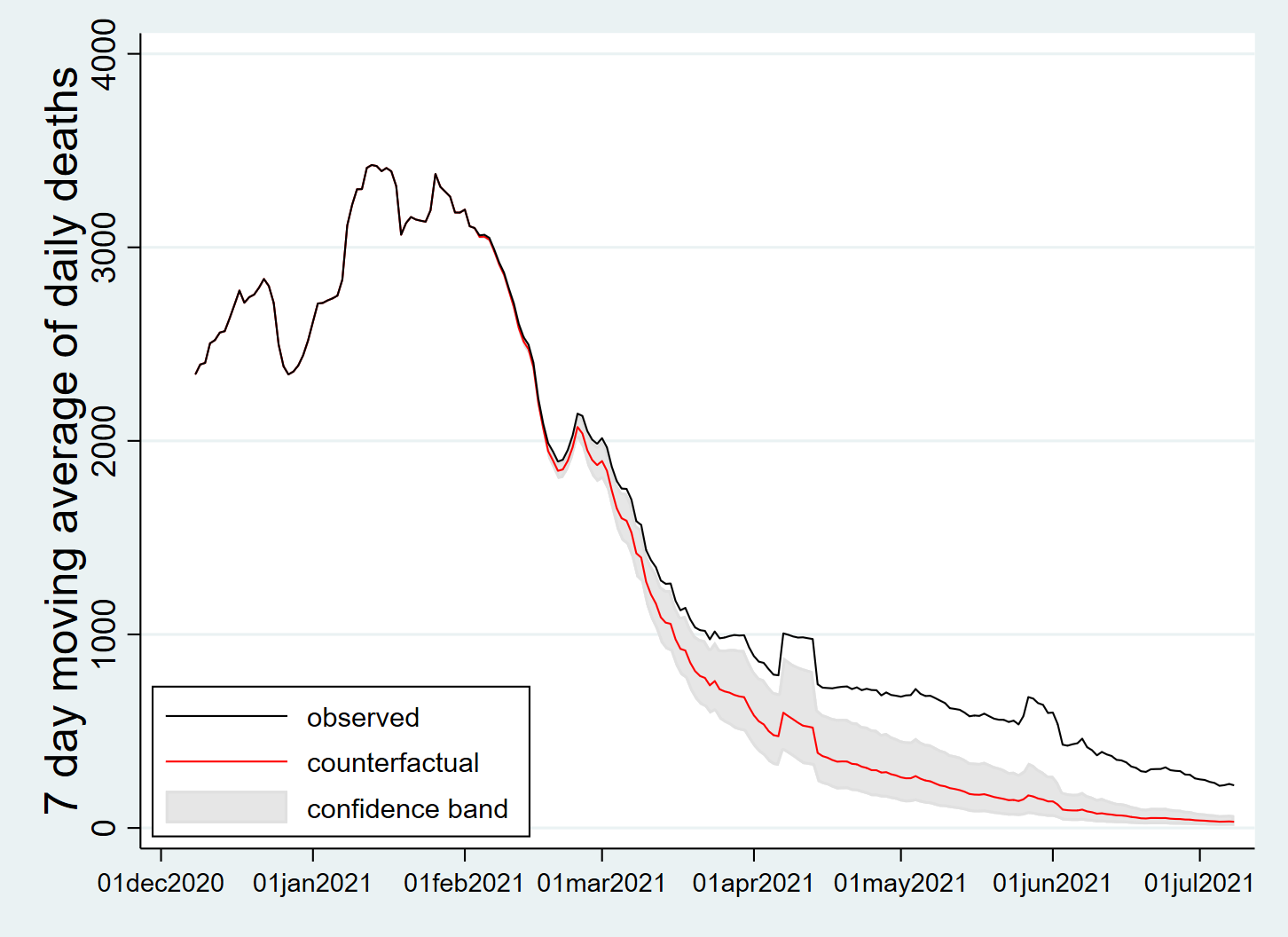}
			\end{subfigure}
		\end{center}
		\footnotesize\textbf{Note}: In the left panels, red solid lines are changes in log of weekly case counts due to altered vaccine schedules. In the right panels, black solid lines are actual case numbers and red solid lines are counterfactual counts. Shaded areas are 90\% confidence bands.
	\end{figure}
			
	Figure \ref{fig: counterfactual case} shows the counterfactual simulations for cases counts. We draw changes in log of weekly case counts due to altered vaccine schedules in the left panels. Actual and counterfactual case numbers are plotted in the right panels. For Canada, the case counts substantially increase compared to their actual counterparts under the hypothetical vaccination schedule. The counterfactual outcomes are significantly different from the actual outcomes. The daily case counts decrease by 274 on average between Feb 1 and Apr 30, 2021 as first-dose progress is slightly faster in the hypothetical vaccination schedule in the earlier period. The case numbers increase on average by 1,014 between May 1 and Jul 8, 2021 as the counterfactual first-dose progress goes below the actual progress. On the contrary, the counterfactual outcomes become significantly lower than the actual values for the US as first-dose progress is much faster in the hypothetical scenario. Between Feb 1 and Jul 8, 2021, the estimated reductions in the daily case numbers are 21,587. In terms of cumulative cases, those are translated into a 7\% increase for Canada and a 43\% reduction for the US between Feb 1 and July 8, 2021.
	
	Similar patterns are observed in the counterfactual death counts as shown in Figure \ref{fig: counterfactual death}. For Canada, the daily death counts initially decrease on average by 1.4 between Feb 1 and Apr 30, 2021 and subsequently increase on average by 6.2 between May 1 and Jul 8, 2021. The average reduction in daily death numbers is 271 for the US from Feb 1 to Jul 8, 2021. In terms of cumulative death counts, those are translated into a 4.6\% increase for Canada and a 25.9\% reduction for the US between Feb 1 and July 8, 2021.
	
	In alternative hypothetical scenarios, we adjust the dosing interval to 6 weeks and 12 weeks. The results are provided in Supplementary Materials \ref{appen: Additional counterfactuals}. For the US, gaps between counterfactuals and observed outcomes widen as the dosing interval becomes longer. When it comes to Canada, the 6-week dosing regimen results in worse health outcomes than the 8-week interval. Under the 12-week dosing regimen, the hypothetical vaccination rates closely follow the actual rates, but the hypothetical first-dose progress is slightly faster in the earlier vaccination stage, as Canada delayed the second dose in March 2021. Hence reductions in cases and deaths are observed when the 12-week dosing interval is imposed from the beginning.
	
	The counterfactual outcomes are calculated under the assumption that government policies remain at their actual values observed in the data. As government policies were eased in response to the decline in new cases, this would be also the case in the counterfactual situations where the case numbers still drop along with vaccination progress. If we adjust the government policy index according to the relative changes in case and death counts, the gaps between the counterfactual outcomes and the actual outcomes will shrink.
	
	\section{Discussion}
	
	\subsection{Vaccination strategy across heterogeneous populations}
	There were elevated concerns about vulnerable sub-populations such as the elderly, healthcare workers, and immunocompromised people when the UK and Canada decided to extend the dosing interval. The COVID-19-related mortality risk has been reported to be higher for the elderly and clinically vulnerable people with underlying medical conditions. Front-line healthcare providers work in direct contact with COVID-19 infected patients as well as other vulnerable populations. Therefore, these groups have to be prioritized for full vaccination if the marginal efficacy gain from the second dose is large. Nonetheless, the UK and Canada both extended the dosing interval for the whole population based on the rationale described in Section 2 (see \cite{nhsletter}, \cite{dhsc}, and \cite{naci}). \cite{skowronskisingle} report that single-dose vaccination provides robust protection against the original strain and the variants of concern for adults 70 years or older using data from British Columbia, Canada. More recently, similar results have been reported by \cite{shrotri2021vaccine} using the UK data. Thus the vulnerable groups can benefit from receiving their first doses more quickly under the extended dosing interval scheme. Furthermore, maximizing the number of vaccinated people can reduce community transmission and overcrowding in hospitals. A larger fraction of vaccinated people in the population lowers the risk of exposure for the vulnerable people, as our panel data analysis suggests. The reduced healthcare burden also helps the elderly and vulnerable people access non-COVID-19-related medical services.
	
	\subsection{Variants of concern}
	Each country suffers from different variants of concern, some of which are known to be more easily transmitted than the original strain. We try to control this by using country fixed effects and country-specific time trends. With or without time trends, our estimates remain with the same signs across different model specifications. In the period of our analysis, a number of countries such as the UK and Israel experienced another wave due to the Delta variant. When we exclude the period of the new wave, all the estimates still remain very similar. As the Delta variant has become the most dominant variant in many countries, there may be a structural break in the evolution of the pandemic. The period after the rise of the Delta variant is an avenue of future research. Provided single-dose vaccination is still highly effective against new variants (although somewhat weaker than against the original strain), the policy implications drawn from our findings remain valid.
	
	The evidence on single-dose effectiveness of approved vaccines against symptomatic infection is mixed, but more recent studies tend to report that the vaccine effectiveness against the Delta variant is solid. \cite{lopez2021effectiveness} report that one dose effectiveness against the Delta variant is much lower for Comirnaty (35.6\%) and Vaxzevria (30\%) than the effectiveness with two doses (Comirnaty: 88\%, Vaxzevria: 67\%) using data until May 16, 2021, from England. However, more recent research (\cite{pouwels2021effect}) from the UK reports that the marginal efficacy gain from the first dose against the Delta variant (Comirnaty: 58\%, Spikevax: 75\%, Vaxzevria: 43\%) is larger.\footnote{The reported vaccine effectiveness 14 days after the second dose is Comirnaty: 82\% and Vaxzevria: 67\%. They have insufficient data for the second dose of Spikevax.} \cite{nasreen2021effectiveness} also find similar vaccine effectiveness with one dose against the Delta variant (Comirnaty: 57\%, Spikevax: 70\%, Vaxzevria: 68\%) using Canadian data. \cite{tang2021bnt162b2} observe that the marginal efficacy gain from the second dose is negligible for mRNA-based vaccines using Qatari data.
	
	Vaccine effectiveness against severe illness (hospitalization or death) caused by the Delta variant is shown to be much higher than for symptomatic infection after partial vaccination. \cite{nasreen2021effectiveness} estimate that vaccine effectiveness against severe outcomes after single-dose inoculation of Comirnaty, Spikevax, and Vaxzevria was 81\%, 90\%, and 91\% respectively. A technical report by Public Health England (\cite{stowe2021uk}) and \cite{tang2021bnt162b2} also strongly agree with the Canadian results on vaccine effectiveness against hospitalization only after dose 1. This implies that maximizing the number of at least partially vaccinated people is an effective strategy to prevent worse health outcomes, which are of primary policy interest.
	
	Of course, worse variants can arise. The Omicron variant is quickly becoming a dominant variant in many countries. There have been elevated concerns that vaccine effectiveness could be substantially lower against the Omicron variant than against the Delta variant (see \cite{karim2021omicron} and \cite{callaway2021bad}). However, information on first-dose vaccine effectiveness against the new variant is very limited.\footnote{\cite{public2021sars} reports vaccine effectiveness against the Omicron variant using the UK data. First-dose effectiveness of Comirnaty against symptomatic infections was slightly lower than against the delta variant. However, the Omicron case number used in this analysis for first-dose effectiveness is very small (28), which resulting in a wide confidence band.} Countries can always modify their vaccination strategy depending upon updated medical evidence on the vaccine effectiveness against new variants and their policy goal (minimizing new infections or severe illness).
	
	\subsection{Waning effects}
	
	Countries with high vaccination rates have repeatedly reported that the vaccine effectiveness after the second dose is decreasing over time.   \cite{andrews2021vaccine} show that the vaccine effectiveness against symptomatic infection peaked a week after the second dose and started to substantially drop over time (Comirnaty: 92.4\% $\to$ 69.7\% (+20 weeks), Vaxzevria: 62.7\% $\to$ 47.3\% (+20 weeks)) using the UK data. \cite{nanduri2021effectiveness} also report that the effectiveness of mRNA vaccines against infection declined from 74.7\% (Mar-May 2021) to 53.1\% (Jun-Jul 2021) among elderly nursing home residents in the US. \cite{chemaitelly2021waning} observe the same pattern in the Qatari data. \cite{goldberg2021waning} provide a detailed breakdown of the waning effectiveness by age group using the Israeli data. The rate ratio for infection among people fully vaccinated in the month when they were first eligible was around 1.6, as compared with individuals fully immunized 2 months later, for all age groups. When it comes to severe illness (hospitalization and death), \cite{andrews2021vaccine} report limited waning in effectiveness. On the contrary,  \cite{chemaitelly2021waning} and \cite{goldberg2021waning} show substantial waning in protection.
	
	It is not clear whether the waning effectiveness comes from decreased immunity or more new infectious variants. It may be from both. An additional confounding factor is that fully vaccinated people tend to increase social interactions, which increases the exposure risk among the vaccinated. Nonetheless, the recent results in the literature clearly suggest the immunity against infection wanes in a few months after receiving the second dose. As far as the first dose itself provides robust protection for 3--4 months, extending the dosing interval could potentially protect vaccinated people longer because the timing of receiving the second dose is delayed, and the delay may produce stronger immune responses.
	
	\section{Concluding Remarks}
	We evaluate the impact of vaccination progress and types of vaccines on the spread of COVID-19 using cross-country data. We exploit variations in vaccination rates over time and across countries using standard time-series and panel data models. We find a significant negative association between vaccination progress and the transmission of COVID-19. The key factor is the share of vaccinated people with at least one dose, which leads to significant reductions in new infections and deaths. Second-dose progress offered no further reductions, at least in the short run. For vaccines from China, we find significant effects on new cases and deaths only after full vaccination. Vaccination progress increases people's mobility, which indirectly partially offsets effects on growth rates of new infections and deaths. Our counterfactual experiments show that extending the interval between two vaccine doses can substantially reduce new cases and deaths.
	
	Our findings support the idea of delaying the second dose. As many advanced countries use the remaining vaccine orders to administer booster shots (the third dose), the supply of mRNA-based vaccines will be likely to be limited for middle- and low-income countries. Given the constraints, a more efficient allocation strategy for available vaccine stocks would be to maximize the number of (at least) partially vaccinated people. However, this does not appear to apply to vaccines from China. Countries relying on those vaccines may need to administer both shots within the recommended interval. Another policy implication is that the indirect effects of vaccination offset its direct effects on health outcomes via increased mobility. Many countries eased public health restrictions in line with the decline in new infections. Lower case numbers, vaccination progress, and eased restrictions all lead to greater mobility. Hence maintaining policy restrictions helps contain the virus by reducing mobility and social interactions. We do not focus on each public health measure and only include a composite index of the overall stringency of government responses in our analysis. As coefficient estimates on this index remain negative and significant across all the specifications considered, our results further support the previous studies on the effectiveness of other policy interventions.
	
	Since the Delta variant emerged, countries with high vaccination rates have experienced a surge in new infections. The vaccine effectiveness of approved vaccines against the new variant is substantially lower than against the original strain. However, the single dose efficacy of mRNA and adenoviral vaccines is still higher than the marginal efficacy gain from the second dose. It is also shown that those vaccines remain highly effective against severe illness, even with only one dose. This implies that delaying the second dose can be an effective strategy to prevent worse health outcomes, hospitalization, and death, amid the rise of the Delta variant. Severe illness caused by the virus is a primary policy concern because it leads to a great burden on the healthcare system. It may not be possible to reduce case counts to zero by vaccination, but the widely provided protection offered by vaccines can make the disease more manageable. The same implications could apply for new variants provided single-dose effectiveness against severe illness remains solid.
	
	We conclude that vaccination has indeed been a very powerful weapon to battle the COVID-19 pandemic. However, it does not mean that other policy interventions can be quickly phased out as vaccination rates reach a high enough level. Containment measures are still valuable tools to supplement the vaccination efforts. We focus on the impact of vaccination on epidemiological outcomes. An interesting research avenue is the impact of vaccination, along with eased restrictions, on economic activities such as retail sales, employment, and economic growth as recently studied in \cite{dave2021statewide}.
	
	\bibliography{covid19-final}

\begin{thebibliography}{}

\bibitem[Andersson et~al., 2021]{anderssonjheforth}
Andersson, O., Campos-Mercade, P., Meier, A.~N., and Wengström, E. (2021).
\newblock Anticipation of {COVID}-19 vaccines reduces willingness to socially
  distance.
\newblock {\em Journal of Health Economics}, 80:102530.

\bibitem[Andrews et~al., 2021]{andrews2021vaccine}
Andrews, N., Tessier, E., Stowe, J., Gower, C., Kirsebom, F., Simmons, R.,
  Gallagher, E., Chand, M., Brown, K., Ladhani, S.~N., Ramsay, M., and Bernal,
  J.~L. (2021).
\newblock Vaccine effectiveness and duration of protection of {C}omirnaty,
  {V}axzevria and {S}pikevax against mild and severe {COVID}-19 in the {UK}.
\newblock {\em medRxiv
  [\url{https://www.medrxiv.org/content/early/2021/10/06/2021.09.15.21263583}]}.

\bibitem[Angyal et~al., 2022]{angyal2021t}
Angyal, A., Longet, S., Moore, S., Payne, R.~P., Harding, A., Tipton, T.,
  Rongkard, P., Ali, M., Hering, L.~M., Meardon, N., et~al. (2022).
\newblock T-cell and antibody responses to first {BNT}162b2 vaccine dose in
  previously infected and {SARS}-{CoV}-2-naive {UK} health-care workers: a
  multicentre prospective cohort study.
\newblock {\em The Lancet Microbe}, 3(1):e21--e31.

\bibitem[Atherton et~al., 2020]{dhsc}
Atherton, F., McBride, M., Smith, G., Whitty, C., and Van-Tam, J. (2020).
\newblock Letter to the profession from the {UK} {C}hief {M}edical {O}fficers
  regarding the uk {COVID}-19 vaccination programmes.
\newblock {\em Department for Health and Social Care
  [\url{https://www.gov.uk/government/publications/letter-to-the-profession-from-the-uk-chief-medical-officers-on-the-uk-covid-19-vaccination-programmes}]}.

\bibitem[Bernal et~al., 2021]{lopez2021effectiveness}
Bernal, J.~L., Andrews, N., Gower, C., Gallagher, E., Simmons, R., Thelwall,
  S., Stowe, J., Tessier, E., Groves, N., Dabrera, G., Myers, R., Campbell,
  C.~N., Amirthalingam, G., Edmunds, M., Zambon, M., Brown, K.~E., Hopkins, S.,
  Chand, M., and Ramsay, M. (2021).
\newblock Effectiveness of {C}ovid-19 vaccines against the {B}.1.617.2 (delta)
  variant.
\newblock {\em New England Journal of Medicine}, 385(7):585--594.

\bibitem[Callaway and Ledford, 2021]{callaway2021bad}
Callaway, E. and Ledford, H. (2021).
\newblock How bad is {O}micron? {W}hat scientists know so far.
\newblock {\em Nature}, 600(7888):197--199.

\bibitem[Carazo et~al., 2021]{carazosingle}
Carazo, S., Talbot, D., Boulianne, N., Brisson, M., Gilca, R., Deceuninck, G.,
  Brousseau, N., Drolet, M., Ouakki, M., Sauvageau, C., Barkati, S., Fortin,
  {\'{E}}., Carignan, A., Wals, P.~D., Skowronski, D.~M., and Serres, G.~D.
  (2021).
\newblock Single-dose messenger {RNA} vaccine effectiveness against severe
  acute respiratory syndrome coronavirus 2 in healthcare workers extending 16
  weeks postvaccination: A test-negative design from {Q}u{\'{e}}bec, {C}anada.
\newblock {\em Clinical Infectious Diseases}, ciab739.

\bibitem[Chemaitelly et~al., 2021]{chemaitelly2021waning}
Chemaitelly, H., Tang, P., Hasan, M.~R., AlMukdad, S., Yassine, H.~M.,
  Benslimane, F.~M., Al~Khatib, H.~A., Coyle, P., Ayoub, H.~H., Al~Kanaani, Z.,
  et~al. (2021).
\newblock Waning of {BNT}162b2 vaccine protection against {SARS}-{C}o{V}-2
  infection in qatar.
\newblock {\em New England Journal of Medicine}, 385(24):e83.

\bibitem[Chernozhukov et~al., 2021]{chernozhukov2021causal}
Chernozhukov, V., Kasahara, H., and Schrimpf, P. (2021).
\newblock Causal impact of masks, policies, behavior on early covid-19 pandemic
  in the {U}.{S}.
\newblock {\em Journal of Econometrics}, 220(1):23--62.
\newblock Pandemic Econometrics.

\bibitem[Dashdorj et~al., 2021]{dashdorj2021direct}
Dashdorj, N.~J., Wirz, O.~F., R{\"o}ltgen, K., Haraguchi, E., Buzzanco~III,
  A.~S., Sibai, M., Wang, H., Miller, J.~A., Solis, D., Sahoo, M.~K., et~al.
  (2021).
\newblock Direct comparison of antibody responses to four {SARS}-{C}o{V}-2
  vaccines in mongolia.
\newblock {\em Cell Host \& Microbe}, 29(12):1738--1743.e4.

\bibitem[Dave et~al., 2021]{dave2021statewide}
Dave, D.~M., Sabia, J.~J., and Safford, S. (2021).
\newblock The limits of reopening policy to alter economic behavior: New
  evidence from texas.
\newblock Working Paper 28804 [\url{http://www.nber.org/papers/w28804}],
  National Bureau of Economic Research.

\bibitem[Goldberg et~al., 2021]{goldberg2021waning}
Goldberg, Y., Mandel, M., Bar-On, Y.~M., Bodenheimer, O., Freedman, L., Haas,
  E.~J., Milo, R., Alroy-Preis, S., Ash, N., and Huppert, A. (2021).
\newblock Waning immunity after the {BNT}162b2 vaccine in israel.
\newblock {\em New England Journal of Medicine}, 385(24):e85.

\bibitem[Hall et~al., 2021]{hall2021covid}
Hall, V.~J., Foulkes, S., Saei, A., Andrews, N., Oguti, B., Charlett, A.,
  Wellington, E., Stowe, J., Gillson, N., Atti, A., et~al. (2021).
\newblock {COVID}-19 vaccine coverage in health-care workers in england and
  effectiveness of {BNT}162b2 {mRNA} vaccine against infection ({SIREN}): a
  prospective, multicentre, cohort study.
\newblock {\em The Lancet}, 397(10286):1725--1735.

\bibitem[Hunter and Brainard, 2021]{hunter2021estimating}
Hunter, P.~R. and Brainard, J. (2021).
\newblock Estimating the effectiveness of the {P}fizer {COVID}-19 {BNT}162b2
  vaccine after a single dose. a reanalysis of a study of
  {\textquoteleft}real-world{\textquoteright} vaccination outcomes from
  {I}srael.
\newblock {\em medRxiv
  [\url{https://www.medrxiv.org/content/early/2021/02/03/2021.02.01.21250957}]}.

\bibitem[Hyndman et~al., 2020]{Rforcast}
Hyndman, R.~J., Athanasopoulos, G., Bergmeir, C., Caceres, G., Chhay, L.,
  O'Hara-Wild, M., Petropoulos, F., Razbash, S., and Wang, E. (2020).
\newblock Package ‘forecast’.
\newblock {\em [\url{https://cran. r-project.
  org/web/packages/forecast/forecast.pdf}]}.

\bibitem[Karaivanov et~al., 2021]{karaivanov2021face}
Karaivanov, A., Lu, S.~E., Shigeoka, H., Chen, C., and Pamplona, S. (2021).
\newblock Face masks, public policies and slowing the spread of {COVID}-19:
  evidence from {C}anada.
\newblock {\em Journal of Health Economics}, 78:102475.

\bibitem[Karim and Karim, 2021]{karim2021omicron}
Karim, S. S.~A. and Karim, Q.~A. (2021).
\newblock Omicron {SARS}-{CoV}-2 variant: a new chapter in the {COVID}-19
  pandemic.
\newblock {\em The Lancet}, 398(10317):2126--2128.

\bibitem[Kitagawa and Wang, 2021]{kitagawa2021should}
Kitagawa, T. and Wang, G. (2021).
\newblock Who should get vaccinated? {I}ndividualized allocation of vaccines
  over {SIR} network.
\newblock {\em Journal of Econometrics}, In press.

\bibitem[Mahase, 2020]{bma_second_dose}
Mahase, E. (2020).
\newblock Covid-19: Order to reschedule and delay second vaccine dose is
  {\textquotedblleft}totally unfair,{\textquotedblright} says {BMA}.
\newblock {\em BMJ}, 371.

\bibitem[Miller, 2021]{cbcnews2021}
Miller, A. (2021).
\newblock Why {C}anada's decision to delay 2nd doses of {COVID}-19 vaccines may
  not work for everyone.
\newblock {\em CBC News
  [\url{https://www.cbc.ca/news/health/vaccine-dose-delay-canada-covid-19-research-1.5965996}]}.

\bibitem[Moghadas et~al., 2021]{moghadas2021evaluation}
Moghadas, S.~M., Vilches, T.~N., Zhang, K., Nourbakhsh, S., Sah, P.,
  Fitzpatrick, M.~C., and Galvani, A.~P. (2021).
\newblock Evaluation of {COVID}-19 vaccination strategies with a delayed second
  dose.
\newblock {\em {PLOS} Biology}, 19(4):e3001211.

\bibitem[Nanduri et~al., 2021]{nanduri2021effectiveness}
Nanduri, S., Pilishvili, T., Derado, G., Soe, M.~M., Dollard, P., Wu, H., Li,
  Q., Bagchi, S., Dubendris, H., Link-Gelles, R., et~al. (2021).
\newblock Effectiveness of {P}fizer-{BioNTech} and moderna vaccines in
  preventing {SARS}-{CoV}-2 infection among nursing home residents before and
  during widespread circulation of the {SARS}-{CoV}-2 b.1.617.2 (delta) variant
  {\textemdash} national healthcare safety network, march 1{\textendash}august
  1, 2021.
\newblock {\em {MMWR}. Morbidity and Mortality Weekly Report},
  70(34):1163--1166.

\bibitem[Nasreen et~al., 2021]{nasreen2021effectiveness}
Nasreen, S., Chung, H., He, S., Brown, K.~A., Gubbay, J.~B., Buchan, S.~A.,
  et~al. (2021).
\newblock Effectiveness of {mRNA} and {ChAdOx1} {COVID}-19 vaccines against
  symptomatic {SARS}-{CoV}-2 infection and severe outcomes with variants of
  concern in {O}ntario.
\newblock {\em medRxiv
  [\url{https://www.medrxiv.org/content/early/2021/09/30/2021.06.28.21259420}]}.

\bibitem[{National Advisory Committee on Immunization}, 2021]{naci}
{National Advisory Committee on Immunization} (2021).
\newblock An advisory committee statement ({ACS}).
\newblock {\em
  [\url{https://www.canada.ca/content/dam/phac-aspc/documents/services/immunization/national-advisory-committee-on-immunization-naci/recommendations-use-covid-19-vaccines/recommendations-use-covid-19-vaccines-en.pdf}]}.

\bibitem[NHS, 2020]{nhsletter}
NHS (2020).
\newblock Letter: {COVID}-19 vaccination – for immediate action – 30
  {D}ecember 2020.
\newblock {\em
  [\url{https://www.england.nhs.uk/coronavirus/wp-content/uploads/sites/52/2020/12/C0994-System-letter-COVID-19-vaccination-deployment-planning-30-December-2020.pdf}]}.

\bibitem[Payne et~al., 2021]{payne2021immunogenicity}
Payne, R.~P., Longet, S., Austin, J.~A., Skelly, D.~T., Dejnirattisai, W.,
  Adele, S., Meardon, N., Faustini, S., Al-Taei, S., Moore, S.~C., et~al.
  (2021).
\newblock Immunogenicity of standard and extended dosing intervals of
  {BNT}162b2 {mRNA} vaccine.
\newblock {\em Cell}, 184(23):5699--5714.e11.

\bibitem[Polack et~al., 2020]{polack2020safety}
Polack, F.~P., Thomas, S.~J., Kitchin, N., Absalon, J., Gurtman, A., Lockhart,
  S., Perez, J.~L., Marc, G.~P., Moreira, E.~D., Zerbini, C., et~al. (2020).
\newblock Safety and efficacy of the {BNT}162b2 {mRNA} covid-19 vaccine.
\newblock {\em New England Journal of Medicine}, 383(27):2603--2615.
\newblock PMID: 33301246.

\bibitem[Pouwels et~al., 2021]{pouwels2021effect}
Pouwels, K.~B., Pritchard, E., Matthews, P.~C., Stoesser, N., Eyre, D.~W.,
  Vihta, K.-D., House, T., Hay, J., Bell, J.~I., Newton, J.~N., Farrar, J.,
  Crook, D., Cook, D., Rourke, E., Studley, R., Peto, T. E.~A., Diamond, I.,
  and Walker, A.~S. (2021).
\newblock Effect of {D}elta variant on viral burden and vaccine effectiveness
  against new {SARS}-{CoV}-2 infections in the {UK}.
\newblock {\em Nature Medicine}, 27(12):2127--2135.

\bibitem[{Public Health England}, 2021]{public2021sars}
{Public Health England} (2021).
\newblock {SARS}-{C}o{V}-2 variants of concern and variants under investigation
  in {E}ngland.
\newblock {\em technical briefing 12
  [\url{https://assets.publishing.service.gov.uk/government/uploads/system/uploads/attachment_data/file/991343/Variants_of_Concern_VOC_Technical_Briefing_14.pdf}]}.

\bibitem[Romero-Brufau et~al., 2021]{romero2021public}
Romero-Brufau, S., Chopra, A., Ryu, A.~J., Gel, E., Raskar, R., Kremers, W.,
  Anderson, K.~S., Subramanian, J., Krishnamurthy, B., Singh, A., Pasupathy,
  K., Dong, Y., O{\textquoteright}Horo, J.~C., Wilson, W.~R., Mitchell, O., and
  Kingsley, T.~C. (2021).
\newblock Public health impact of delaying second dose of {BNT}162b2 or
  {mRNA}-1273 covid-19 vaccine: simulation agent based modeling study.
\newblock {\em BMJ}, 373.

\bibitem[Saad-Roy et~al., 2021]{Saad_Roy_2021}
Saad-Roy, C.~M., Morris, S.~E., Metcalf, C. J.~E., Mina, M.~J., Baker, R.~E.,
  Farrar, J., Holmes, E.~C., Pybus, O.~G., Graham, A.~L., Levin, S.~A.,
  Grenfell, B.~T., and Wagner, C.~E. (2021).
\newblock Epidemiological and evolutionary considerations of {SARS}-{CoV}-2
  vaccine dosing regimes.
\newblock {\em Science}, 372(6540):363--370.

\bibitem[Samanovic et~al., 2021]{samanovic2021poor}
Samanovic, M.~I., Cornelius, A.~R., Gray-Gaillard, S.~L., Allen, J.~R.,
  Karmacharya, T., Wilson, J.~P., Hyman, S.~W., Tuen, M., Koralov, S.~B.,
  Mulligan, M.~J., and Herati, R.~S. (2021).
\newblock Robust immune responses are observed after one dose of {BNT}162b2
  {mRNA} vaccine dose in {SARS}-{CoV}-2 experienced individuals.
\newblock {\em Science Translational Medicine}, first release:eabi8961.

\bibitem[Shrotri et~al., 2021]{shrotri2021vaccine}
Shrotri, M., Krutikov, M., Palmer, T., Giddings, R., Azmi, B., Subbarao, S.,
  Fuller, C., Irwin-Singer, A., Davies, D., Tut, G., {Lopez Bernal}, J., Moss,
  P., Hayward, A., Copas, A., and Shallcross, L. (2021).
\newblock Vaccine effectiveness of the first dose of {ChAdOx1} {nCoV}-19 and
  {BNT}162b2 against {SARS}-{CoV}-2 infection in residents of long-term care
  facilities in {E}ngland ({VIVALDI}): a prospective cohort study.
\newblock {\em The Lancet Infectious Diseases}, 21(11):1529--1538.

\bibitem[Skowronski and De~Serres, 2021]{NEJMc2036242}
Skowronski, D.~M. and De~Serres, G. (2021).
\newblock Safety and efficacy of the {BNT}162b2 {mRNA} covid-19 vaccine.
\newblock {\em New England Journal of Medicine}, 384(16):1576--1578.
\newblock PMID: 33596348.

\bibitem[Skowronski et~al., 2021]{skowronskisingle}
Skowronski, D.~M., Setayeshgar, S., Zou, M., Prystajecky, N., Tyson, J.~R.,
  Galanis, E., Naus, M., Patrick, D.~M., Sbihi, H., Adam, S.~E., Henry, B.,
  Hoang, L. M.~N., Sadarangani, M., Jassem, A.~N., and Krajden, M. (2021).
\newblock Single-dose {mRNA} vaccine effectiveness against severe acute
  respiratory syndrome coronavirus 2 ({SARS}-{CoV}-2), including alpha and
  gamma variants: {A} test-negative design in adults 70 years and older in
  {B}ritish {C}olumbia, {C}anada.
\newblock {\em Clinical Infectious Diseases}, ciab616.

\bibitem[Stamatatos et~al., 2021]{stamatatos2021mrna}
Stamatatos, L., Czartoski, J., Wan, Y.-H., Homad, L.~J., Rubin, V., Glantz, H.,
  Neradilek, M., Seydoux, E., Jennewein, M.~F., MacCamy, A.~J., Feng, J., Mize,
  G., Rosa, S. C.~D., Finzi, A., Lemos, M.~P., Cohen, K.~W., Moodie, Z.,
  McElrath, M.~J., and McGuire, A.~T. (2021).
\newblock {mRNA} vaccination boosts cross-variant neutralizing antibodies
  elicited by {SARS}-{CoV}-2 infection.
\newblock {\em Science}, 372(6549):1413--1418.

\bibitem[Stowe et~al., 2021]{stowe2021uk}
Stowe, J., Andrews, N., Gower, C., Gallagher, E., Utsi, L., Simmons, R.,
  Thelwall, S., Tessier, E., Groves, N., Dabrera, G., Myers, R., Campbell, C.,
  Amirthalingam, G., Edmunds, M., Zambon, M., Brown, K., Hopkins, S., Chand,
  M., Ramsay, M., , and Bernal, J.~L. (2021).
\newblock Effectiveness of {COVID}-19 vaccines against hospital admission with
  the {D}elta ({B}.1.617.2) variant.
\newblock {\em
  [\url{https://khub.net/web/phe-national/public-library/-/document_library/v2WsRK3ZlEig/view/479607266}]}.

\bibitem[Tang et~al., 2021]{tang2021bnt162b2}
Tang, P., Hasan, M.~R., Chemaitelly, H., Yassine, H.~M., Benslimane, F.~M.,
  Al~Khatib, H.~A., AlMukdad, S., Coyle, P., Ayoub, H.~H., Al~Kanaani, Z.,
  et~al. (2021).
\newblock {BNT}162b2 and {mRNA}-1273 {COVID}-19 vaccine effectiveness against
  the {SARS}-{CoV}-2 {D}elta variant in {Q}atar.
\newblock {\em Nature Medicine}, 27(12):2136--2143.

\bibitem[Tauzin et~al., 2021]{tauzin2021strong}
Tauzin, A., Gong, S.~Y., Beaudoin-Bussi{\`e}res, G., V{\'e}zina, D., Gasser,
  R., Nault, L., Marchitto, L., Benlarbi, M., Chatterjee, D., Nayrac, M.,
  et~al. (2021).
\newblock Strong humoral immune responses against {SARS}-{CoV}-2 {S}pike after
  {BNT}162b2 {mRNA} vaccination with a 16-week interval between doses.
\newblock {\em Cell Host \& Microbe}, In press.

\bibitem[Tuite et~al., 2021]{tuite2021alternative}
Tuite, A.~R., Zhu, L., Fisman, D.~N., and Salomon, J.~A. (2021).
\newblock Alternative dose allocation strategies to increase benefits from
  constrained {COVID}-19 vaccine supply.
\newblock {\em Annals of Internal Medicine}, 174(4):570--572.

\bibitem[Voysey et~al., 2021]{voysey2021safety}
Voysey, M., Clemens, S. A.~C., Madhi, S.~A., Weckx, L.~Y., Folegatti, P.~M.,
  Aley, P.~K., Angus, B., Baillie, V.~L., Barnabas, S.~L., Bhorat, Q.~E.,
  et~al. (2021).
\newblock Safety and efficacy of the {ChAdOx1} {nCoV}-19 vaccine ({AZD}1222)
  against {SARS}-{CoV}-2: an interim analysis of four randomised controlled
  trials in {B}razil, {S}outh {A}frica, and the {UK}.
\newblock {\em The Lancet}, 397(10269):99--111.

\bibitem[{World Health Organization}, 2021]{world2021background}
{World Health Organization} (2021).
\newblock Background document on the inactivated vaccine {S}inovac-{CoronaVac}
  against {COVID}-19: background document to the {WHO} interim recommendations
  for use of the inactivated {COVID}-19 vaccine, {CoronaVac}, developed by
  {S}inovac, 24 {M}ay 2021.
\newblock {\em
  [\url{https://apps.who.int/iris/bitstream/handle/10665/341455/WHO-2019-nCoV-vaccines-SAGE-recommendation-Sinovac-CoronaVac-background-2021.1-eng.pdf?sequence=1}]}.

\end{thebibliography}
	\addcontentsline{toc}{chapter}{References}
	\bibliographystyle{apalike}
	
	\newpage
	\appendix
	\setcounter{page}{1}
	\setcounter{table}{0}
	\setcounter{figure}{0}
	\renewcommand{\thetable}{A\arabic{table}}
	\renewcommand{\thefigure}{A\arabic{figure}}
	
	\includepdf[pages={1}]{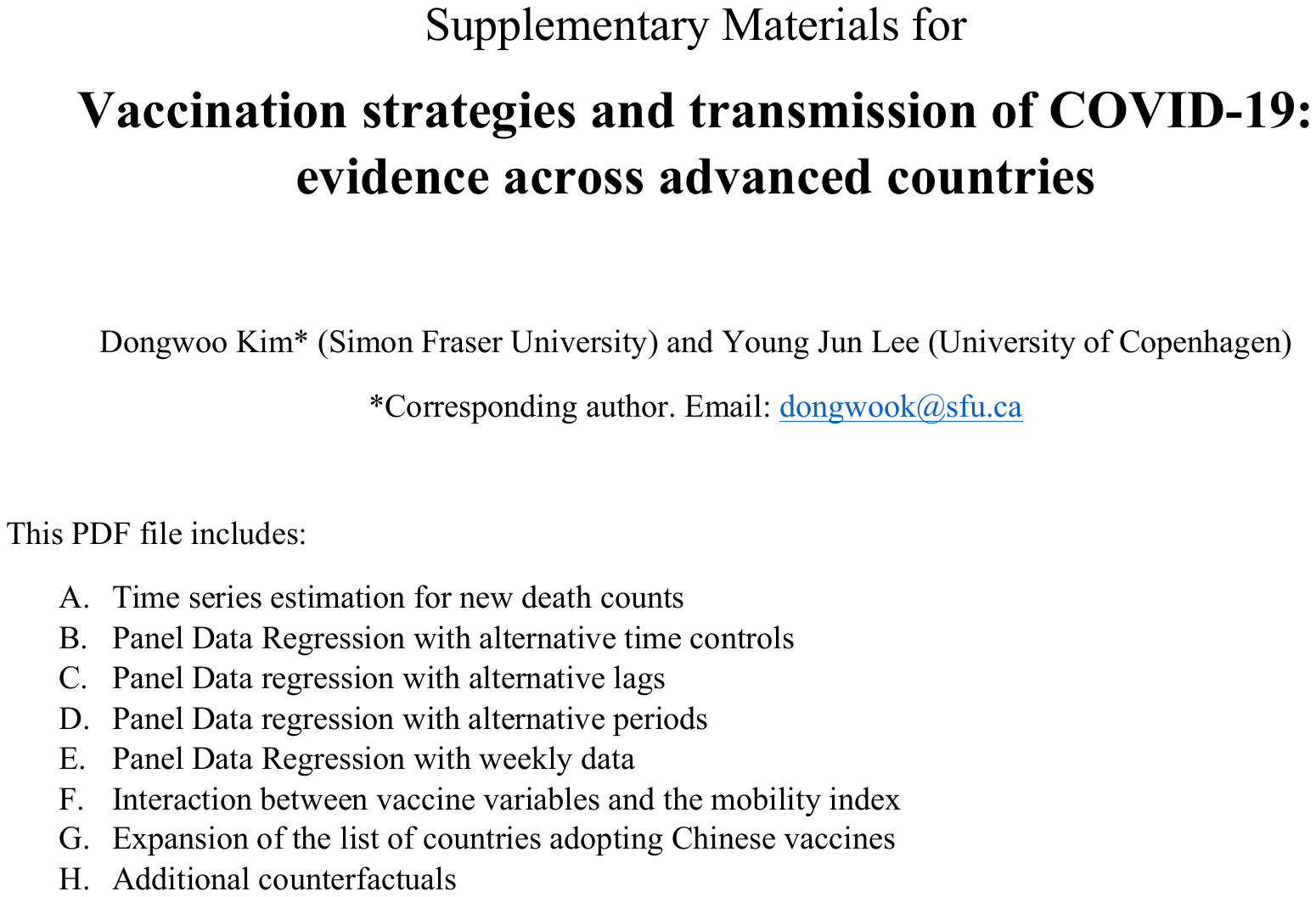}

	\section{Time series estimation for new death counts}\label{appen: time-series death}
	
	\begin{table}[h!]
		\caption{ARIMA model results for new deaths - countries relying on vaccines developed in the US and Europe}
		\label{tab: ARIMAX estimation death - Canada, Israel, UK, US}
		\begin{center}
		\small
		\begin{tabular}{@{\extracolsep{5pt}}lcccc}
			\hline
			\hline \\[-1.8ex]
			&  Canada & Israel & US & UK \\
			\cline{2-5}
			\\[-1.8ex] $(p,d,q)$& $(1,1,1)$ & $(2,1,2)$ & $(6,1,2)$ & $(4,1,2)$\\
			\hline \\[-1.8ex]
			$\Delta V1_{t-35}$ & $-$0.037$^{*}$ & 0.020 & $-$0.015 & $-$0.131$^{***}$ \\
			& (0.022) & (0.024) & (0.058) & (0.035) \\
			$\Delta V2_{t-21}$ & $-$0.015 & $-$0.061$^{**}$ & $-$0.044 & 0.057 \\
			& (0.057) & (0.025) & (0.066) & (0.056) \\
			$wkd_t$ & $-$0.194$^{***}$ & 0.003 & $-$0.611$^{***}$ & $-$0.705$^{***}$ \\
			& (0.047) & (0.025) & (0.124) & (0.054) \\
			$\Delta P_{t-28}$ & $-$0.019 & $-$0.008$^{*}$ & $-$0.002 & $-$0.002 \\
			& (0.029) & (0.004) & (0.012) & (0.007) \\
			$\Delta M_{t-28}$ & 0.003 & $-$0.005 & 0.005 & $-$0.003 \\
			& (0.013) & (0.006) & (0.010) & (0.008) \\
			\hline \\[-1.8ex]
		\end{tabular}
		\end{center}
	\footnotesize \textbf{Note:} Time period -- June 1, 2020 to July 8, 2021. The results are produced using daily data. We use lags of included regressors that are 14 days behind those in the case equation following \cite{karaivanov2021face}. $(p,d,q)$ is the optimal ARIMA order chosen by \texttt{auto.arima} in \textsf{R}. Standard errors are in parentheses. $^{***}$, $^{**}$, and $^{*}$ denote the 99\%, 95\% and 90\% confidence level respectively.
	\end{table}
	
	\newpage
	\begin{table}[h!]
		\begin{center}
			\small
			\caption{ARIMA model results for new deaths - countries relying on vaccines from China}
			\label{tab: ARIMAX estimation death - Chinese vaccine}
			\begin{tabular}{@{\extracolsep{5pt}}lcccc}
				\hline
				\hline \\[-1.8ex]
				& Chile & Uruguay & UAE & Bahrain\\
				\cline{2-5}
				\\[-1.8ex] $(p,d,q)$& $(4,1,2)$ & $(2,1,2)$ & $(2,1,1)$ & $(0,1,1)$\\
				\hline \\[-1.8ex]
				$\Delta V1_{t-35}$ & $-$0.007 & $-$0.017 &  & 0.020 \\
				& (0.021) & (0.025) &  & (0.033) \\
				$\Delta V2_{t-21}$ & 0.017 & 0.006 &  & $-$0.015 \\
				& (0.022) & (0.033) &  & (0.024) \\
				$\Delta V_{t-35}$ &  &  & $-$0.016$^{**}$ &  \\
				&  &  & (0.007) &  \\
				$wkd_t$ & $-$0.004 & 0.016 & $-$0.033 & 0.001 \\
				& (0.141) & (0.025) & (0.040) & (0.022) \\
				$\Delta P_{t-28}$ & 0.002 & 0.003 & 0.020 & $-$0.043$^{**}$ \\
				& (0.015) & (0.006) & (0.014) & (0.018) \\
				$\Delta M_{t-28}$ & $-$0.005 & $-$0.001 & $-$0.021 & $-$0.038 \\
				& (0.006) & (0.005) & (0.018) & (0.025) \\
				\hline \\[-1.8ex]
			\end{tabular}
		\end{center}
	\footnotesize \textbf{Note:} Time period -- June 1, 2020 to July 8, 2021. The results are produced using daily data. We use lags of included regressors that are 14 days behind those in the case equation following \cite{karaivanov2021face}. $(p,d,q)$ is the optimal ARIMA order chosen by \texttt{auto.arima} in \textsf{R}. Standard errors are in parentheses. $^{***}$, $^{**}$, and $^{*}$ denote the 99\%, 95\% and 90\% confidence level respectively. The UAE does not publicly announce first and second dose progress separately so we include total doses administered ($V$) instead.
	\end{table}
	\newpage
	\section{Panel Data Regression with alternative time controls}\label{appen: alternative time trends}
	\begin{table}[h!]
		\global\long\def\sym#1{\ifmmode^{#1}\else$^{#1}$\fi}%
		\caption{The direct effects of vaccination, policy, and behavior on new cases}
		\begin{center}
		\begin{tabular}{lr@{\extracolsep{0pt}.}lr@{\extracolsep{0pt}.}lr@{\extracolsep{0pt}.}lr@{\extracolsep{0pt}.}l}
			\hline\hline
			\multicolumn{1}{l}{} & \multicolumn{8}{c}{Dependent variable: $\Delta\log\Delta C_{t}$}\tabularnewline
			\cline{2-9}
			& \multicolumn{2}{c}{(1)} & \multicolumn{2}{c}{(2)} & \multicolumn{2}{c}{(3)} & \multicolumn{2}{c}{(4)}\tabularnewline
			\midrule
			$V1_{t-21}$ & -0&0176\sym{{*}{*}{*}} & -0&0189\sym{{*}{*}{*}} & -0&0176\sym{{*}{*}{*}} & -0&0227\sym{{*}{*}{*}}\tabularnewline
			& (0&0036) & (0&0041) & (0&0046) & (0&0042)\tabularnewline
			\addlinespace
			$V2_{t-7}$ & 0&0066 & 0&0046 & 0&0047 & 0&0025\tabularnewline
			& (0&0041) & (0&0051) & (0&0065) & (0&0058)\tabularnewline
			$V1_{t-21}^{CHN}$ & \multicolumn{2}{c}{} & 0&0179\sym{{*}{*}{*}} & \multicolumn{2}{c}{} & 0&0377\sym{{*}{*}{*}}\tabularnewline
			& \multicolumn{2}{c}{} & (0&0059) & \multicolumn{2}{c}{} & (0&0081)\tabularnewline
			\addlinespace
			$V2_{t-7}^{CHN}$ & \multicolumn{2}{c}{} & -0&0100\sym{*} & \multicolumn{2}{c}{} & 0&0024\tabularnewline
			& \multicolumn{2}{c}{} & (0&0059) & \multicolumn{2}{c}{} & (0&0063)\tabularnewline
			\addlinespace
			$P_{t-14}$ & -0&0035\sym{*} & -0&0037\sym{*} & -0&0057\sym{{*}{*}} & -0&0060\sym{{*}{*}}\tabularnewline
			& (0&0021) & (0&0021) & (0&0022) & (0&0023)\tabularnewline
			\addlinespace
			$M_{t-14}$ & 0&0058\sym{{*}{*}{*}} & 0&0065\sym{{*}{*}{*}} & 0&0098\sym{{*}{*}{*}} & 0&0100\sym{{*}{*}{*}}\tabularnewline
			& (0&0016) & (0&0015) & (0&0013) & (0&0013)\tabularnewline
			$\Delta\log\Delta T_{t}$ & 0&4860\sym{{*}{*}{*}} & 0&4804\sym{{*}{*}{*}} & 0&3882\sym{{*}{*}} & 0&3847\sym{{*}{*}}\tabularnewline
			& (0&1712) & (0&1696) & (0&1442) & (0&1427)\tabularnewline
			$\Delta\log\Delta C_{t-14}$ & 0&1022\sym{{*}{*}{*}} & 0&1022\sym{{*}{*}{*}} & 0&0735\sym{{*}{*}} & 0&0656\sym{*}\tabularnewline
			& (0&0373) & (0&0372) & (0&0346) & (0&0354)\tabularnewline
			\addlinespace
			$\log\Delta C_{t-14}$ & -0&0744\sym{{*}{*}{*}} & -0&0831\sym{{*}{*}{*}} & -0&1628\sym{{*}{*}{*}} & -0&1622\sym{{*}{*}{*}}\tabularnewline
			& (0&0136) & (0&0132) & (0&0208) & (0&0214)\tabularnewline
			\addlinespace			\midrule
			Country fixed effects & \multicolumn{2}{c}{yes} & \multicolumn{2}{c}{yes} & \multicolumn{2}{c}{yes} & \multicolumn{2}{c}{yes}\tabularnewline
			Country specific trend in days & \multicolumn{2}{c}{linear} & \multicolumn{2}{c}{linear} & \multicolumn{2}{c}{cubic} & \multicolumn{2}{c}{cubic}\tabularnewline
			\midrule
			${\rm R}^{2}$ & 0&3219 & 0&3266 & 0&4285 & 0&4346\tabularnewline
			Adjusted ${\rm R}^{2}$ & 0&3197 & 0&3244 & 0&4237 & 0&4298\tabularnewline
			Number of countries & \multicolumn{2}{c}{37} & \multicolumn{2}{c}{37} & \multicolumn{2}{c}{37} & \multicolumn{2}{c}{37}\tabularnewline
			Obs. per country & \multicolumn{2}{c}{382} & \multicolumn{2}{c}{382} & \multicolumn{2}{c}{382} & \multicolumn{2}{c}{382}\tabularnewline
			\midrule
		\end{tabular}
	\end{center}
	\footnotesize \textbf{Note:} Time period -- June 1, 2020 to July 8, 2021. The results are produced using daily data. Standard errors in parentheses are clustered at the country level. $^{***}$, $^{**}$, and $^{*}$ denote the 99\%, 95\% and 90\% confidence level respectively.
	\end{table}
	
	\begin{table}[htbp]
		\global\long\def\sym#1{\ifmmode^{#1}\else$^{#1}$\fi}%
		\caption{The direct effects of vaccination, policy, and behavior on new deaths}
		\begin{center}
		\begin{tabular}{lr@{\extracolsep{0pt}.}lr@{\extracolsep{0pt}.}lr@{\extracolsep{0pt}.}lr@{\extracolsep{0pt}.}l}
			\hline\hline
			\multicolumn{1}{l}{} & \multicolumn{8}{c}{Dependent variable: $\Delta\log\Delta D_{t}$}\tabularnewline
			\cline{2-9}
			& \multicolumn{2}{c}{(1)} & \multicolumn{2}{c}{(2)} & \multicolumn{2}{c}{(3)} & \multicolumn{2}{c}{(4)}\tabularnewline
			\midrule
			$V1_{t-35}$ & -0&0202\sym{{*}{*}{*}} & -0&0206\sym{{*}{*}{*}} & -0&0112\sym{{*}{*}} & -0&0133\sym{{*}{*}{*}}\tabularnewline
			& (0&0040) & (0&0047) & (0&0043) & (0&0041)\tabularnewline
			\addlinespace
			$V2_{t-21}$ & 0&0075 & 0&0027 & 0&0083 & 0&0081\tabularnewline
			& (0&0053) & (0&0069) & (0&0050) & (0&0054)\tabularnewline
			$V1_{t-35}^{CHN}$ & \multicolumn{2}{c}{} & 0&0155\sym{{*}{*}} & \multicolumn{2}{c}{} & 0&0139\sym{{*}{*}}\tabularnewline
			& \multicolumn{2}{c}{} & (0&0074) & \multicolumn{2}{c}{} & (0&0060)\tabularnewline
			$V2_{t-21}^{CHN}$ & \multicolumn{2}{c}{} & -0&0017 & \multicolumn{2}{c}{} & -0&0016\tabularnewline
			& \multicolumn{2}{c}{} & (0&0083) & \multicolumn{2}{c}{} & (0&0064)\tabularnewline
			\addlinespace
			$P_{t-28}$ & -0&0035\sym{*} & -0&0037\sym{*} & -0&0038\sym{{*}{*}} & -0&0039\sym{{*}{*}}\tabularnewline
			& (0&0019) & (0&0019) & (0&0018) & (0&0018)\tabularnewline
			\addlinespace
			$M_{t-28}$ & 0&0053\sym{{*}{*}{*}} & 0&0059\sym{{*}{*}{*}} & 0&0087\sym{{*}{*}{*}} & 0&0088\sym{{*}{*}{*}}\tabularnewline
			& (0&0019) & (0&0017) & (0&0014) & (0&0014)\tabularnewline
			$\Delta\log\Delta D_{t-28}$ & 0&0656\sym{{*}{*}} & 0&0681\sym{{*}{*}} & 0&0520\sym{*} & 0&0506\sym{*}\tabularnewline
			& (0&0293) & (0&0289) & (0&0262) & (0&0262)\tabularnewline
			\addlinespace
			$\log\Delta D_{t-28}$ & -0&0817\sym{{*}{*}{*}} & -0&0942\sym{{*}{*}{*}} & -0&1263\sym{{*}{*}{*}} & -0&1249\sym{{*}{*}{*}}\tabularnewline
			& (0&0121) & (0&0121) & (0&0127) & (0&0127)\tabularnewline
			\addlinespace
			\midrule
			Country fixed effects & \multicolumn{2}{c}{yes} & \multicolumn{2}{c}{yes} & \multicolumn{2}{c}{yes} & \multicolumn{2}{c}{yes}\tabularnewline
			Country specific trend in days & \multicolumn{2}{c}{linear} & \multicolumn{2}{c}{linear} & \multicolumn{2}{c}{cubic} & \multicolumn{2}{c}{cubic}\tabularnewline
			\midrule
			${\rm R}^{2}$ & 0&1271 & 0&1324 & 0&1860 & 0&1864\tabularnewline
			Adjusted ${\rm R}^{2}$ & 0&1244 & 0&1295 & 0&1789 & 0&1792\tabularnewline
			Number of countries & \multicolumn{2}{c}{37} & \multicolumn{2}{c}{37} & \multicolumn{2}{c}{37} & \multicolumn{2}{c}{37}\tabularnewline
			Obs. per country & \multicolumn{2}{c}{368} & \multicolumn{2}{c}{368} & \multicolumn{2}{c}{368} & \multicolumn{2}{c}{368}\tabularnewline
			\midrule
		\end{tabular}
	\end{center}
	\footnotesize \textbf{Note:} Time period -- June 1, 2020 to July 8, 2021. The results are produced using daily data. Standard errors in 	parentheses are clustered at the country level. $^{***}$, $^{**}$, and $^{*}$ denote the 99\%, 95\% and 90\% confidence level respectively.
	\end{table}
	
	\begin{table}[htbp]
		\global\long\def\sym#1{\ifmmode^{#1}\else$^{#1}$\fi}%
		\caption{The direct effects of vaccination, policy and information on mobility}
		\small
		\begin{center}
		\begin{tabular}{lr@{\extracolsep{0pt}.}lr@{\extracolsep{0pt}.}lr@{\extracolsep{0pt}.}lr@{\extracolsep{0pt}.}l}
			\hline\hline
			\multicolumn{1}{l}{} & \multicolumn{8}{c}{Dependent variable: $M_{t}$}\tabularnewline
			\cline{2-9}
			& \multicolumn{2}{c}{(1)} & \multicolumn{2}{c}{(2)} & \multicolumn{2}{c}{(3)} & \multicolumn{2}{c}{(4)}\tabularnewline
			\midrule
			$\Delta V1_{t}$ & 0&4922\sym{{*}{*}{*}} & 0&6020\sym{{*}{*}{*}} & 0&4196\sym{{*}{*}{*}} & 0&4985\sym{{*}{*}{*}}\tabularnewline
			& (0&1103) & (0&1472) & (0&1150) & (0&1548)\tabularnewline
			$\Delta V2_{t}$ & 0&1511 & 0&2597 & 0&1710 & 0&2787\tabularnewline
			& (0&1291) & (0&1579) & (0&1365) & (0&1798)\tabularnewline
			\addlinespace
			$V1_{t-7}$ & 0&0619\sym{*} & 0&0549 & 0&0650\sym{{*}{*}} & 0&0590\tabularnewline
			& (0&0356) & (0&0397) & (0&0306) & (0&0353)\tabularnewline
			\addlinespace
			$V2_{t-7}$ & -0&0244 & -0&0130 & -0&0023 & 0&0071\tabularnewline
			& (0&0382) & (0&0450) & (0&0302) & (0&0345)\tabularnewline
			$\Delta V1_{t}^{CHN}$ & \multicolumn{2}{c}{} & -0&4470\sym{{*}{*}} & \multicolumn{2}{c}{} & -0&3085\sym{*}\tabularnewline
			& \multicolumn{2}{c}{} & (0&1658) & \multicolumn{2}{c}{} & (0&1761)\tabularnewline
			$\Delta V2_{t}^{CHN}$ & \multicolumn{2}{c}{} & -0&3209 & \multicolumn{2}{c}{} & -0&3235\tabularnewline
			& \multicolumn{2}{c}{} & (0&2853) & \multicolumn{2}{c}{} & (0&3159)\tabularnewline
			$V1_{t-7}^{CHN}$ & \multicolumn{2}{c}{} & -0&0517 & \multicolumn{2}{c}{} & -0&0405\tabularnewline
			& \multicolumn{2}{c}{} & (0&0965) & \multicolumn{2}{c}{} & (0&0937)\tabularnewline
			$V2_{t-7}^{CHN}$ & \multicolumn{2}{c}{} & 0&0493 & \multicolumn{2}{c}{} & 0&0364\tabularnewline
			& \multicolumn{2}{c}{} & (0&0759) & \multicolumn{2}{c}{} & (0&0654)\tabularnewline
			$\Delta P_{t}$ & -0&3474\sym{{*}{*}{*}} & -0&3491\sym{{*}{*}{*}} & -0&3576\sym{{*}{*}{*}} & -0&3584\sym{{*}{*}{*}}\tabularnewline
			& (0&0452) & (0&0449) & (0&0432) & (0&0430)\tabularnewline
			$P_{t-7}$ & -0&1744\sym{{*}{*}{*}} & -0&1765\sym{{*}{*}{*}} & -0&1546\sym{{*}{*}{*}} & -0&1572\sym{{*}{*}{*}}\tabularnewline
			& (0&0203) & (0&0199) & (0&0187) & (0&0185)\tabularnewline
			$\Delta\log\Delta C_{t}$ & 0&7572\sym{{*}{*}{*}} & 0&7607\sym{{*}{*}{*}} & \multicolumn{2}{c}{} & \multicolumn{2}{c}{}\tabularnewline
			& (0&2356) & (0&2389) & \multicolumn{2}{c}{} & \multicolumn{2}{c}{}\tabularnewline
			\addlinespace
			$\log\Delta C_{t}$ & -1&0037\sym{{*}{*}{*}} & -0&9634\sym{{*}{*}{*}} & \multicolumn{2}{c}{} & \multicolumn{2}{c}{}\tabularnewline
			& (0&1176) & (0&1190) & \multicolumn{2}{c}{} & \multicolumn{2}{c}{}\tabularnewline
			$\Delta\log\Delta D_{t}$ & \multicolumn{2}{c}{} & \multicolumn{2}{c}{} & 0&2439\sym{*} & 0&2339\sym{*}\tabularnewline
			& \multicolumn{2}{c}{} & \multicolumn{2}{c}{} & (0&1361) & (0&1333)\tabularnewline
			$\log\Delta D_{t}$ & \multicolumn{2}{c}{} & \multicolumn{2}{c}{} & -1&1954\sym{{*}{*}{*}} & -1&1445\sym{{*}{*}{*}}\tabularnewline
			& \multicolumn{2}{c}{} & \multicolumn{2}{c}{} & (0&1565) & (0&1588)\tabularnewline
			$M_{t-7}$ & 0&7065\sym{{*}{*}{*}} & 0&7034\sym{{*}{*}{*}} & 0&6735\sym{{*}{*}{*}} & 0&6721\sym{{*}{*}{*}}\tabularnewline
			& (0&0230) & (0&0234) & (0&0242) & (0&0243)\tabularnewline
			\midrule
			Country fixed effects & \multicolumn{2}{c}{yes} & \multicolumn{2}{c}{yes} & \multicolumn{2}{c}{yes} & \multicolumn{2}{c}{yes}\tabularnewline
			Country specific trend in days & \multicolumn{2}{c}{linear} & \multicolumn{2}{c}{linear} & \multicolumn{2}{c}{linear} & \multicolumn{2}{c}{linear}\tabularnewline
			\midrule
			${\rm R}^{2}$ & 0&8432 & 0&8438 & 0&8443 & 0&8447\tabularnewline
			Adjusted ${\rm R}^{2}$ & 0&8427 & 0&8433 & 0&8438 & 0&8442\tabularnewline
			Number of countries & \multicolumn{2}{c}{37} & \multicolumn{2}{c}{37} & \multicolumn{2}{c}{37} & \multicolumn{2}{c}{37}\tabularnewline
			Obs. per country & \multicolumn{2}{c}{396} & \multicolumn{2}{c}{396} & \multicolumn{2}{c}{396} & \multicolumn{2}{c}{396}\tabularnewline
			\midrule
		\end{tabular}
	\end{center}
	\footnotesize \textbf{Note:} Time period -- June 1, 2020 to July 8, 2021. The results are produced using daily data. Standard errors in parentheses are clustered at the country level. $^{***}$, $^{**}$, and $^{*}$ denote the 99\%, 95\% and 90\% confidence level respectively.
	\end{table}
	
	\newpage
	
	\section{Panel Data regression with alternative lags}\label{appen: alternative lags}
		\begin{figure}[h!]
		\begin{center}
			\caption{Alternative lags for vaccination variables}\label{fig: alternative lags}
			\vspace{-3mm}
			\begin{subfigure}[b]{\textwidth}
				\centering
				\caption[]%
				{{\small No time trends}}
				\includegraphics[width=0.75\textwidth]{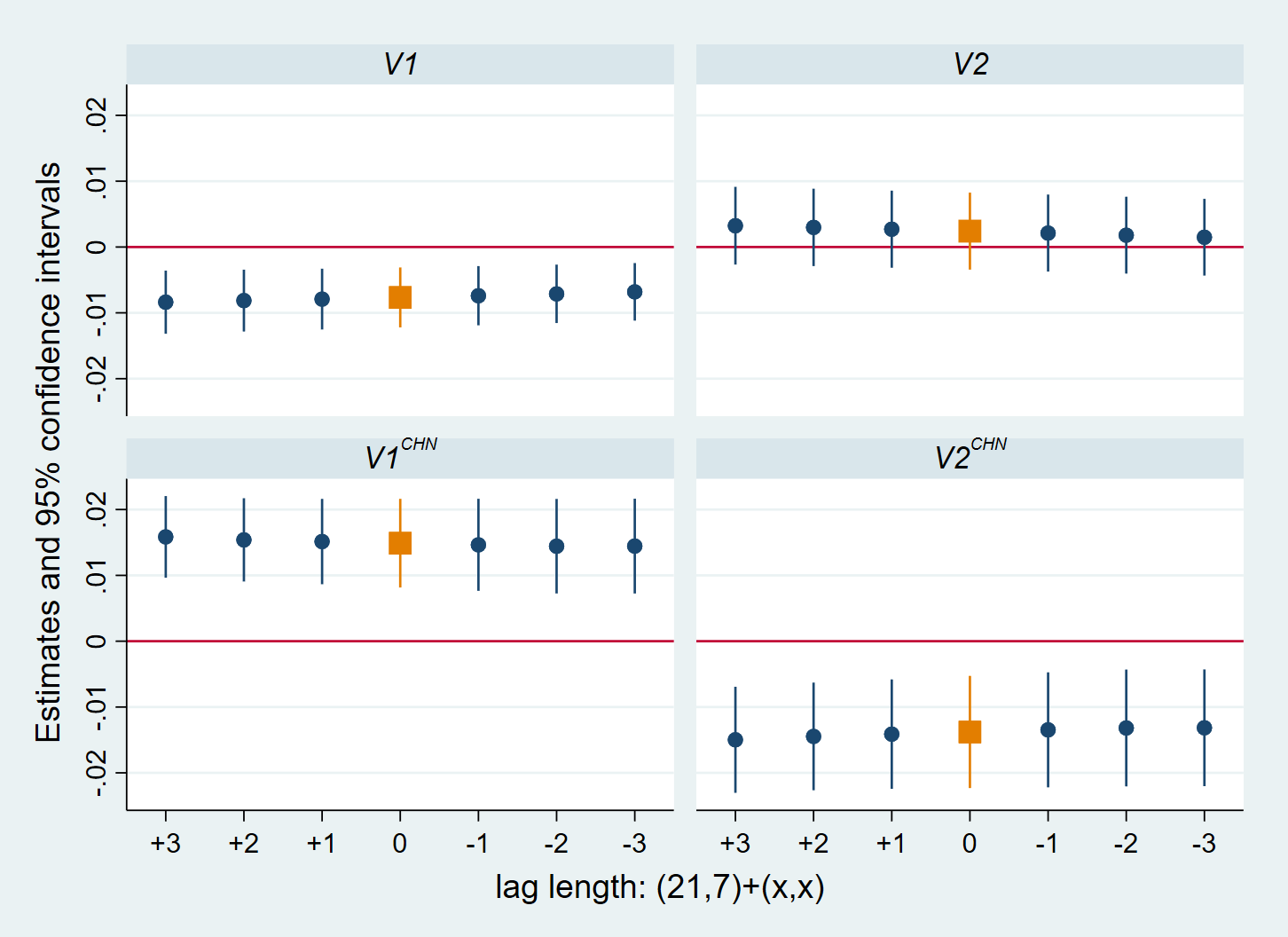}
			\end{subfigure}
			\vspace{1mm}
			\begin{subfigure}[b]{\textwidth}
				\centering
				\caption[]%
				{{\small Cubic time trends}}
				\includegraphics[width=0.75\textwidth]{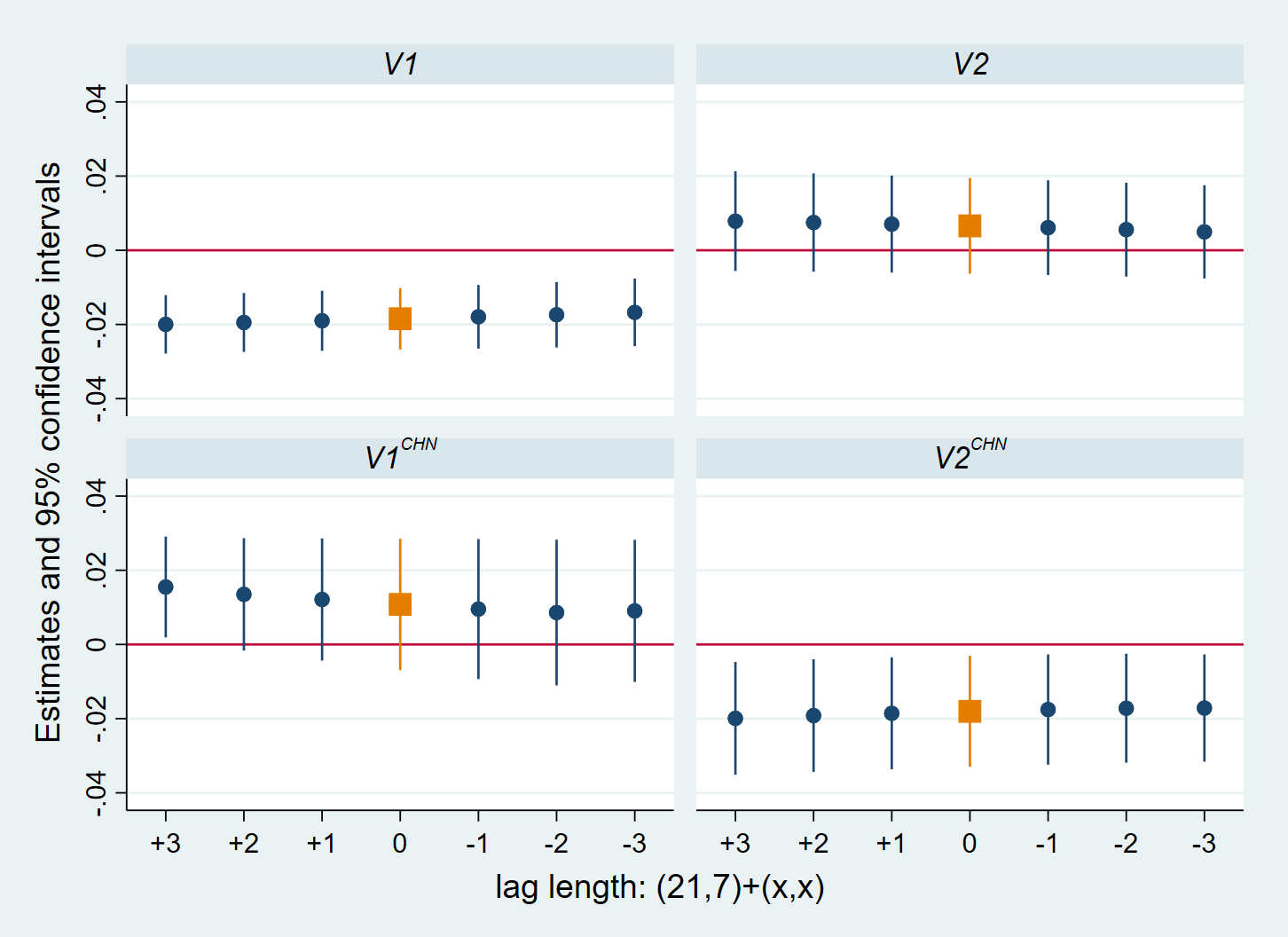}		
			\end{subfigure}
		\end{center}
		\footnotesize \textbf{Note:} The orange squares are estimates from the baseline specifications. The dark-blue dots are estimates from alternative lag specifications. The solid lines are the 95\% confidence intervals.   
	\end{figure}

	\newpage
	
	\section{Panel Data regression with alternative periods}\label{appen: alternative periods}
	\begin{table}[h!]
		\global\long\def\sym#1{\ifmmode^{#1}\else$^{#1}$\fi}%
		\caption{Early and late sub-periods}\label{tab: Early and late sub-periods}
		\begin{center}
		\begin{tabular}{lr@{\extracolsep{0pt}.}lr@{\extracolsep{0pt}.}lr@{\extracolsep{0pt}.}lr@{\extracolsep{0pt}.}l}
			\hline\hline
			\multicolumn{1}{l}{} & \multicolumn{8}{c}{Dependent variable: $\Delta\log\Delta C_{t}$}\tabularnewline
			\cmidrule(lr){2-5}\cmidrule(lr){6-9}
			& \multicolumn{4}{c}{July 1, 2020 - July 8, 2021} & \multicolumn{4}{c}{June 1, 2020 - May 31, 2021}\tabularnewline
			\midrule
			$V1_{t-21}$ & -0&0164\sym{{*}{*}{*}} & -0&0185\sym{{*}{*}{*}} & -0&0132\sym{{*}{*}{*}} & -0&0139\sym{{*}{*}{*}}\tabularnewline
			& (0&0042) & (0&0041) & (0&0043) & (0&0049)\tabularnewline
			\addlinespace
			$V2_{t-7}$ & 0&0035 & 0&0058 & 0&0006 & 0&0006\tabularnewline
			& (0&0055) & (0&0059) & (0&0038) & (0&0053)\tabularnewline
			$V1_{t-21}^{CHN}$ & \multicolumn{2}{c}{} & 0&0203\sym{{*}{*}{*}} & \multicolumn{2}{c}{} & 0&0059\tabularnewline
			& \multicolumn{2}{c}{} & (0&0055) & \multicolumn{2}{c}{} & (0&0083)\tabularnewline
			$V2_{t-7}^{CHN}$ & \multicolumn{2}{c}{} & -0&0199\sym{{*}{*}{*}} & \multicolumn{2}{c}{} & -0&0029\tabularnewline
			& \multicolumn{2}{c}{} & (0&0067) & \multicolumn{2}{c}{} & (0&0063)\tabularnewline
			\addlinespace
			$P_{t-14}$ & -0&0031\sym{{*}{*}} & -0&0030\sym{{*}{*}} & -0&0027 & -0&0027\tabularnewline
			& (0&0014) & (0&0015) & (0&0019) & (0&0019)\tabularnewline
			\addlinespace
			$M_{t-14}$ & 0&0106\sym{{*}{*}{*}} & 0&0106\sym{{*}{*}{*}} & 0&0104\sym{{*}{*}{*}} & 0&0105\sym{{*}{*}{*}}\tabularnewline
			& (0&0012) & (0&0012) & (0&0012) & (0&0012)\tabularnewline
			\addlinespace
			$\Delta\log\Delta C_{t-14}$ & 0&0950\sym{{*}{*}{*}} & 0&0939\sym{{*}{*}{*}} & 0&0927\sym{{*}{*}} & 0&0918\sym{{*}{*}}\tabularnewline
			& (0&0318) & (0&0318) & (0&0379) & (0&0376)\tabularnewline
			\addlinespace
			$\log\Delta C_{t-14}$ & -0&1412\sym{{*}{*}{*}} & -0&1412\sym{{*}{*}{*}} & -0&1672\sym{{*}{*}{*}} & -0&1665\sym{{*}{*}{*}}\tabularnewline
			& (0&0204) & (0&0208) & (0&0210) & (0&0213)\tabularnewline
			\addlinespace
			$\Delta\log\Delta T_{t}$ & 0&3942\sym{{*}{*}} & 0&3939\sym{{*}{*}} & 0&3569\sym{{*}{*}} & 0&3569\sym{{*}{*}}\tabularnewline
			& (0&1467) & (0&1467) & (0&1377) & (0&1376)\tabularnewline
			\midrule
			Country fixed effects & \multicolumn{2}{c}{yes} & \multicolumn{2}{c}{yes} & \multicolumn{2}{c}{yes} & \multicolumn{2}{c}{yes}\tabularnewline
			Quadratic time trends & \multicolumn{2}{c}{yes} & \multicolumn{2}{c}{yes} & \multicolumn{2}{c}{yes} & \multicolumn{2}{c}{yes}\tabularnewline
			\midrule
			${\rm R}^{2}$ & 0&4054 & 0&4067 & 0&3950 & 0&3951\tabularnewline
			Adjusted ${\rm R}^{2}$ & 0&4018 & 0&4032 & 0&3911 & 0&3911\tabularnewline
			Number of countries & \multicolumn{2}{c}{37} & \multicolumn{2}{c}{37} & \multicolumn{2}{c}{37} & \multicolumn{2}{c}{37}\tabularnewline
			Obs. per country & \multicolumn{2}{c}{373} & \multicolumn{2}{c}{373} & \multicolumn{2}{c}{344} & \multicolumn{2}{c}{344}\tabularnewline
			\midrule
		\end{tabular}
	\end{center}
	\footnotesize \textbf{Note:} The results are produced using daily data. Standard errors in parentheses are clustered at the country level. $^{***}$, $^{**}$, and $^{*}$ denote the 99\%, 95\% and 90\% confidence level respectively.
	\end{table}
	\newpage
	
	\section{Panel Data Regression with weekly data}\label{appen: panel weekly frequency estimation}
	\begin{table}[h!]
		\global\long\def\sym#1{\ifmmode^{#1}\else$^{#1}$\fi}%
		\caption{The direct effects of vaccination, policy and behavior on new cases}\label{tab: panel weekly cases}
		\begin{center}
		\begin{tabular}{lr@{\extracolsep{0pt}.}lr@{\extracolsep{0pt}.}lr@{\extracolsep{0pt}.}lr@{\extracolsep{0pt}.}l}
			\hline\hline
			\multicolumn{1}{l}{} & \multicolumn{8}{c}{Dependent variable: $\Delta\log\Delta C_{t}$}\tabularnewline
			\cline{2-9}
			& \multicolumn{2}{c}{(1)} & \multicolumn{2}{c}{(2)} & \multicolumn{2}{c}{(3)} & \multicolumn{2}{c}{(4)}\tabularnewline
			\midrule
			$V1_{t-21}$ & -0&0077\sym{{*}{*}{*}} & -0&0078\sym{{*}{*}{*}} & -0&0166\sym{{*}{*}{*}} & -0&0180\sym{{*}{*}{*}}\tabularnewline
			& (0&0020) & (0&0021) & (0&0044) & (0&0043)\tabularnewline
			\addlinespace
			$V2_{t-7}$ & 0&0025 & 0&0015 & 0&0011 & 0&0032\tabularnewline
			& (0&0023) & (0&0027) & (0&0057) & (0&0062)\tabularnewline
			$V1_{t-21}^{CHN}$ & \multicolumn{2}{c}{} & 0&0164\sym{{*}{*}{*}} & \multicolumn{2}{c}{} & 0&0152\sym{{*}{*}}\tabularnewline
			& \multicolumn{2}{c}{} & (0&0032) & \multicolumn{2}{c}{} & (0&0073)\tabularnewline
			$V2_{t-7}^{CHN}$ & \multicolumn{2}{c}{} & -0&0145\sym{{*}{*}{*}} & \multicolumn{2}{c}{} & -0&0171\sym{{*}{*}}\tabularnewline
			& \multicolumn{2}{c}{} & (0&0042) & \multicolumn{2}{c}{} & (0&0072)\tabularnewline
			\addlinespace
			$P_{t-14}$ & -0&0024\sym{*} & -0&0022\sym{*} & -0&0034\sym{{*}{*}} & -0&0033\sym{{*}{*}}\tabularnewline
			& (0&0012) & (0&0013) & (0&0015) & (0&0015)\tabularnewline
			\addlinespace
			$M_{t-14}$ & 0&0054\sym{{*}{*}{*}} & 0&0055\sym{{*}{*}{*}} & 0&0108\sym{{*}{*}{*}} & 0&0108\sym{{*}{*}{*}}\tabularnewline
			& (0&0012) & (0&0012) & (0&0015) & (0&0015)\tabularnewline
			\addlinespace
			$\Delta\log\Delta C_{t-14}$ & 0&0787 & 0&0732 & 0&0522 & 0&0515\tabularnewline
			& (0&0517) & (0&0518) & (0&0469) & (0&0467)\tabularnewline
			\addlinespace
			$\log\Delta C_{t-14}$ & -0&0329\sym{{*}{*}{*}} & -0&0365\sym{{*}{*}{*}} & -0&1399\sym{{*}{*}{*}} & -0&1403\sym{{*}{*}{*}}\tabularnewline
			& (0&0076) & (0&0073) & (0&0224) & (0&0226)\tabularnewline
			\addlinespace
			$\Delta\log\Delta T_{t}$ & 0&5589\sym{{*}{*}{*}} & 0&5543\sym{{*}{*}{*}} & 0&4391\sym{{*}{*}{*}} & 0&4387\sym{{*}{*}{*}}\tabularnewline
			& (0&1434) & (0&1426) & (0&1244) & (0&1243)\tabularnewline
			\midrule
			Country fixed effects & \multicolumn{2}{c}{yes} & \multicolumn{2}{c}{yes} & \multicolumn{2}{c}{yes} & \multicolumn{2}{c}{yes}\tabularnewline
			Country specific trend in weeks & \multicolumn{2}{c}{no} & \multicolumn{2}{c}{no} & \multicolumn{2}{c}{quadratic} & \multicolumn{2}{c}{quadratic}\tabularnewline
			\midrule
			${\rm R}^{2}$ & 0&2837 & 0&2879 & 0&3980 & 0&3987\tabularnewline
			Adjusted ${\rm R}^{2}$ & 0&2813 & 0&2848 & 0&3726 & 0&3728\tabularnewline
			Number of countries & \multicolumn{2}{c}{41} & \multicolumn{2}{c}{41} & \multicolumn{2}{c}{41} & \multicolumn{2}{c}{41}\tabularnewline
			Obs. per country & \multicolumn{2}{c}{54} & \multicolumn{2}{c}{54} & \multicolumn{2}{c}{54} & \multicolumn{2}{c}{54}\tabularnewline
			\midrule
		\end{tabular}
	\end{center}
	\footnotesize \textbf{Note:} Time period -- June 1, 2020 to July 8, 2021. The results are produced using weekly data. Standard errors in parentheses are clustered at the country level. $^{***}$, $^{**}$, and $^{*}$ denote the 99\%, 95\% and 90\% confidence level respectively.
	\end{table}
	
	\begin{table}[htbp]
		\global\long\def\sym#1{\ifmmode^{#1}\else$^{#1}$\fi}%
		\caption{The direct effects of vaccination, policy and behavior on new deaths}\label{tab: panel weekly deaths}
		\begin{center}
		\begin{tabular}{lr@{\extracolsep{0pt}.}lr@{\extracolsep{0pt}.}lr@{\extracolsep{0pt}.}lr@{\extracolsep{0pt}.}l}
			\hline\hline
			\multicolumn{1}{l}{} & \multicolumn{8}{c}{Dependent variable: $\Delta\log\Delta D_{t}$}\tabularnewline
			\cline{2-9}
			& \multicolumn{2}{c}{(1)} & \multicolumn{2}{c}{(2)} & \multicolumn{2}{c}{(3)} & \multicolumn{2}{c}{(4)}\tabularnewline
			\midrule
			$V1_{t-35}$ & -0&0097\sym{{*}{*}{*}} & -0&0094\sym{{*}{*}{*}} & -0&0123\sym{{*}{*}{*}} & -0&0130\sym{{*}{*}{*}}\tabularnewline
			& (0&0022) & (0&0023) & (0&0038) & (0&0039)\tabularnewline
			\addlinespace
			$V2_{t-21}$ & 0&0039 & 0&0018 & 0&0044 & 0&0068\tabularnewline
			& (0&0030) & (0&0032) & (0&0042) & (0&0044)\tabularnewline
			$V1_{t-35}^{CHN}$ & \multicolumn{2}{c}{} & 0&0172\sym{{*}{*}{*}} & \multicolumn{2}{c}{} & 0&0035\tabularnewline
			& \multicolumn{2}{c}{} & (0&0038) & \multicolumn{2}{c}{} & (0&0069)\tabularnewline
			$V2_{t-21}^{CHN}$ & \multicolumn{2}{c}{} & -0&0121\sym{{*}{*}} & \multicolumn{2}{c}{} & -0&0098\sym{*}\tabularnewline
			& \multicolumn{2}{c}{} & (0&0055) & \multicolumn{2}{c}{} & (0&0055)\tabularnewline
			\addlinespace
			$P_{t-28}$ & -0&0010 & -0&0007 & -0&0016 & -0&0016\tabularnewline
			& (0&0014) & (0&0015) & (0&0013) & (0&0013)\tabularnewline
			\addlinespace
			$M_{t-28}$ & 0&0056\sym{{*}{*}{*}} & 0&0055\sym{{*}{*}{*}} & 0&0083\sym{{*}{*}{*}} & 0&0081\sym{{*}{*}{*}}\tabularnewline
			& (0&0016) & (0&0016) & (0&0017) & (0&0016)\tabularnewline
			$\Delta\log\Delta D_{t-28}$ & 0&0636\sym{{*}{*}{*}} & 0&0611\sym{{*}{*}{*}} & 0&0639\sym{{*}{*}{*}} & 0&0633\sym{{*}{*}{*}}\tabularnewline
			& (0&0223) & (0&0224) & (0&0227) & (0&0228)\tabularnewline
			\addlinespace
			$\log\Delta D_{t-28}$ & -0&0448\sym{{*}{*}{*}} & -0&0492\sym{{*}{*}{*}} & -0&1317\sym{{*}{*}{*}} & -0&1319\sym{{*}{*}{*}}\tabularnewline
			& (0&0107) & (0&0100) & (0&0121) & (0&0122)\tabularnewline
			\addlinespace
			\midrule
			Country fixed effects & \multicolumn{2}{c}{yes} & \multicolumn{2}{c}{yes} & \multicolumn{2}{c}{yes} & \multicolumn{2}{c}{yes}\tabularnewline
			Country specific trend in weeks & \multicolumn{2}{c}{no} & \multicolumn{2}{c}{no} & \multicolumn{2}{c}{quadratic} & \multicolumn{2}{c}{quadratic}\tabularnewline
			\midrule
			${\rm R}^{2}$ & 0&1017 & 0&1063 & 0&1852 & 0&1855\tabularnewline
			Adjusted ${\rm R}^{2}$ & 0&0990 & 0&1028 & 0&1500 & 0&1494\tabularnewline
			Number of countries & \multicolumn{2}{c}{41} & \multicolumn{2}{c}{41} & \multicolumn{2}{c}{41} & \multicolumn{2}{c}{41}\tabularnewline
			Obs. per country & \multicolumn{2}{c}{52} & \multicolumn{2}{c}{52} & \multicolumn{2}{c}{52} & \multicolumn{2}{c}{52}\tabularnewline
			\midrule
		\end{tabular}
	\end{center}
	\footnotesize \textbf{Note:} Time period -- June 1, 2020 to July 8, 2021. The results are produced using weekly data. Standard errors in parentheses are clustered at the country level. $^{***}$, $^{**}$, and $^{*}$ denote the 99\%, 95\% and 90\% confidence level respectively.
	\end{table}
	
	\begin{table}[htbp]
		\global\long\def\sym#1{\ifmmode^{#1}\else$^{#1}$\fi}%
		\caption{The direct effects of vaccination, policy and information on mobility}\label{tab: panel weekly mobility}
		\small
		\begin{center}
		\begin{tabular}{lr@{\extracolsep{0pt}.}lr@{\extracolsep{0pt}.}lr@{\extracolsep{0pt}.}lr@{\extracolsep{0pt}.}l}
			\hline\hline
			\multicolumn{1}{l}{} & \multicolumn{8}{c}{Dependent variable: $M_{t}$}\tabularnewline
			\cline{2-9}
			& \multicolumn{2}{c}{(1)} & \multicolumn{2}{c}{(2)} & \multicolumn{2}{c}{(3)} & \multicolumn{2}{c}{(4)}\tabularnewline
			\midrule
			$\Delta V1_{t}$ & 0&5894\sym{{*}{*}{*}} & 0&6601\sym{{*}{*}{*}} & 0&5188\sym{{*}{*}{*}} & 0&5778\sym{{*}{*}{*}}\tabularnewline
			& (0&1216) & (0&1504) & (0&1223) & (0&1521)\tabularnewline
			$\Delta V2_{t}$ & 0&2504\sym{*} & 0&3451\sym{{*}{*}} & 0&2876\sym{{*}{*}} & 0&3860\sym{{*}{*}}\tabularnewline
			& (0&1277) & (0&1528) & (0&1388) & (0&1725)\tabularnewline
			$V1_{t-7}$ & 0&0531 & 0&0390 & 0&0424 & 0&0292\tabularnewline
			& (0&0327) & (0&0371) & (0&0316) & (0&0364)\tabularnewline
			$V2_{t-7}$ & -0&0133 & -0&0032 & 0&0072 & 0&0163\tabularnewline
			& (0&0361) & (0&0396) & (0&0320) & (0&0355)\tabularnewline
			$\Delta V1_{t}^{CHN}$ & \multicolumn{2}{c}{} & -0&3402 & \multicolumn{2}{c}{} & -0&2865\tabularnewline
			& \multicolumn{2}{c}{} & (0&2269) & \multicolumn{2}{c}{} & (0&2347)\tabularnewline
			$\Delta V2_{t}^{CHN}$ & \multicolumn{2}{c}{} & -0&3465 & \multicolumn{2}{c}{} & -0&3598\tabularnewline
			& \multicolumn{2}{c}{} & (0&2477) & \multicolumn{2}{c}{} & (0&2670)\tabularnewline
			$V1_{t-7}^{CHN}$ & \multicolumn{2}{c}{} & 0&0410 & \multicolumn{2}{c}{} & 0&0395\tabularnewline
			& \multicolumn{2}{c}{} & (0&0645) & \multicolumn{2}{c}{} & (0&0597)\tabularnewline
			$V2_{t-7}^{CHN}$ & \multicolumn{2}{c}{} & -0&0176 & \multicolumn{2}{c}{} & -0&0180\tabularnewline
			& \multicolumn{2}{c}{} & (0&0731) & \multicolumn{2}{c}{} & (0&0625)\tabularnewline
			$\Delta P_{t}$ & -0&2323\sym{{*}{*}{*}} & -0&2320\sym{{*}{*}{*}} & -0&2559\sym{{*}{*}{*}} & -0&2547\sym{{*}{*}{*}}\tabularnewline
			& (0&0505) & (0&0507) & (0&0485) & (0&0489)\tabularnewline
			$P_{t-7}$ & -0&1182\sym{{*}{*}{*}} & -0&1204\sym{{*}{*}{*}} & -0&1096\sym{{*}{*}{*}} & -0&1120\sym{{*}{*}{*}}\tabularnewline
			& (0&0179) & (0&0182) & (0&0183) & (0&0186)\tabularnewline
			$\Delta\log\Delta C_{t}$ & 0&2696 & 0&2993 & \multicolumn{2}{c}{} & \multicolumn{2}{c}{}\tabularnewline
			& (0&2639) & (0&2663) & \multicolumn{2}{c}{} & \multicolumn{2}{c}{}\tabularnewline
			$\log\Delta C_{t}$ & -0&7097\sym{{*}{*}{*}} & -0&6992\sym{{*}{*}{*}} & \multicolumn{2}{c}{} & \multicolumn{2}{c}{}\tabularnewline
			& (0&0909) & (0&0976) & \multicolumn{2}{c}{} & \multicolumn{2}{c}{}\tabularnewline
			$\Delta\log\Delta D_{t}$ & \multicolumn{2}{c}{} & \multicolumn{2}{c}{} & -0&1759 & -0&1645\tabularnewline
			& \multicolumn{2}{c}{} & \multicolumn{2}{c}{} & (0&2216) & (0&2211)\tabularnewline
			$\log\Delta D_{t}$ & \multicolumn{2}{c}{} & \multicolumn{2}{c}{} & -0&7710\sym{{*}{*}{*}} & -0&7585\sym{{*}{*}{*}}\tabularnewline
			& \multicolumn{2}{c}{} & \multicolumn{2}{c}{} & (0&1279) & (0&1334)\tabularnewline
			$M_{t-7}$ & 0&7547\sym{{*}{*}{*}} & 0&7547\sym{{*}{*}{*}} & 0&7402\sym{{*}{*}{*}} & 0&7405\sym{{*}{*}{*}}\tabularnewline
			& (0&0248) & (0&0249) & (0&0272) & (0&0272)\tabularnewline
			\midrule
			Country fixed effects & \multicolumn{2}{c}{yes} & \multicolumn{2}{c}{yes} & \multicolumn{2}{c}{yes} & \multicolumn{2}{c}{yes}\tabularnewline
			Time effects & \multicolumn{2}{c}{no} & \multicolumn{2}{c}{no} & \multicolumn{2}{c}{no} & \multicolumn{2}{c}{no}\tabularnewline
			\midrule
			${\rm R}^{2}$ & 0&8232 & 0&8235 & 0&8232 & 0&8234\tabularnewline
			Adjusted ${\rm R}^{2}$ & 0&8225 & 0&8225 & 0&8225 & 0&8224\tabularnewline
			Number of countries & \multicolumn{2}{c}{41} & \multicolumn{2}{c}{41} & \multicolumn{2}{c}{41} & \multicolumn{2}{c}{41}\tabularnewline
			Obs. per country & \multicolumn{2}{c}{56} & \multicolumn{2}{c}{56} & \multicolumn{2}{c}{56} & \multicolumn{2}{c}{56}\tabularnewline
			\midrule
		\end{tabular}
	\end{center}
	\footnotesize \textbf{Note:} Time period -- June 1, 2020 to July 8, 2021. The results are produced using weekly data. Standard errors in parentheses are clustered at the country level. $^{***}$, $^{**}$, and $^{*}$ denote the 99\%, 95\% and 90\% confidence level respectively.
	\end{table}
	
	\newpage{}
	
	\section{Interaction between vaccine variables and the mobility index}\label{appen: interaction}

	\begin{table}[h!]
		\global\long\def\sym#1{\ifmmode^{#1}\else$^{#1}$\fi}%
		\caption{The direct effects of vaccination, policy, behavior and their interactions
			on cases}
		\label{fig: interaction_cases}
		\small%
		\begin{center}
		\begin{tabular}{lr@{\extracolsep{0pt}.}lr@{\extracolsep{0pt}.}lr@{\extracolsep{0pt}.}lr@{\extracolsep{0pt}.}l}
			\hline\hline
			\multicolumn{1}{l}{} & \multicolumn{8}{c}{Dependent variable: $\Delta\log\Delta C_{t}$}\tabularnewline
			\cline{2-9}
			& \multicolumn{2}{c}{(1)} & \multicolumn{2}{c}{(2)} & \multicolumn{2}{c}{(3)} & \multicolumn{2}{c}{(4)}\tabularnewline
			\midrule
			$V1_{t-14}$ & -0&0085\sym{{*}{*}} & -0&0086\sym{{*}{*}} & -0&0156\sym{{*}{*}{*}} & -0&0173\sym{{*}{*}{*}}\tabularnewline
			& (0&0038) & (0&0038) & (0&0054) & (0&0061)\tabularnewline
			$V2_{t-14}$ & 0&0040 & 0&0041 & 0&0035 & 0&0072\tabularnewline
			& (0&0055) & (0&0056) & (0&0083) & (0&0098)\tabularnewline
			$V1_{t-14}^{CHN}$ & \multicolumn{2}{c}{} & 0&0185\sym{{*}{*}{*}} & \multicolumn{2}{c}{} & 0&0130\tabularnewline
			& \multicolumn{2}{c}{} & (0&0061) & \multicolumn{2}{c}{} & (0&0080)\tabularnewline
			\addlinespace
			$V2_{t-14}^{CHN}$ & \multicolumn{2}{c}{} & -0&0255\sym{{*}{*}{*}} & \multicolumn{2}{c}{} & -0&0311\sym{{*}{*}{*}}\tabularnewline
			& \multicolumn{2}{c}{} & (0&0080) & \multicolumn{2}{c}{} & (0&0100)\tabularnewline
			\addlinespace
			$P_{t-14}$ & -0&0015 & -0&0016 & -0&0016 & -0&0013\tabularnewline
			& (0&0011) & (0&0011) & (0&0013) & (0&0012)\tabularnewline
			\addlinespace
			$M_{t-14}$ & 0&0059\sym{{*}{*}{*}} & 0&0055\sym{{*}{*}{*}} & 0&0099\sym{{*}{*}{*}} & 0&0103\sym{{*}{*}{*}}\tabularnewline
			& (0&0015) & (0&0015) & (0&0017) & (0&0017)\tabularnewline
			$V1_{t-14}\times M_{t-14}$ & -0&0002 & -0&0001 & -0&0002 & -0&0002\tabularnewline
			& (0&0002) & (0&0002) & (0&0003) & (0&0003)\tabularnewline
			\addlinespace
			$V2_{t-14}\times M_{t-14}$ & 0&0001 & 0&0002 & 0&0002 & 0&0003\tabularnewline
			& (0&0004) & (0&0004) & (0&0004) & (0&0004)\tabularnewline
			$V1_{t-14}^{CHN}\times M_{t-14}$ & \multicolumn{2}{c}{} & 0&0003 & \multicolumn{2}{c}{} & 0&0002\tabularnewline
			& \multicolumn{2}{c}{} & (0&0003) & \multicolumn{2}{c}{} & (0&0003)\tabularnewline
			$V2_{t-14}^{CHN}\times M_{t-14}$ & \multicolumn{2}{c}{} & -0&0007 & \multicolumn{2}{c}{} & -0&0008\sym{*}\tabularnewline
			& \multicolumn{2}{c}{} & (0&0005) & \multicolumn{2}{c}{} & (0&0004)\tabularnewline
			\addlinespace
			$\Delta\log\Delta C_{t-14}$ & 0&1022\sym{{*}{*}} & 0&1004\sym{{*}{*}} & 0&0824\sym{{*}{*}} & 0&0826\sym{{*}{*}}\tabularnewline
			& (0&0390) & (0&0391) & (0&0336) & (0&0335)\tabularnewline
			$\log\Delta C_{t-14}$ & -0&0369\sym{{*}{*}{*}} & -0&0391\sym{{*}{*}{*}} & -0&1562\sym{{*}{*}{*}} & -0&1572\sym{{*}{*}{*}}\tabularnewline
			& (0&0105) & (0&0106) & (0&0217) & (0&0219)\tabularnewline
			\addlinespace
			$\Delta\log\Delta T_{t}$ & 0&5145\sym{{*}{*}{*}} & 0&5107\sym{{*}{*}{*}} & 0&4054\sym{{*}{*}{*}} & 0&4025\sym{{*}{*}{*}}\tabularnewline
			& (0&1754) & (0&1745) & (0&1461) & (0&1450)\tabularnewline
			\midrule
			Country fixed effects & \multicolumn{2}{c}{yes} & \multicolumn{2}{c}{yes} & \multicolumn{2}{c}{yes} & \multicolumn{2}{c}{yes}\tabularnewline
			Country specific trend in days & \multicolumn{2}{c}{no} & \multicolumn{2}{c}{no} & \multicolumn{2}{c}{quadratic} & \multicolumn{2}{c}{quadratic}\tabularnewline
			\midrule
			${\rm R}^{2}$ & 0&2707 & 0&2743 & 0&3930 & 0&3962\tabularnewline
			Adjusted ${\rm R}^{2}$ & 0&2702 & 0&2737 & 0&3895 & 0&3926\tabularnewline
			\midrule
		\end{tabular}
	\end{center}
	\footnotesize \textbf{Note:} Time period -- June 1, 2020 to July 8, 2021. The results are produced using daily data. Standard errors in parentheses are clustered at the country level. $^{***}$, $^{**}$, and $^{*}$ denote the 99\%, 95\% and 90\% confidence level respectively.
	\end{table}
	
	\begin{table}[htbp]
		\global\long\def\sym#1{\ifmmode^{#1}\else$^{#1}$\fi}%
		\caption{The direct effects of vaccination, policy, behavior and their interactions
			on deaths}
		\label{fig: interaction_deaths}
		\small%
		\begin{center}
		\begin{tabular}{lr@{\extracolsep{0pt}.}lr@{\extracolsep{0pt}.}lr@{\extracolsep{0pt}.}lr@{\extracolsep{0pt}.}l}
			\hline\hline
			\multicolumn{1}{l}{} & \multicolumn{8}{c}{Dependent variable: $\Delta\log\Delta D_{t}$}\tabularnewline
			\cline{2-9}
			& \multicolumn{2}{c}{(1)} & \multicolumn{2}{c}{(2)} & \multicolumn{2}{c}{(3)} & \multicolumn{2}{c}{(4)}\tabularnewline
			\midrule
			$V1_{t-28}$ & -0&0084\sym{{*}{*}{*}} & -0&0083\sym{{*}{*}{*}} & -0&0077\sym{{*}{*}} & -0&0094\sym{{*}{*}}\tabularnewline
			& (0&0027) & (0&0028) & (0&0032) & (0&0036)\tabularnewline
			$V2_{t-28}$ & -0&0013 & -0&0016 & -0&0062 & -0&0019\tabularnewline
			& (0&0042) & (0&0047) & (0&0052) & (0&0060)\tabularnewline
			$V1_{t-28}^{CHN}$ & \multicolumn{2}{c}{} & 0&0314\sym{{*}{*}{*}} & \multicolumn{2}{c}{} & 0&0045\tabularnewline
			& \multicolumn{2}{c}{} & (0&0040) & \multicolumn{2}{c}{} & (0&0091)\tabularnewline
			\addlinespace
			$V2_{t-28}^{CHN}$ & \multicolumn{2}{c}{} & -0&0357\sym{{*}{*}{*}} & \multicolumn{2}{c}{} & -0&0151\sym{*}\tabularnewline
			& \multicolumn{2}{c}{} & (0&0057) & \multicolumn{2}{c}{} & (0&0076)\tabularnewline
			\addlinespace
			$P_{t-28}$ & -0&0006 & -0&0005 & -0&0007 & -0&0006\tabularnewline
			& (0&0013) & (0&0013) & (0&0010) & (0&0010)\tabularnewline
			\addlinespace
			$M_{t-28}$ & 0&0075\sym{{*}{*}{*}} & 0&0069\sym{{*}{*}{*}} & 0&0081\sym{{*}{*}{*}} & 0&0082\sym{{*}{*}{*}}\tabularnewline
			& (0&0017) & (0&0016) & (0&0015) & (0&0015)\tabularnewline
			$V1_{t-28}\times M_{t-28}$ & -0&0000 & 0&0000 & 0&0002 & 0&0001\tabularnewline
			& (0&0001) & (0&0001) & (0&0001) & (0&0001)\tabularnewline
			\addlinespace
			$V2_{t-28}\times M_{t-28}$ & -0&0003 & -0&0003 & -0&0004\sym{{*}{*}} & -0&0004\tabularnewline
			& (0&0003) & (0&0003) & (0&0002) & (0&0002)\tabularnewline
			$V1_{t-28}^{CHN}\times M_{t-28}$ & \multicolumn{2}{c}{} & 0&0009\sym{{*}{*}{*}} & \multicolumn{2}{c}{} & 0&0001\tabularnewline
			& \multicolumn{2}{c}{} & (0&0002) & \multicolumn{2}{c}{} & (0&0002)\tabularnewline
			$V2_{t-28}^{CHN}\times M_{t-28}$ & \multicolumn{2}{c}{} & -0&0012\sym{{*}{*}{*}} & \multicolumn{2}{c}{} & -0&0003\tabularnewline
			& \multicolumn{2}{c}{} & (0&0003) & \multicolumn{2}{c}{} & (0&0003)\tabularnewline
			$\Delta\log\Delta D_{t-28}$ & 0&0718\sym{{*}{*}} & 0&0715\sym{{*}{*}} & 0&0703\sym{{*}{*}} & 0&0691\sym{{*}{*}}\tabularnewline
			& (0&0274) & (0&0274) & (0&0266) & (0&0267)\tabularnewline
			$\log\Delta D_{t-28}$ & -0&0346\sym{{*}{*}{*}} & -0&0388\sym{{*}{*}{*}} & -0&1282\sym{{*}{*}{*}} & -0&1273\sym{{*}{*}{*}}\tabularnewline
			& (0&0096) & (0&0091) & (0&0117) & (0&0118)\tabularnewline
			\addlinespace
			\midrule
			Country fixed effects & \multicolumn{2}{c}{yes} & \multicolumn{2}{c}{yes} & \multicolumn{2}{c}{yes} & \multicolumn{2}{c}{yes}\tabularnewline
			Country specific trend in days & \multicolumn{2}{c}{no} & \multicolumn{2}{c}{no} & \multicolumn{2}{c}{quadratic} & \multicolumn{2}{c}{quadratic}\tabularnewline
			\midrule
			${\rm R}^{2}$ & 0&0956 & 0&0987 & 0&1728 & 0&1733\tabularnewline
			Adjusted ${\rm R}^{2}$ & 0&0951 & 0&0979 & 0&1679 & 0&1681\tabularnewline
			\midrule
		\end{tabular}
	\end{center}
		\footnotesize \textbf{Note:} Time period -- June 1, 2020 to July 8, 2021. The results are produced using daily data. Standard errors in parentheses are clustered at the country level. $^{***}$, $^{**}$, and $^{*}$ denote the 99\%, 95\% and 90\% confidence level respectively.
	\end{table}
	
	\newpage
	
	\section{Expansion of the list of countries adopting Chinese vaccines}\label{appen: CHN}
	\begin{table}[h!]
		\global\long\def\sym#1{\ifmmode^{#1}\else$^{#1}$\fi}%
		\caption{The direct effects of vaccination, policy, and behavior on new cases}
		\label{fig: CHN}
		\begin{center}
		\begin{tabular}{lr@{\extracolsep{0pt}.}lr@{\extracolsep{0pt}.}lr@{\extracolsep{0pt}.}lr@{\extracolsep{0pt}.}l}
			\hline\hline
			\multicolumn{1}{l}{} & \multicolumn{8}{c}{Dependent variable: $\Delta\log\Delta C_{t}$}\tabularnewline
			\cline{2-9}
			& \multicolumn{2}{c}{(1)} & \multicolumn{2}{c}{(2)} & \multicolumn{2}{c}{(3)} & \multicolumn{2}{c}{(4)}\tabularnewline
			\midrule
			$V1_{t-21}$ & -0&0074\sym{{*}{*}{*}} & -0&0079\sym{{*}{*}{*}} & -0&0173\sym{{*}{*}{*}} & -0&0192\sym{{*}{*}{*}}\tabularnewline
			& (0&0021) & (0&0023) & (0&0043) & (0&0040)\tabularnewline
			\addlinespace
			$V2_{t-7}$ & 0&0029 & 0&0033 & 0&0038 & 0&0082\tabularnewline
			& (0&0025) & (0&0029) & (0&0057) & (0&0063)\tabularnewline
			$V1_{t-21}^{CHN}$ & \multicolumn{2}{c}{} & 0&0085 & \multicolumn{2}{c}{} & 0&0135\sym{*}\tabularnewline
			& \multicolumn{2}{c}{} & (0&0051) & \multicolumn{2}{c}{} & (0&0068)\tabularnewline
			$V2_{t-7}^{CHN}$ & \multicolumn{2}{c}{} & -0&0093\sym{*} & \multicolumn{2}{c}{} & -0&0230\sym{{*}{*}{*}}\tabularnewline
			& \multicolumn{2}{c}{} & (0&0051) & \multicolumn{2}{c}{} & (0&0082)\tabularnewline
			\addlinespace
			$P_{t-14}$ & -0&0020 & -0&0020 & -0&0027\sym{*} & -0&0028\sym{*}\tabularnewline
			& (0&0012) & (0&0013) & (0&0015) & (0&0015)\tabularnewline
			\addlinespace
			$M_{t-14}$ & 0&0052\sym{{*}{*}{*}} & 0&0052\sym{{*}{*}{*}} & 0&0101\sym{{*}{*}{*}} & 0&0098\sym{{*}{*}{*}}\tabularnewline
			& (0&0012) & (0&0012) & (0&0014) & (0&0013)\tabularnewline
			\addlinespace
			$\Delta\log\Delta C_{t-14}$ & 0&1102\sym{{*}{*}{*}} & 0&1092\sym{{*}{*}{*}} & 0&0901\sym{{*}{*}} & 0&0912\sym{{*}{*}}\tabularnewline
			& (0&0395) & (0&0398) & (0&0350) & (0&0347)\tabularnewline
			\addlinespace
			$\log\Delta C_{t-14}$ & -0&0374\sym{{*}{*}{*}} & -0&0386\sym{{*}{*}{*}} & -0&1496\sym{{*}{*}{*}} & -0&1518\sym{{*}{*}{*}}\tabularnewline
			& (0&0077) & (0&0079) & (0&0214) & (0&0213)\tabularnewline
			\addlinespace
			$\Delta\log\Delta T_{t}$ & 0&5094\sym{{*}{*}{*}} & 0&5086\sym{{*}{*}{*}} & 0&4090\sym{{*}{*}{*}} & 0&4075\sym{{*}{*}{*}}\tabularnewline
			& (0&1775) & (0&1774) & (0&1499) & (0&1494)\tabularnewline
			\midrule
			Country fixed effects & \multicolumn{2}{c}{yes} & \multicolumn{2}{c}{yes} & \multicolumn{2}{c}{yes} & \multicolumn{2}{c}{yes}\tabularnewline
			Country specific trend in weeks & \multicolumn{2}{c}{no} & \multicolumn{2}{c}{no} & \multicolumn{2}{c}{quadratic} & \multicolumn{2}{c}{quadratic}\tabularnewline
			\midrule
			${\rm R}^{2}$ & 0&2838 & 0&2845 & 0&3987 & 0&4011\tabularnewline
			Adjusted ${\rm R}^{2}$ & 0&2834 & 0&2841 & 0&3953 & 0&3976\tabularnewline
			Number of countries & \multicolumn{2}{c}{37} & \multicolumn{2}{c}{37} & \multicolumn{2}{c}{37} & \multicolumn{2}{c}{37}\tabularnewline
			Obs. per country & \multicolumn{2}{c}{382} & \multicolumn{2}{c}{382} & \multicolumn{2}{c}{382} & \multicolumn{2}{c}{382}\tabularnewline
			\midrule
		\end{tabular}
	\end{center}
		\footnotesize \textbf{Note:} Time period -- June 1, 2020 to July 8, 2021. The results are produced using daily data. Standard errors in parentheses are clustered at the country level. $^{***}$, $^{**}$, and $^{*}$ denote the 99\%, 95\% and 90\% confidence level respectively.
	\end{table}
	
	\begin{table}[h!]
		\global\long\def\sym#1{\ifmmode^{#1}\else$^{#1}$\fi}%
		\caption{The direct effects of vaccination, policy, and behavior on new deaths}
		\begin{center}
		\begin{tabular}{lr@{\extracolsep{0pt}.}lr@{\extracolsep{0pt}.}lr@{\extracolsep{0pt}.}lr@{\extracolsep{0pt}.}l}
			\hline\hline
			\multicolumn{1}{l}{} & \multicolumn{8}{c}{Dependent variable: $\Delta\log\Delta D_{t}$}\tabularnewline
			\cline{2-9}
			& \multicolumn{2}{c}{(1)} & \multicolumn{2}{c}{(2)} & \multicolumn{2}{c}{(3)} & \multicolumn{2}{c}{(4)}\tabularnewline
			\midrule
			$V1_{t-35}$ & -0&0096\sym{{*}{*}{*}} & -0&0094\sym{{*}{*}{*}} & -0&0125\sym{{*}{*}{*}} & -0&0134\sym{{*}{*}{*}}\tabularnewline
			& (0&0021) & (0&0022) & (0&0040) & (0&0040)\tabularnewline
			\addlinespace
			$V2_{t-21}$ & 0&0036 & 0&0022 & 0&0040 & 0&0074\sym{*}\tabularnewline
			& (0&0029) & (0&0031) & (0&0043) & (0&0043)\tabularnewline
			$V1_{t-35}^{CHN}$ & \multicolumn{2}{c}{} & 0&0061 & \multicolumn{2}{c}{} & 0&0050\tabularnewline
			& \multicolumn{2}{c}{} & (0&0078) & \multicolumn{2}{c}{} & (0&0065)\tabularnewline
			$V2_{t-21}^{CHN}$ & \multicolumn{2}{c}{} & -0&0032 & \multicolumn{2}{c}{} & -0&0126\sym{{*}{*}}\tabularnewline
			& \multicolumn{2}{c}{} & (0&0077) & \multicolumn{2}{c}{} & (0&0062)\tabularnewline
			\addlinespace
			$P_{t-28}$ & -0&0008 & -0&0006 & -0&0013 & -0&0014\tabularnewline
			& (0&0014) & (0&0014) & (0&0013) & (0&0013)\tabularnewline
			\addlinespace
			$M_{t-28}$ & 0&0057\sym{{*}{*}{*}} & 0&0056\sym{{*}{*}{*}} & 0&0077\sym{{*}{*}{*}} & 0&0075\sym{{*}{*}{*}}\tabularnewline
			& (0&0016) & (0&0015) & (0&0016) & (0&0015)\tabularnewline
			$\Delta\log\Delta D_{t-28}$ & 0&0584\sym{{*}{*}} & 0&0584\sym{{*}{*}} & 0&0574\sym{{*}{*}} & 0&0564\sym{{*}{*}}\tabularnewline
			& (0&0281) & (0&0284) & (0&0258) & (0&0258)\tabularnewline
			\addlinespace
			$\log\Delta D_{t-28}$ & -0&0429\sym{{*}{*}{*}} & -0&0464\sym{{*}{*}{*}} & -0&1308\sym{{*}{*}{*}} & -0&1308\sym{{*}{*}{*}}\tabularnewline
			& (0&0096) & (0&0094) & (0&0112) & (0&0113)\tabularnewline
			\addlinespace
			\midrule
			Country fixed effects & \multicolumn{2}{c}{yes} & \multicolumn{2}{c}{yes} & \multicolumn{2}{c}{yes} & \multicolumn{2}{c}{yes}\tabularnewline
			Country specific trend in weeks & \multicolumn{2}{c}{no} & \multicolumn{2}{c}{no} & \multicolumn{2}{c}{quadratic} & \multicolumn{2}{c}{quadratic}\tabularnewline
			\midrule
			${\rm R}^{2}$ & 0&0966 & 0&0979 & 0&1747 & 0&1752\tabularnewline
			Adjusted ${\rm R}^{2}$ & 0&0962 & 0&0973 & 0&1698 & 0&1702\tabularnewline
			Number of countries & \multicolumn{2}{c}{37} & \multicolumn{2}{c}{37} & \multicolumn{2}{c}{37} & \multicolumn{2}{c}{37}\tabularnewline
			Obs. per country & \multicolumn{2}{c}{368} & \multicolumn{2}{c}{368} & \multicolumn{2}{c}{368} & \multicolumn{2}{c}{368}\tabularnewline
			\midrule
		\end{tabular}
	\end{center}
		\footnotesize \textbf{Note:} Time period -- June 1, 2020 to July 8, 2021. The results are produced using daily data. Standard errors in parentheses are clustered at the country level. $^{***}$, $^{**}$, and $^{*}$ denote the 99\%, 95\% and 90\% confidence level respectively.
	\end{table}

	\clearpage
		
	\section{Additional counterfactuals}\label{appen: Additional counterfactuals}
	\begin{figure}[h!]
		\begin{centering}
			\par\end{centering}
		\begin{centering}
			\caption{Canada (Left: 6 weeks, Right: 12 weeks)}
			\par\end{centering}
		\begin{centering}
			\includegraphics[width=0.45\textwidth,height=0.16\textheight]{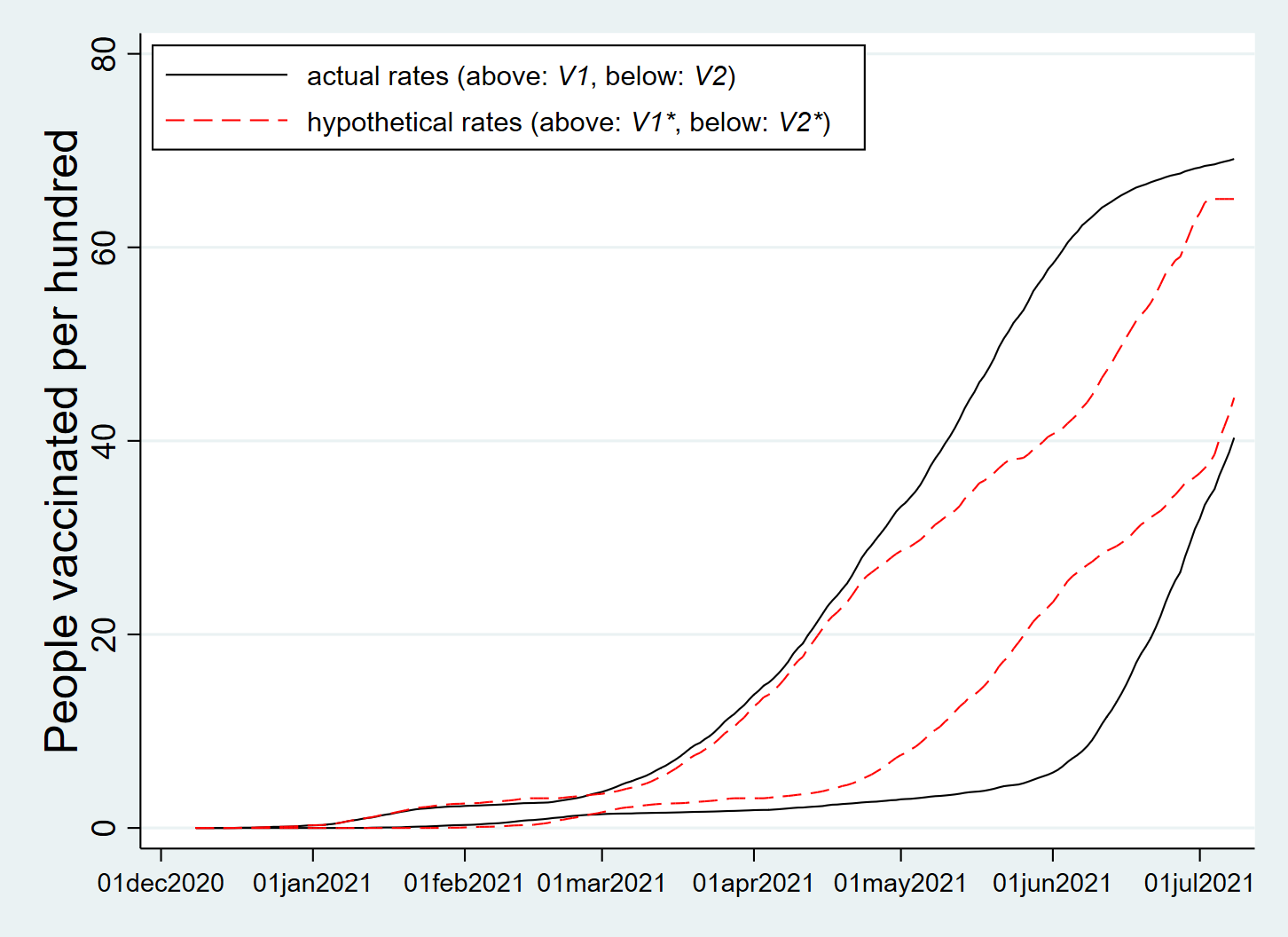}\includegraphics[width=0.45\textwidth,height=0.16\textheight]{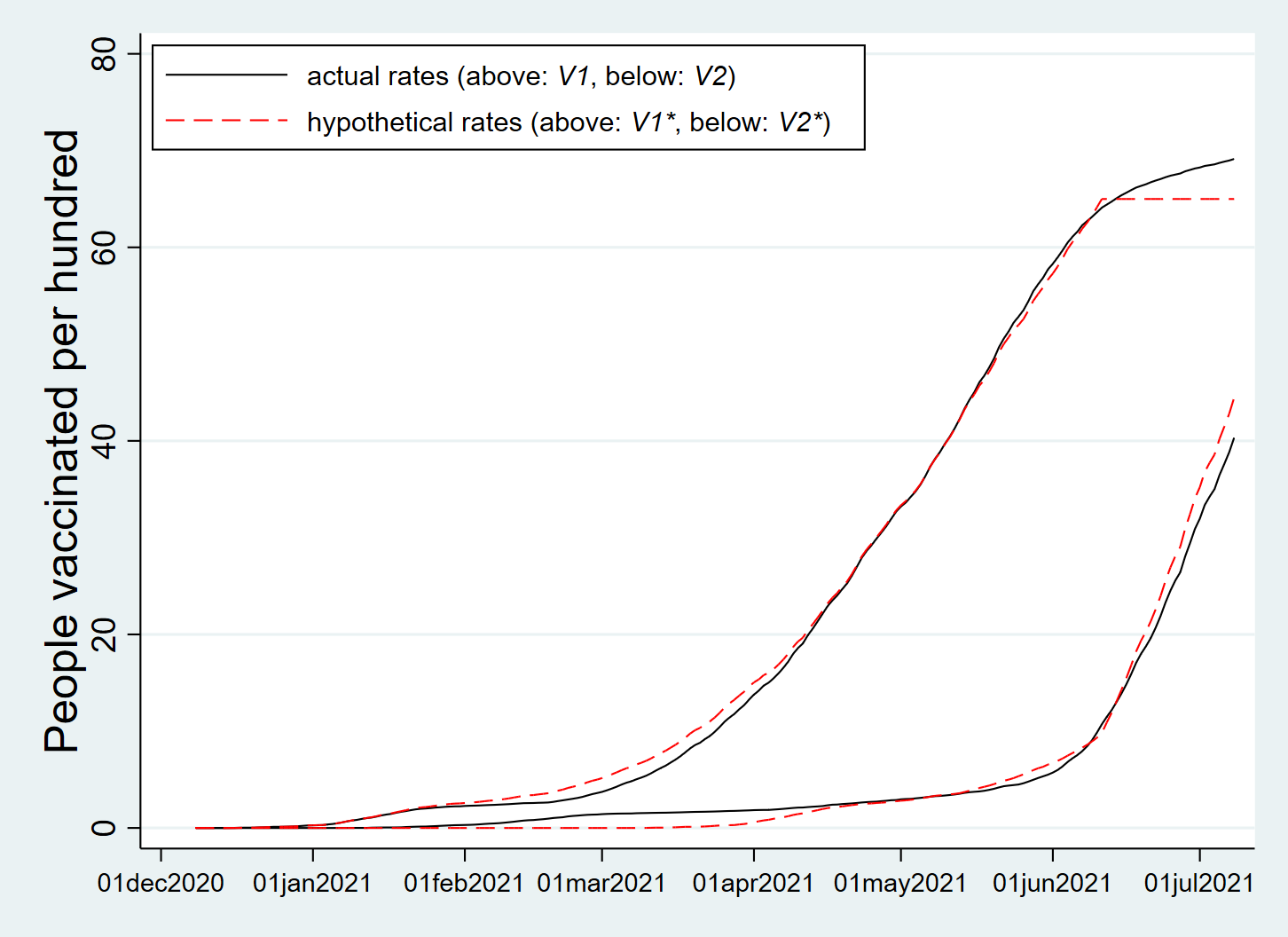}
			\par\end{centering}
		\begin{centering}
			\includegraphics[width=0.45\textwidth,height=0.16\textheight]{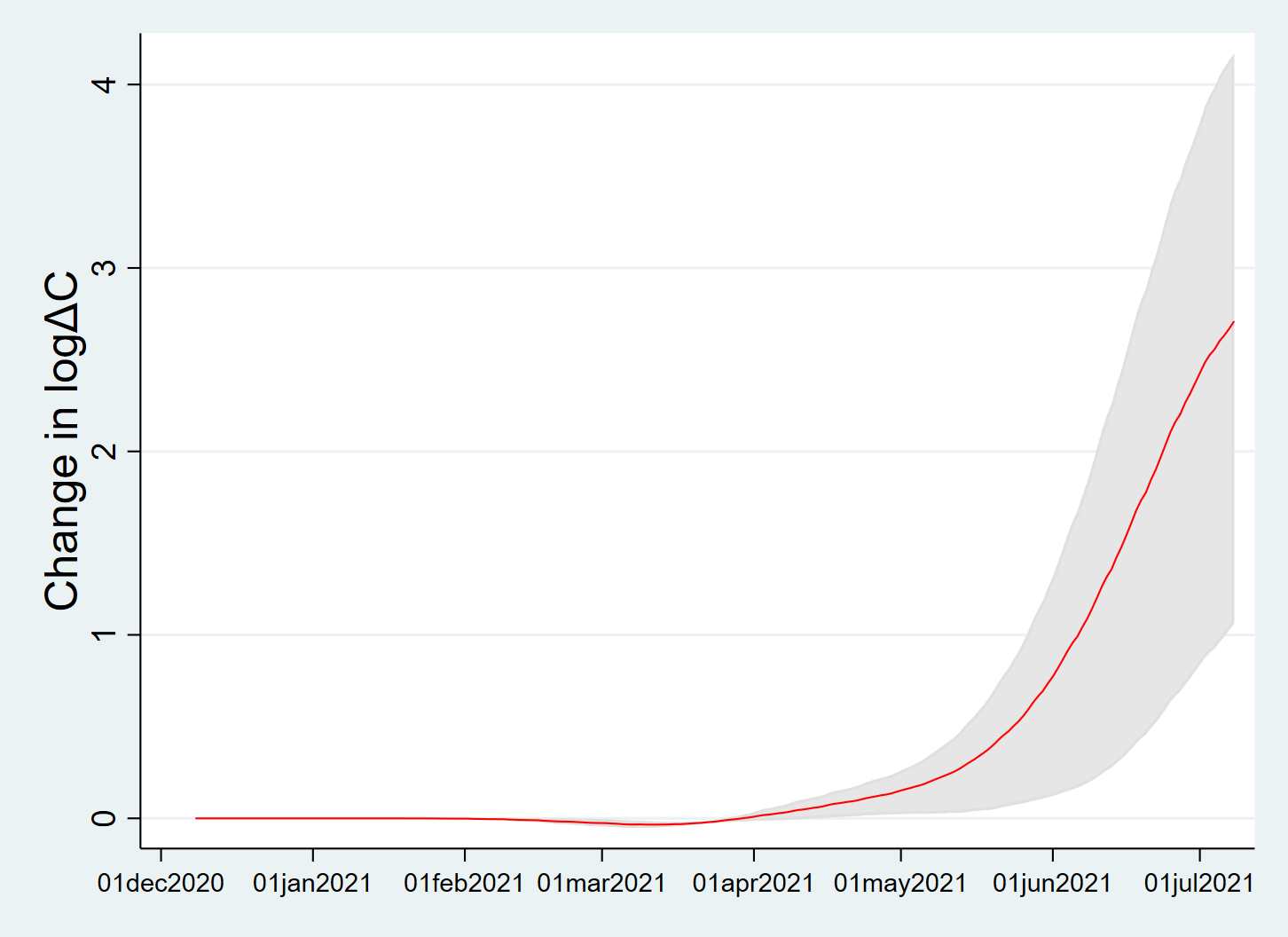}\includegraphics[width=0.45\textwidth,height=0.16\textheight]{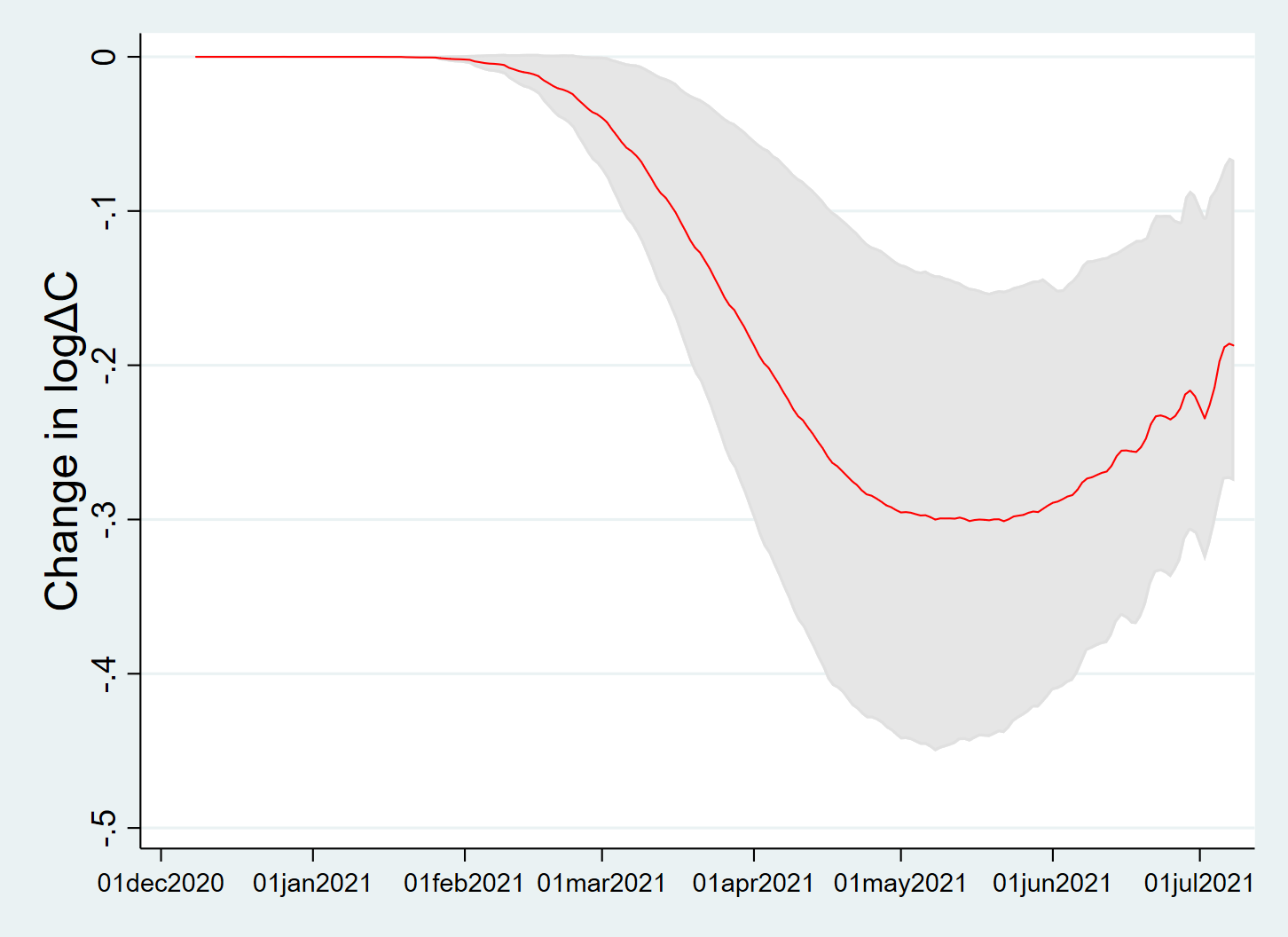}
			\par\end{centering}
		\begin{centering}
			\includegraphics[width=0.45\textwidth,height=0.16\textheight]{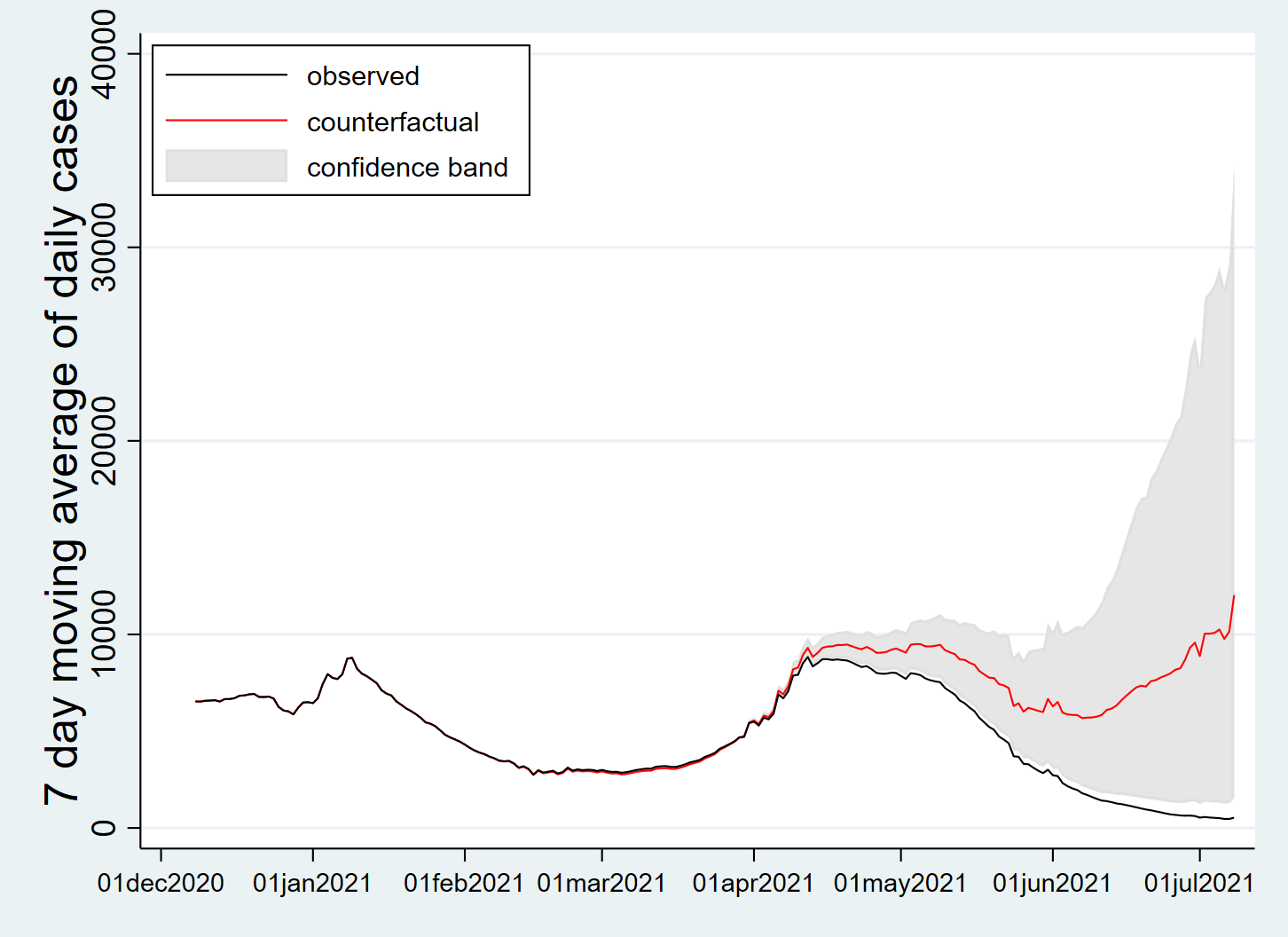}\includegraphics[width=0.45\textwidth,height=0.16\textheight]{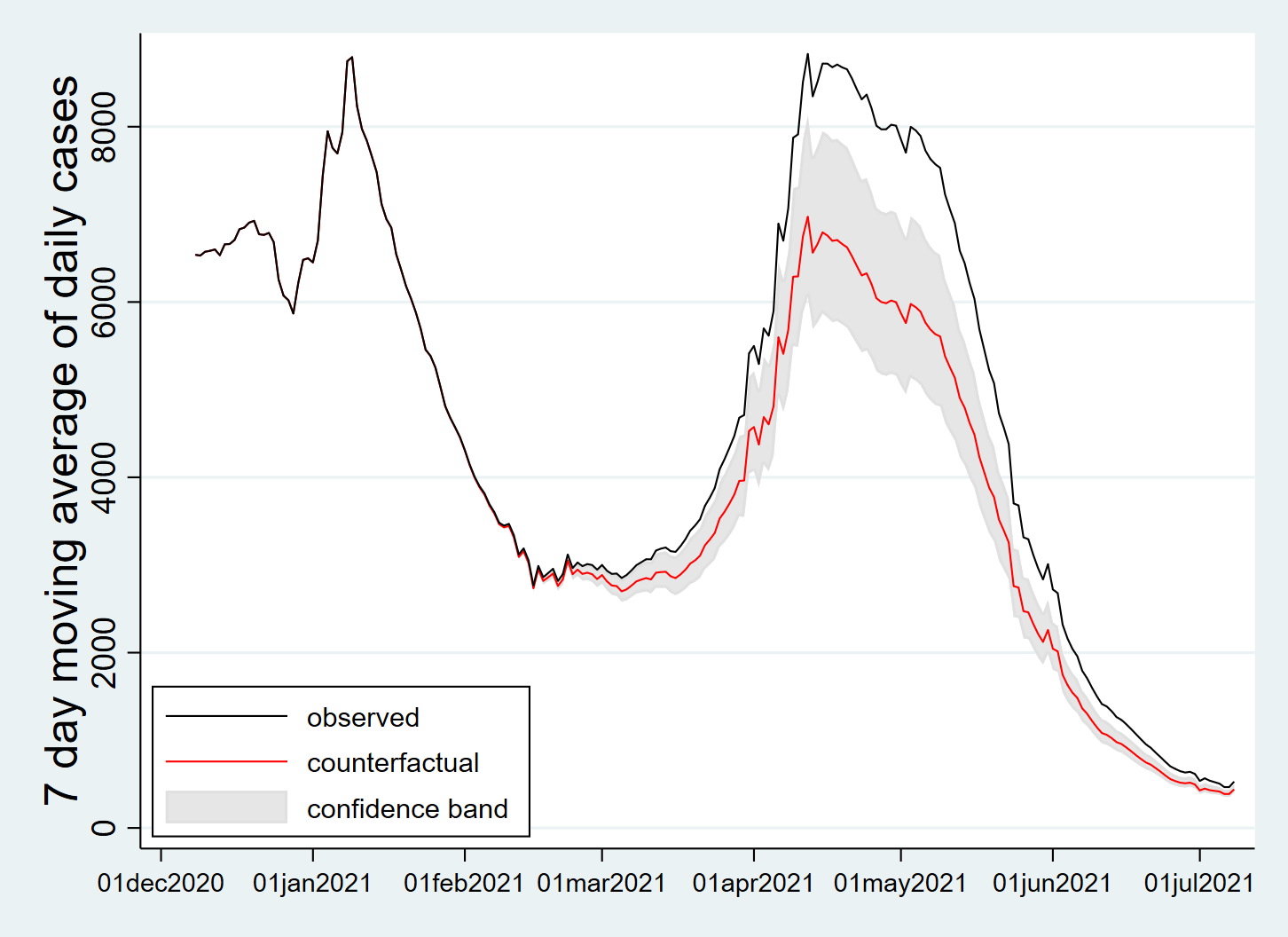}
			\par\end{centering}
		\begin{centering}
			\includegraphics[width=0.45\textwidth,height=0.16\textheight]{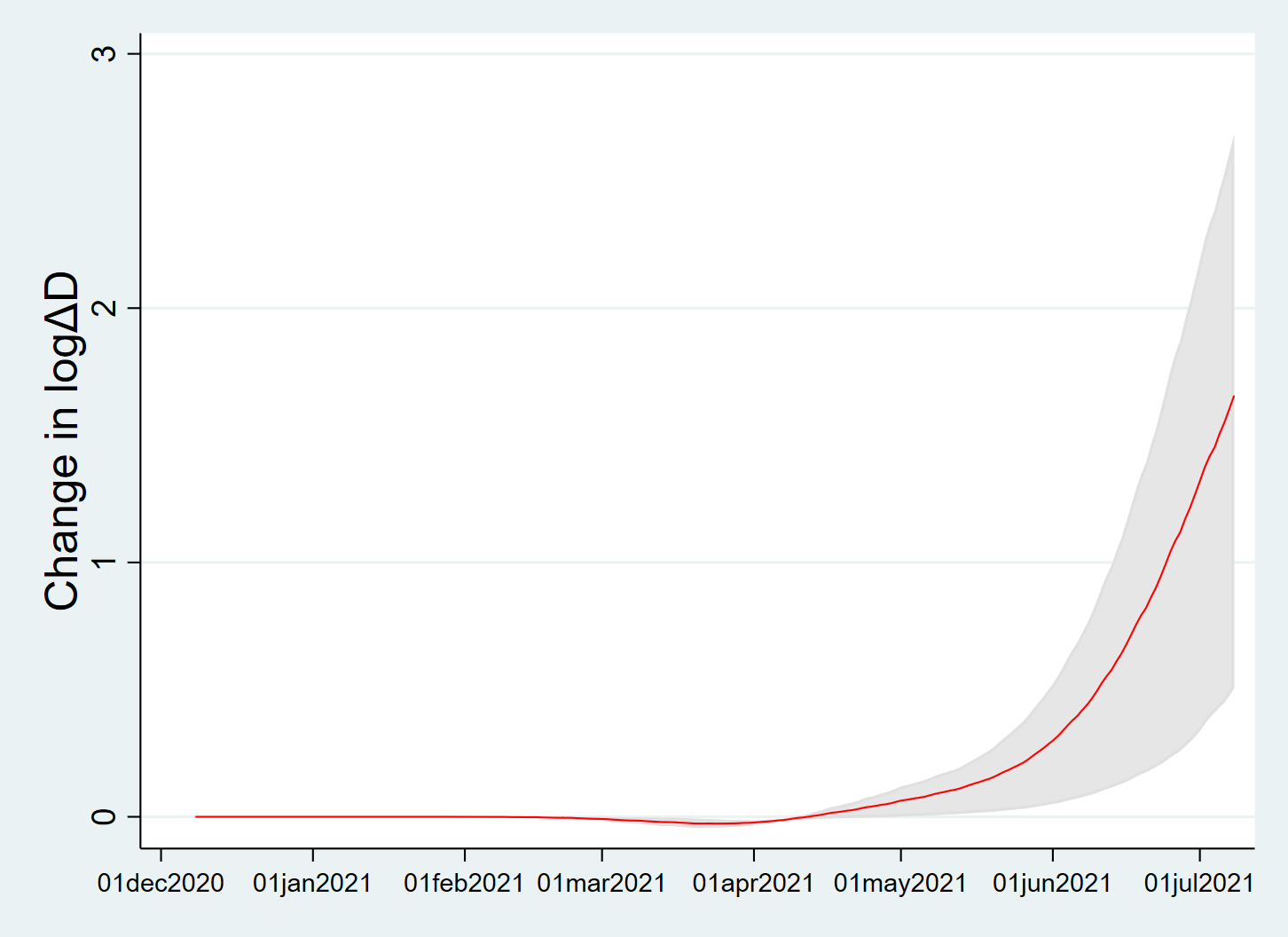}\includegraphics[width=0.45\textwidth,height=0.16\textheight]{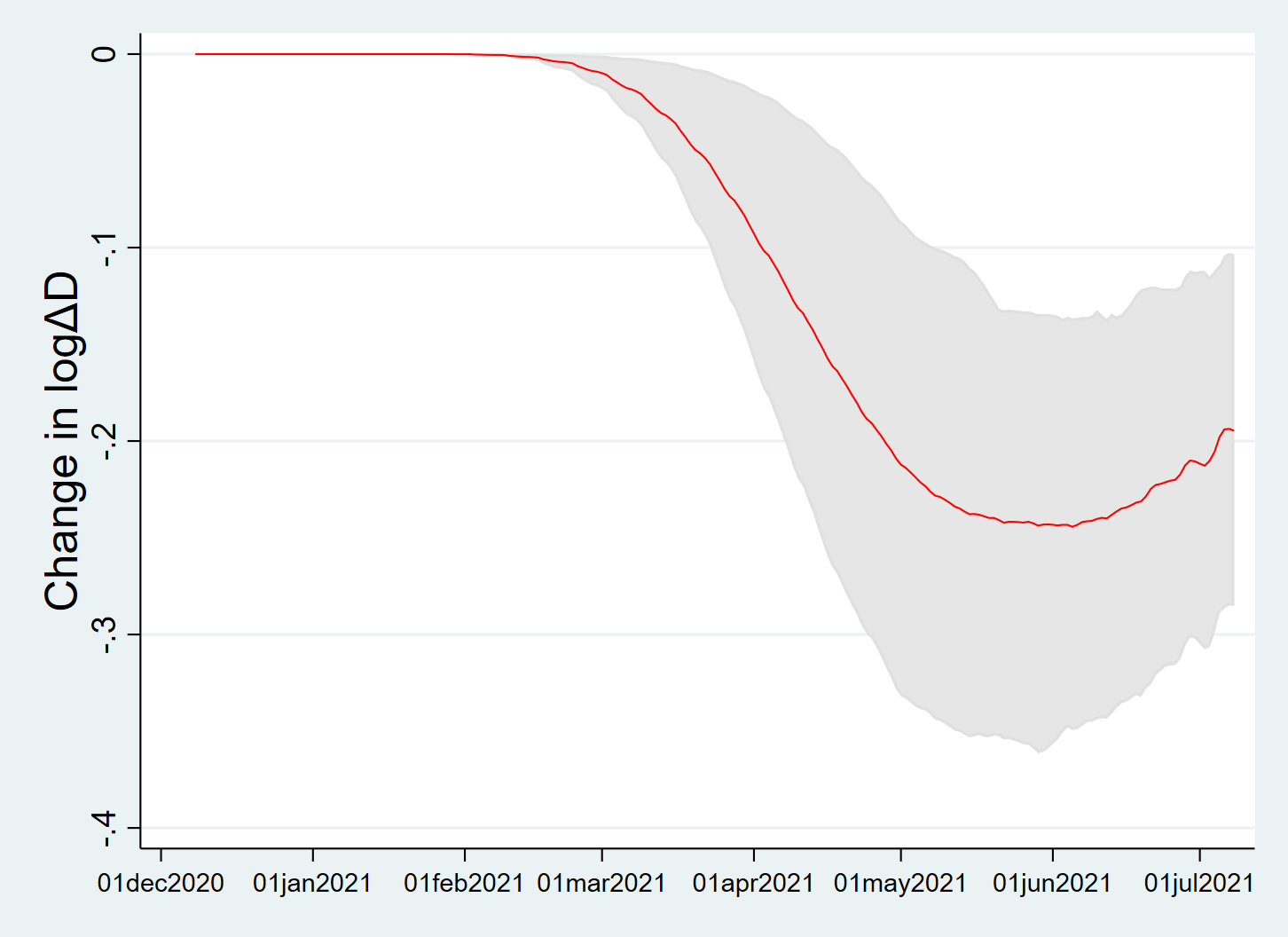}
			\par\end{centering}
		\begin{centering}
			\includegraphics[width=0.45\textwidth,height=0.16\textheight]{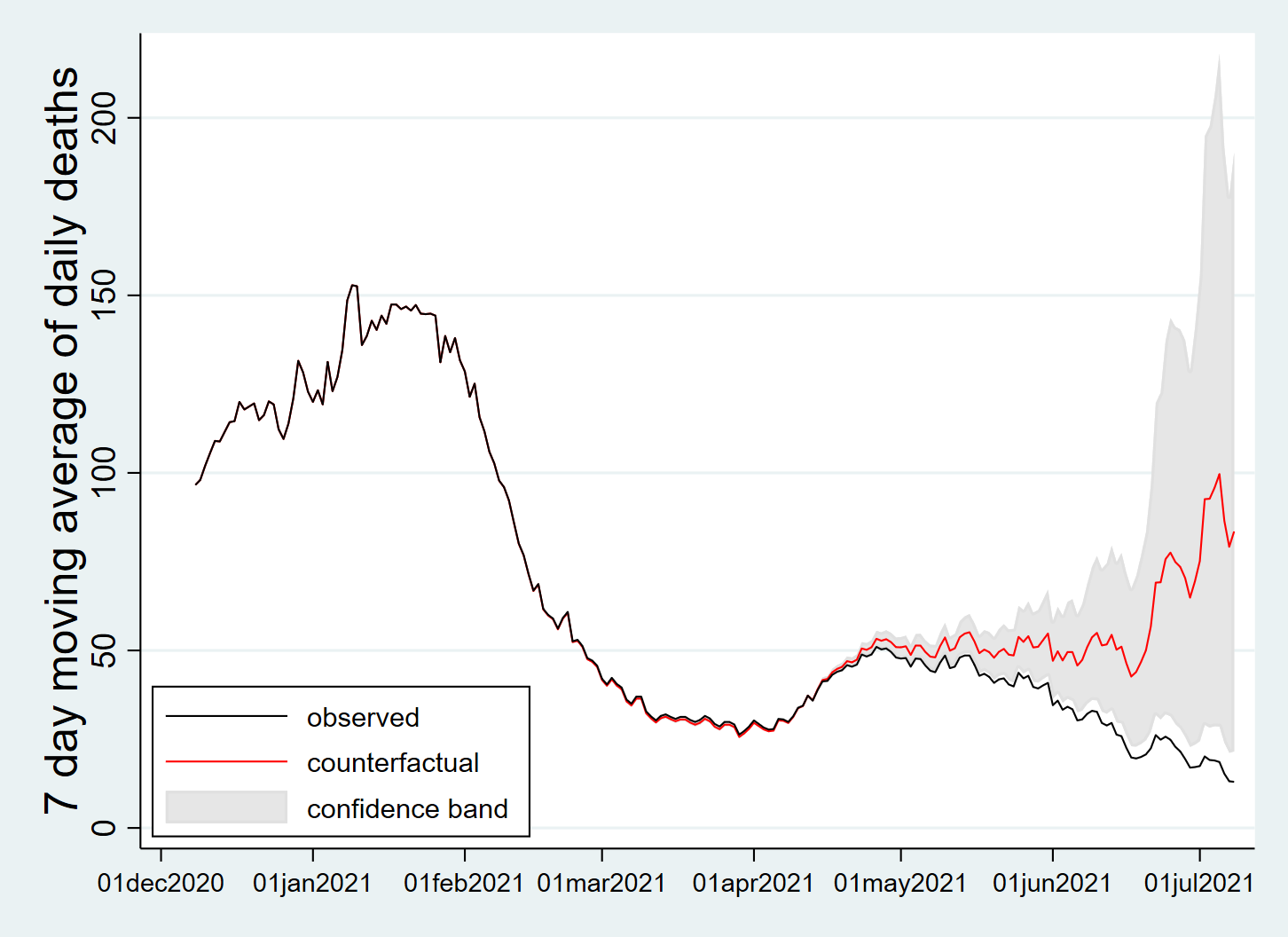}\includegraphics[width=0.45\textwidth,height=0.16\textheight]{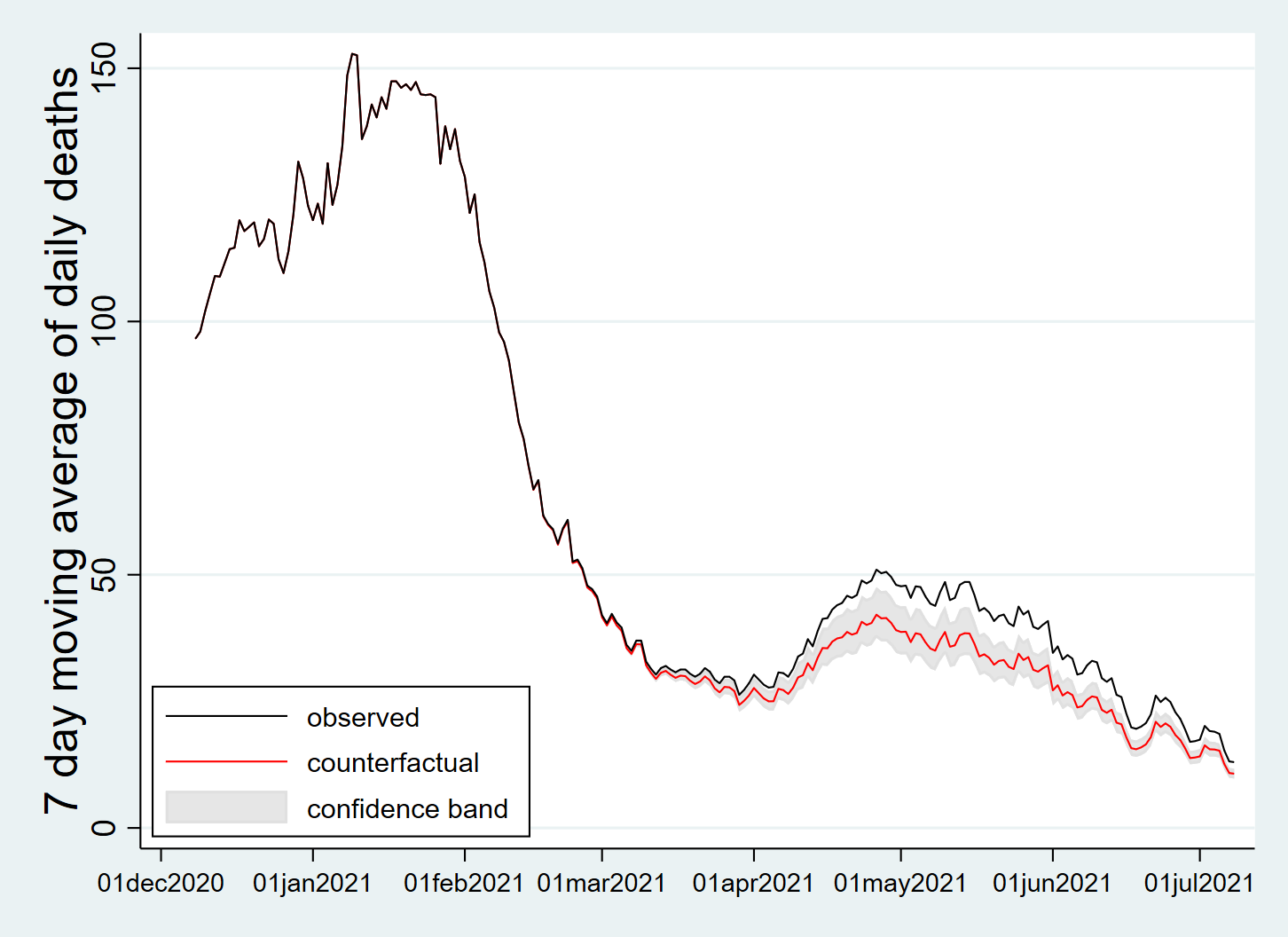}
			\par\end{centering}
		\footnotesize\textbf{Note}: In the first row of panels, black solid lines are actual vaccination rates, and red dotted lines are hypothetical rates we consider in our counterfactual experiments. In the second and fourth row, red lines are changes in the log of weekly case/death counts, respectively. In the third and fifth row of panels, black lines are actual case/death numbers, and red lines are counterfactual counts. Shaded areas are 90\% confidence bands.
	\end{figure}

	\begin{figure}
		\begin{centering}
			\par\end{centering}
		\begin{centering}
			\caption{US (Left: 6 weeks, Right: 12 weeks)}
			\par\end{centering}
		\begin{centering}
			\includegraphics[width=0.45\textwidth,height=0.16\textheight]{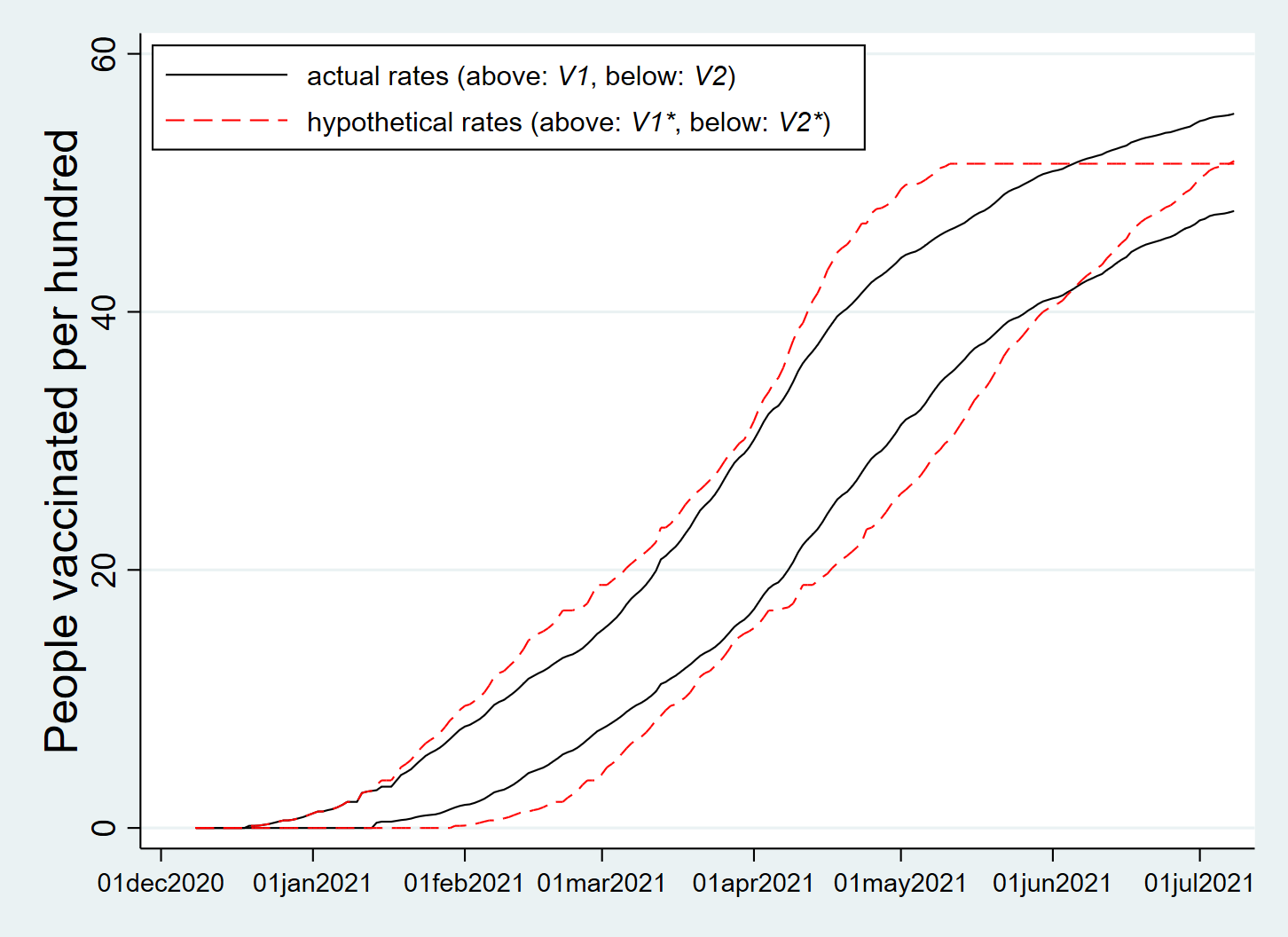}\includegraphics[width=0.45\textwidth,height=0.16\textheight]{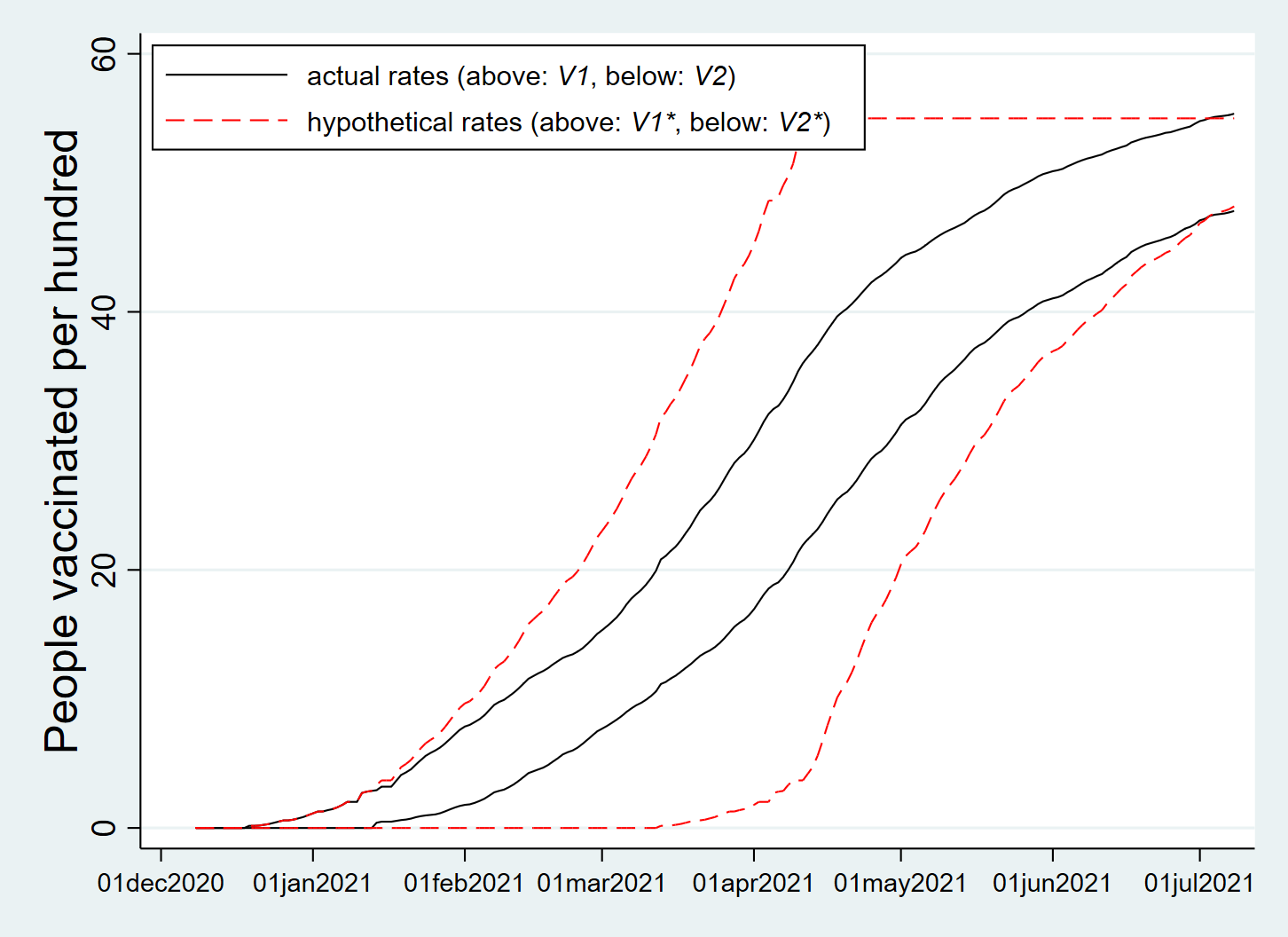}
			\par\end{centering}
		\begin{centering}
			\includegraphics[width=0.45\textwidth,height=0.16\textheight]{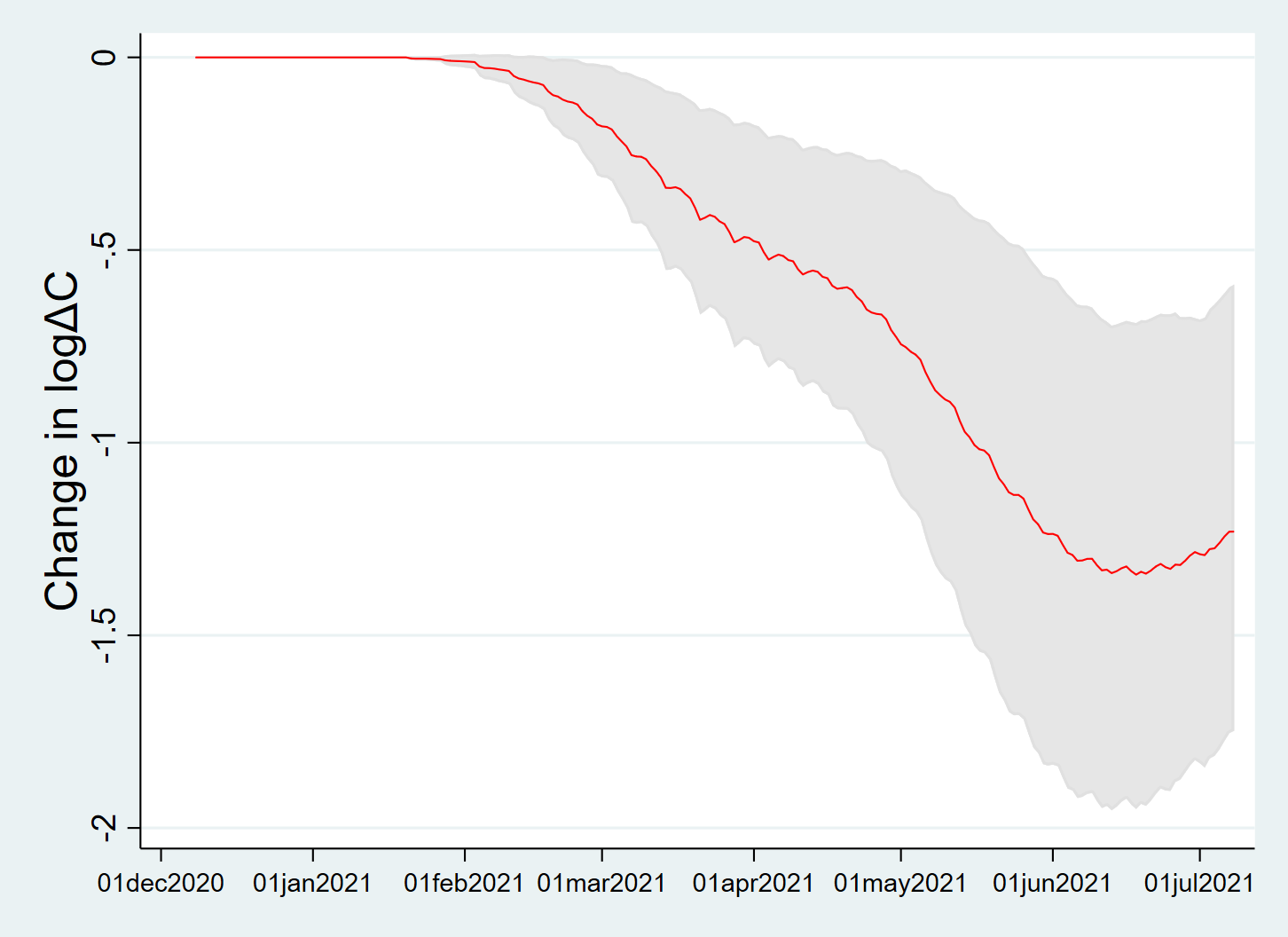}\includegraphics[width=0.45\textwidth,height=0.16\textheight]{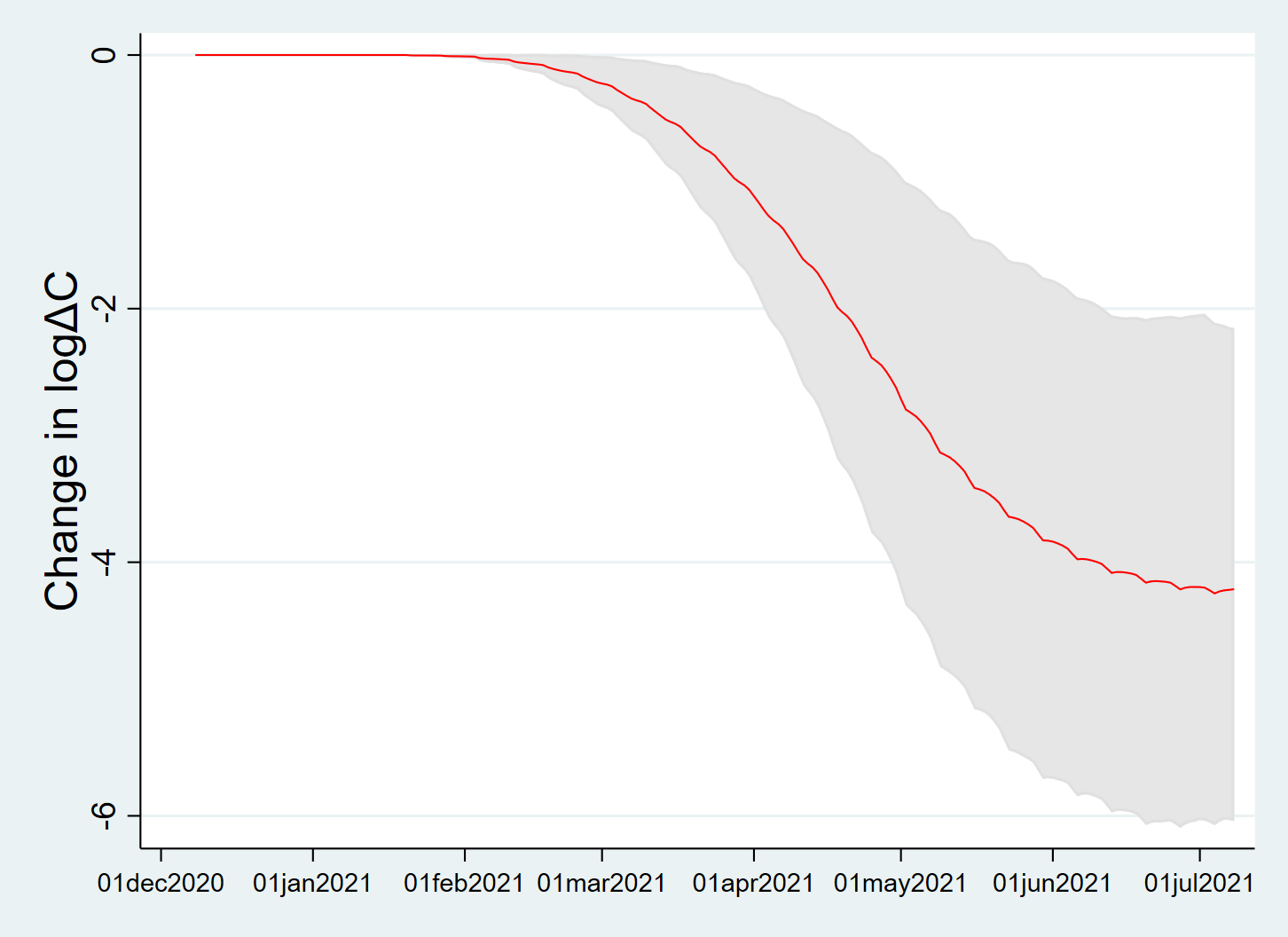}
			\par\end{centering}
		\begin{centering}
			\includegraphics[width=0.45\textwidth,height=0.16\textheight]{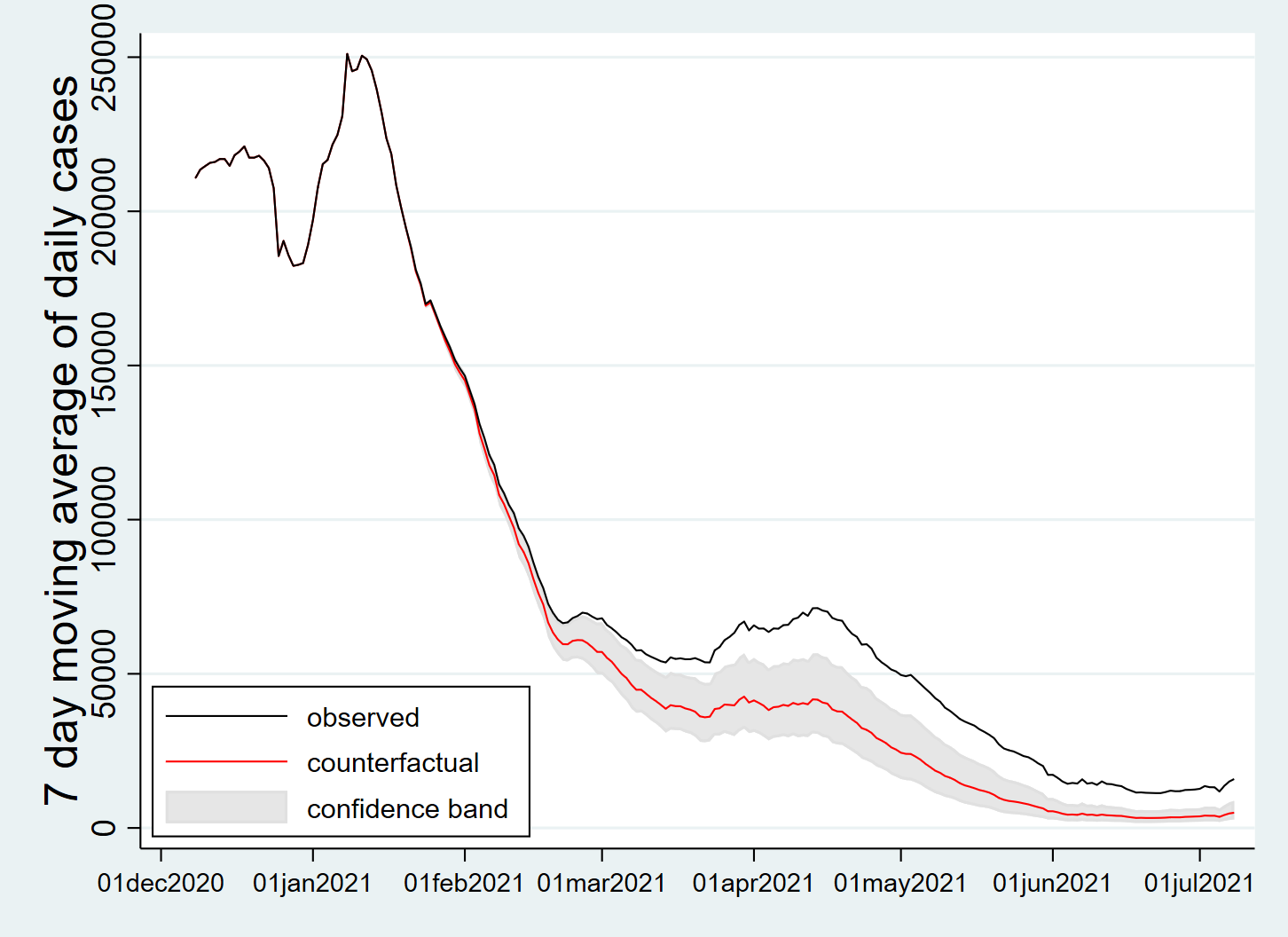}\includegraphics[width=0.45\textwidth,height=0.16\textheight]{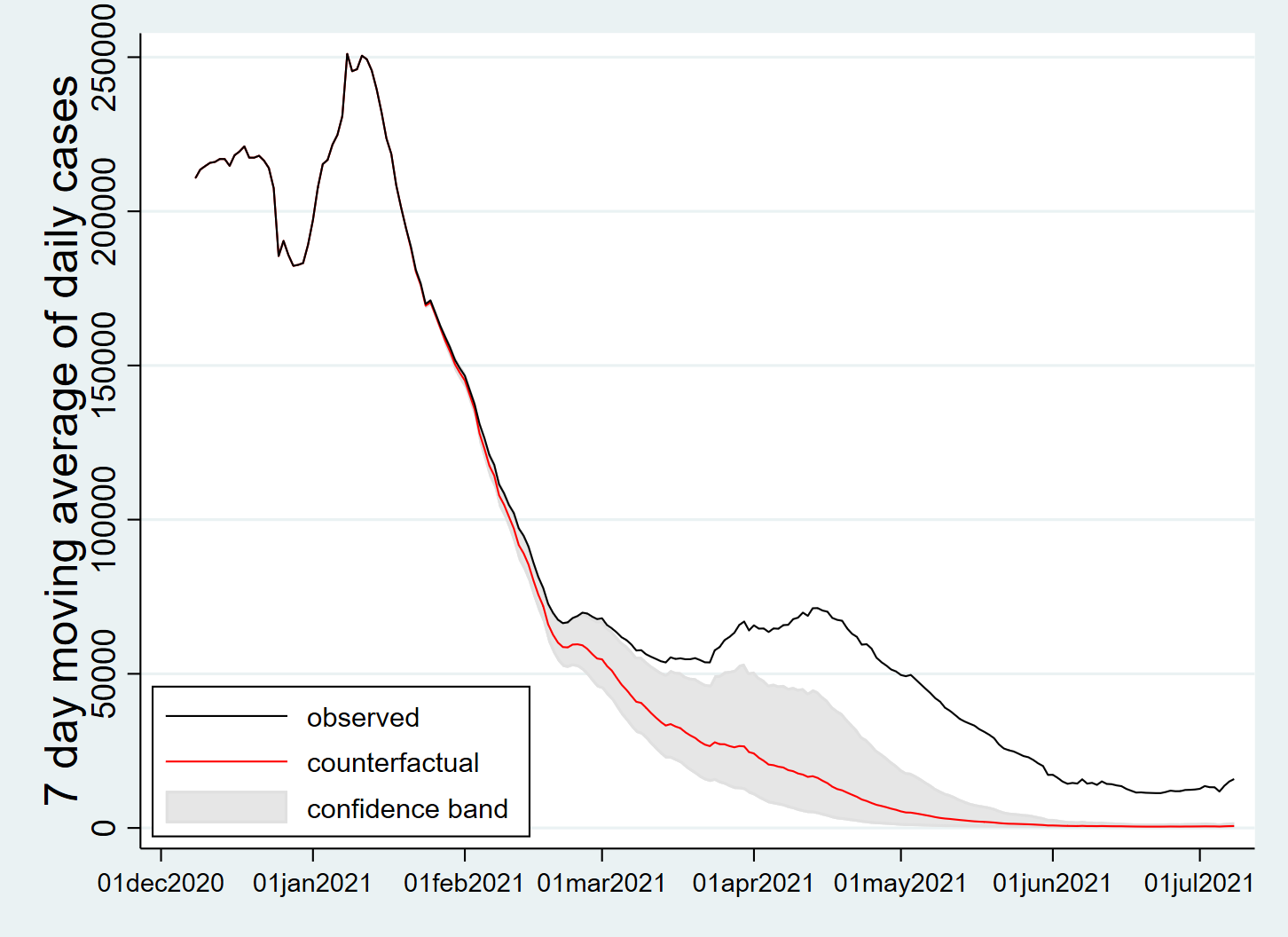}
			\par\end{centering}
		\begin{centering}
			\includegraphics[width=0.45\textwidth,height=0.16\textheight]{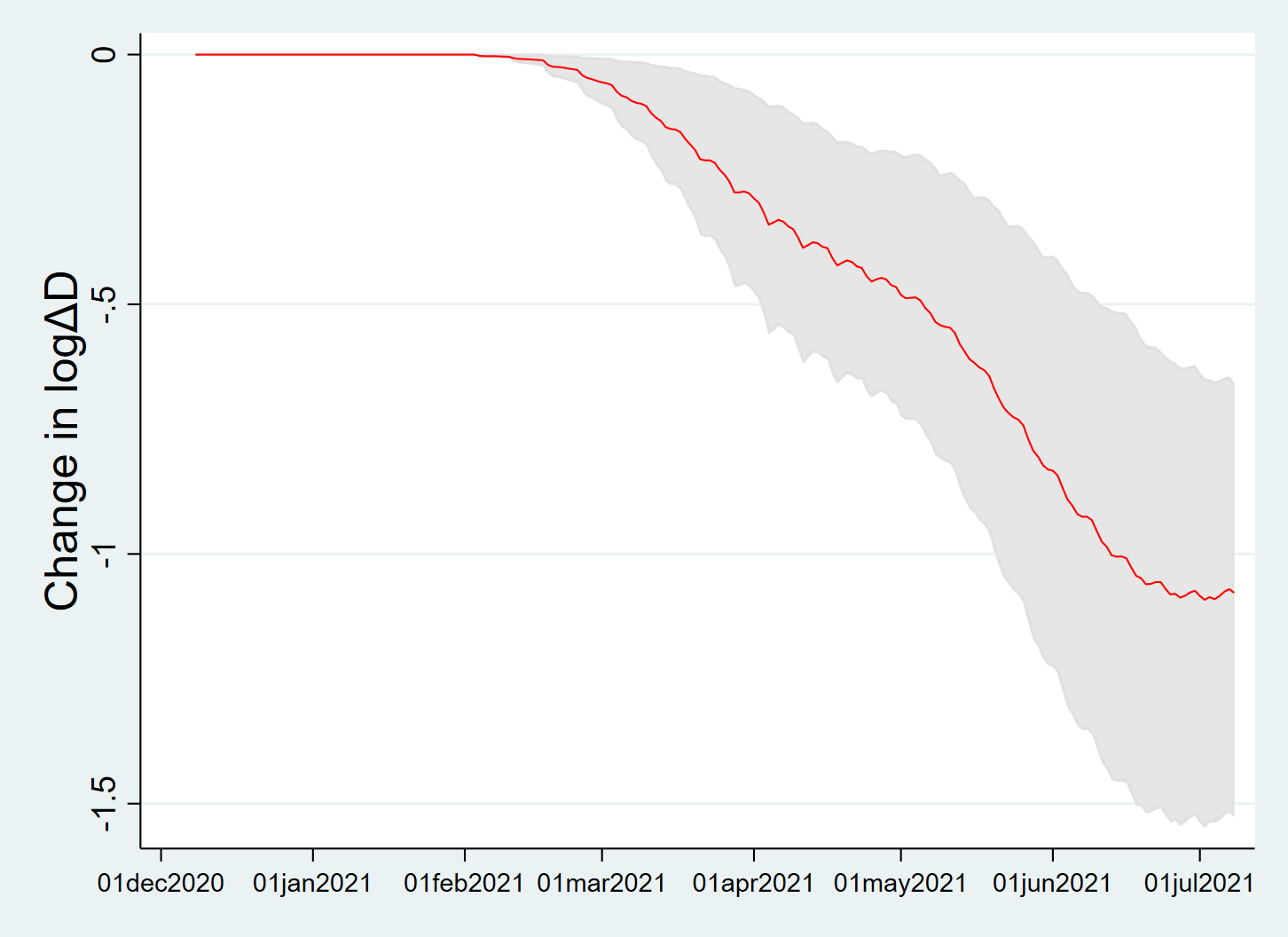}\includegraphics[width=0.45\textwidth,height=0.16\textheight]{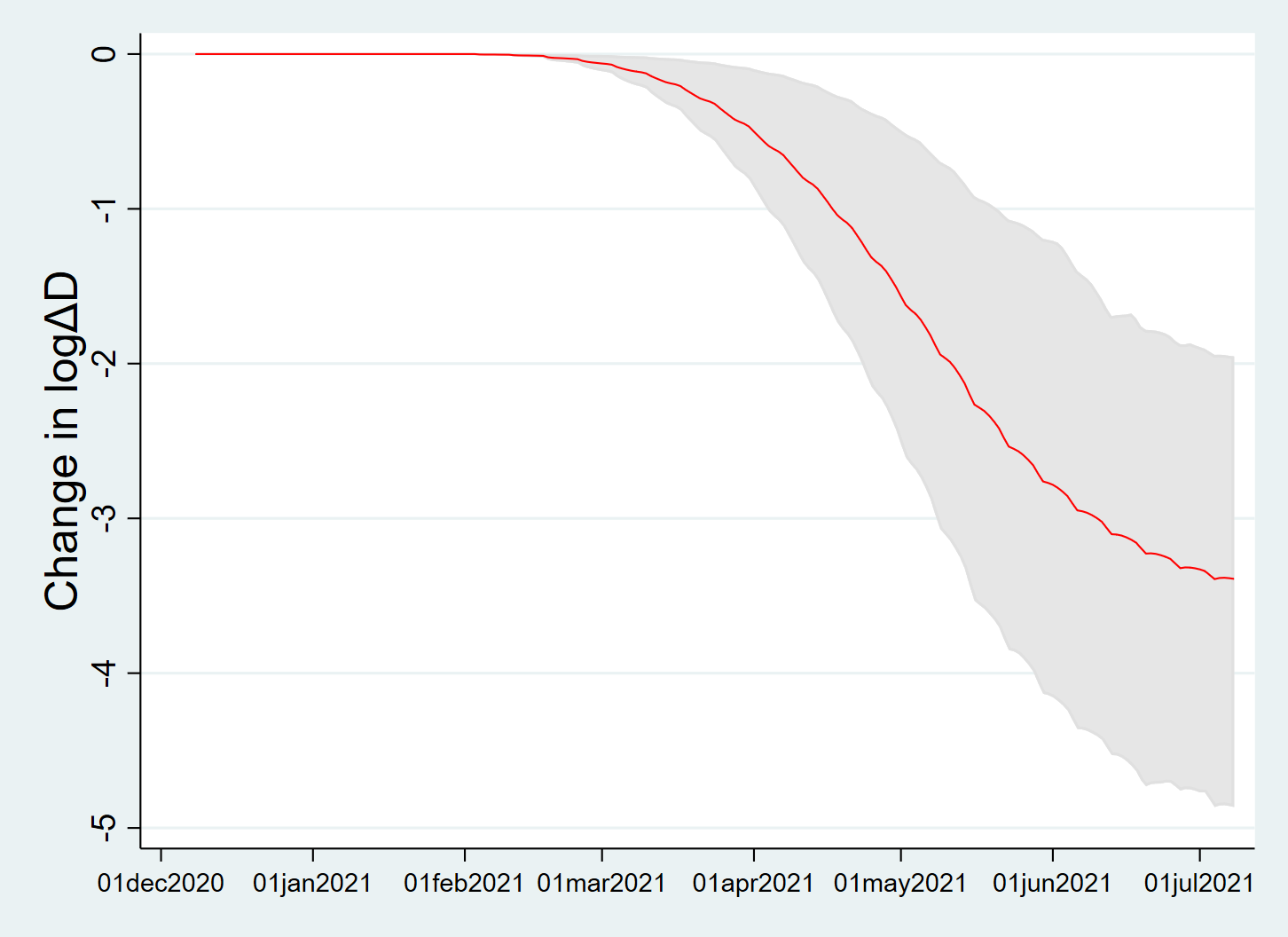}
			\par\end{centering}
		\begin{centering}
			\includegraphics[width=0.45\textwidth,height=0.16\textheight]{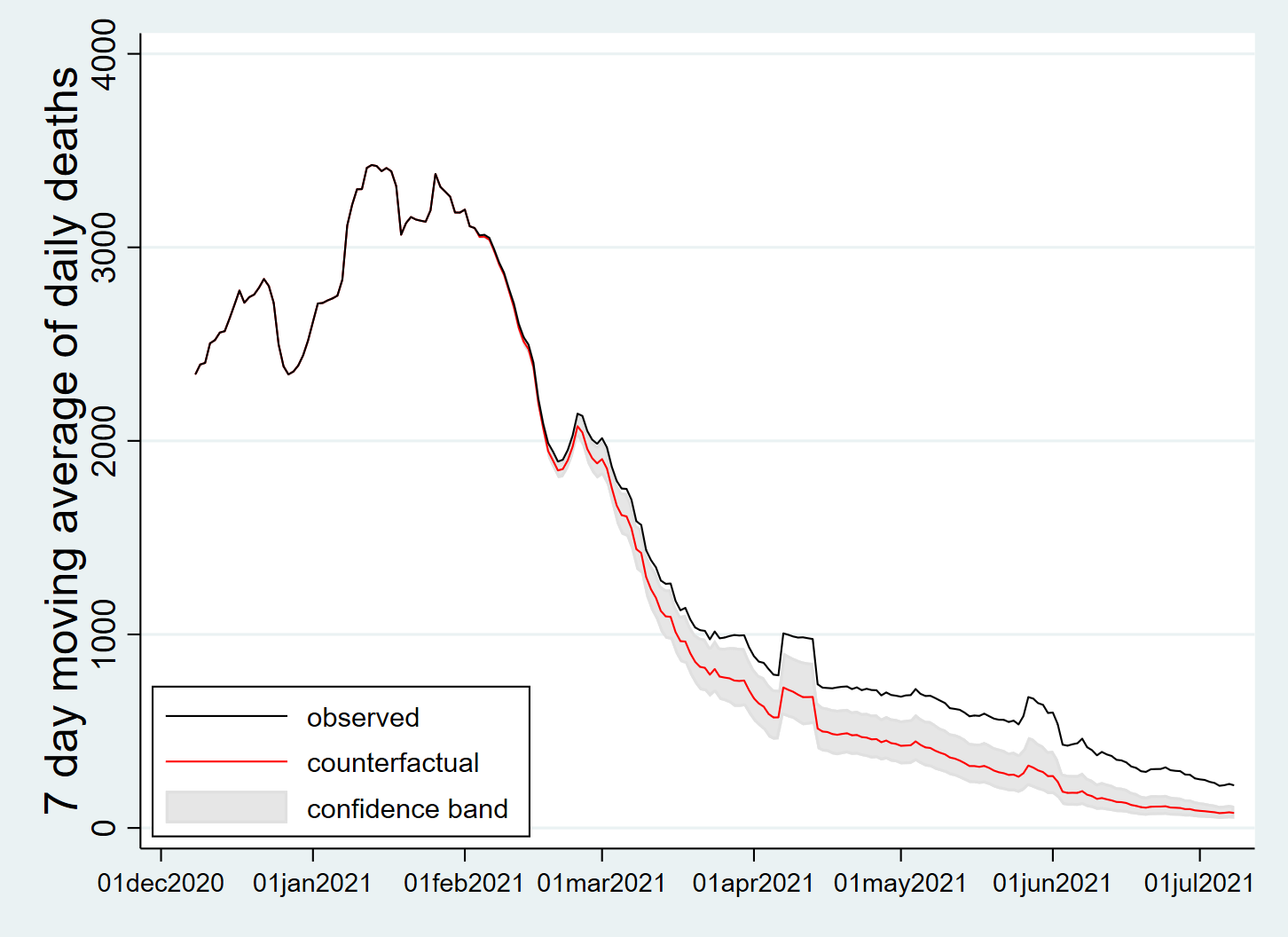}\includegraphics[width=0.45\textwidth,height=0.16\textheight]{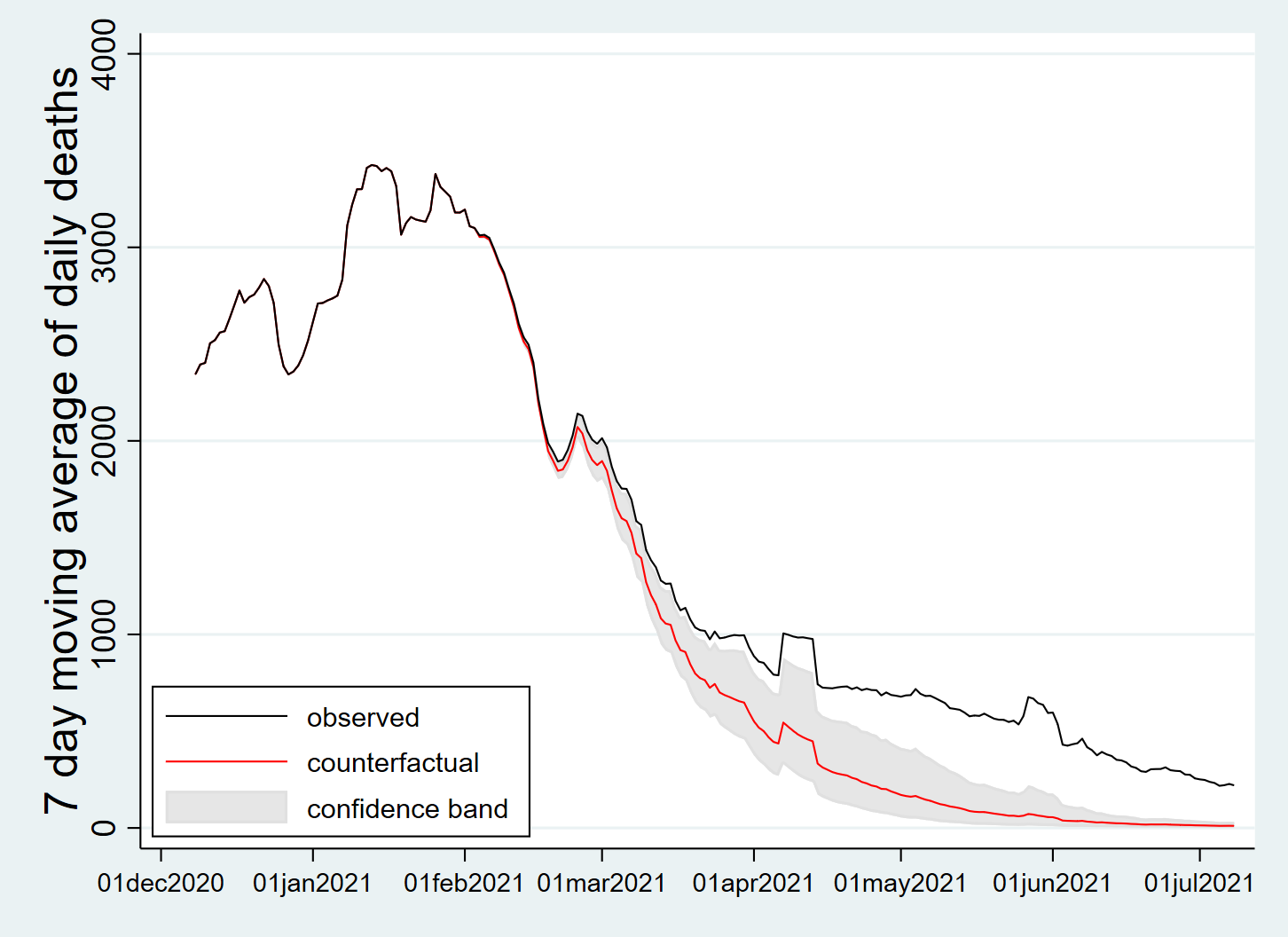}
			\par\end{centering}
		\footnotesize\textbf{Note}: In the first row of panels, black solid lines are actual vaccination rates, and red dotted lines are hypothetical rates we consider in our counterfactual experiments. In the second and fourth row, red lines are changes in the log of weekly case/death counts, respectively. In the third and fifth row of panels, black lines are actual case/death numbers, and red lines are counterfactual counts. Shaded areas are 90\% confidence bands.
	\end{figure}
\end{document}